\documentclass[useAMS,usenatbib,usegraphicx]{mn2e}
\usepackage{times}
\usepackage{float}
\usepackage{graphicx}
\usepackage{pdfpages}
\usepackage{adjustbox}
\usepackage{amsmath}
\usepackage{amssymb}
\usepackage[utf8]{inputenc}
\usepackage{newunicodechar}
\usepackage{caption}

\newcommand{\ms}{M$_{\odot}$}
\newcommand{\rs}{R$_{\odot}$}

\newcommand{\virg}{``}

\title[Binding energy in common envelope simulations]{The effect of binding energy and resolution in simulations of the common envelope binary interaction}
\author[Iaconi et al.]{Roberto Iaconi $^{1,2}$\thanks{E-mail: roberto.iaconi@mq.edu.au}, Orsola De Marco $^{1,2}$, Jean-Claude Passy $^{3,4}$\thanks{Alexander-von-Humboldt fellow} and Jan Staff $^{5}$\\
$^{1}$Department of Physics \& Astronomy, Macquarie University, Sydney, Australia\\
$^{2}$Astronomy, Astrophysics and Astrophotonics Research Centre, Macquarie University, Sydney, Australia\\
$^{3}$Argelander Institute f\"{u}r Astronomie, Bonn Universit\"{a}t, Bonn, Germany\\
$^{4}$Max-Planck-Institut f\"{u}r Intelligente Systeme, T\"{u}bingen, Germany\\
$^{5}$College of Science \& Mathematics, University of the Virgin Islands, USVI, USA}

\begin{document}

\date{}

\pagerange{\pageref{firstpage}--\pageref{lastpage}} \pubyear{\the\year}

\maketitle

\label{firstpage}

\begin{abstract}
The common envelope binary interaction remains one of the least understood phases in the evolution of compact binaries, including those that result in Type Ia supernovae and in mergers that emit detectable gravitational waves. In this work we continue the detailed and systematic analysis of 3D hydrodynamic simulations of the common envelope interaction aimed at understanding the reliability of the results. Our first set of simulations replicate the 5 simulations of Passy et al. (a 0.88~\ms, 90~\rs\ RGB primary with companions in the range 0.1 to 0.9~\ms) using a new AMR gravity solver implemented on our modified version of the hydrodynamic code {\sc Enzo}. Despite smaller final separations obtained, these more resolved simulations do not alter the nature of the conclusions that are drawn. We also carry out 5 identical simulations but with a 2.0~\ms\ primary RGB star with the same core mass as the Passy et al. simulations, isolating the effect of the envelope binding energy. With a more bound envelope all the companions in-spiral faster and deeper though relatively less gas is unbound. Even at the highest resolution, the final separation attained by simulations with a heavier primary is similar to the size of the smoothed potential even if we account for the loss of some angular momentum by the simulation. As a result we suggest that a $\sim 2.0$~\ms\ RGB primary may possibly end in a merger with companions as massive as 0.6~\ms, something that would not be deduced using analytical arguments based on energy conservation.
\end{abstract}

\section{Introduction}
\label{sec:intro}
The common envelope (CE) interaction (\citealt{Paczynski1976}, \citealt{Ivanova2013}) is a binary interaction that leaves behind a compact binary or a stellar merger. After Roche lobe overflow the mass transfer becomes unstable and the companion is engulfed in the envelope of the primary after which the orbital distance is quickly and greatly reduced. The CE interaction is likely the main channel of formation of most compact binaries. Compact binary white dwarfs, neutron stars and black holes have likely gone through one or more CE events during their evolution (but see \citealt{Hirai2017} for a possible alternative scenario). These systems can merge at a later time, generating Type Ia supernovae, gamma ray bursts (\citealt{Fryer2007}) and the emission of detectable gravitational waves \citep{Abbott2016}.

Many observed transients may be due to CE events. \citet{Ivanova2013b} were first to identify the \virg Red Nova'' type transient V1309 Sco (\citealt{Tylenda2011}) as a common envelope merger event. This was followed by two more transients \citep[e.g.,][]{Smith2016,MacLeod2017,Blagorodnova2017}, plausibly identified as CE events. Explaining these transients necessitates a reasonable theoretical description of the CE interaction and CE observations in turn provide a validation of the simulations \citep{Galaviz2017}. This synergy of simulations and observations has only become recently possible with the advent of wide and deep time-resolved surveys (e.g., the Palomar Transient Factory, \citealt{Law2009}, or the Catalina Real time Transient Survey, \citealt{Drake2009}), which have started to detect these fast and elusive events. 

Reconciling observed rates of these phenomena with our theoretical understanding of what causes them is in the hand of population synthesis models (e.g, \citealt{Belczynski2016}, \citealt{Ablimit2016}). These models use prescriptions of how the CE interactions transform binaries into compact systems. 
The prescriptions provide, for example, the post-CE separation reached at the end of the rapid in-spiral. The post-CE orbital separation is a start point for future in-spiral at the hand of different mechanisms and it impacts the time-scale within which the object eventually merges. These prescriptions are mostly {\it ad hoc}. A {\it holy grail} of CE simulations is therefore to provide prescriptions on how final orbital separation depends on system parameters, but simulations have not been sufficiently accurate to do so. 

The CE rapid in-spiral is an intrinsically 3D interaction very difficult to study analytically or with 1D numerical models. Various numerical work have tried to tackle this difficult problem (e.g., \citealt{Terman1994}, \citealt{Sandquist1998}, \citealt{Sandquist2000}, \citealt{Ricker2012}, \citealt{Passy2012}, \citealt{Nandez2015}, \citealt{Nandez2016}, \citealt{Kuruwita2016}, \citealt{Ohlmann2016}, \citealt{Ohlmann2016b}, \citealt{Iaconi2017}) using different numerical techniques (e.g., grid-codes, SPH, unstructured mesh), stellar and orbital setups and including a range of physical ingredients (though by necessity very little physics is included in these simulations).  \citet{Iaconi2017} compared and contextualised different past simulations though the diversity of the techniques and the lack of sufficient coverage of parameter space has made their conclusions partial. Here we continue the effort of comparing simulations by adding to the corpus of simulations that can inherently be compared to each other and that can reveal numerical effects such as those due to resolution, or the effect of using point masses instead of bodies with a given size. These effects plague all simulations and tend, in our opinion, to be under-reported in simulation papers.
 
One of the main problems with CE simulations is that they are by and large unable to unbind the CE (e.g., \citealt{Sandquist1998}, P12,  \citealt{Ohlmann2016}), something that at least some systems in Nature must be able to do since we observe post-CE close binaries. Recently, simulations including the effect of recombination energy of hydrogen and helium have been able to unbind the envelope in some cases \citep{Nandez2016}. 

However, although 1D numerical simulations studying in detail the effect of recombination, such as those performed by \citet{Ivanova2015}, can give us important insights on the energetics and dynamic of the process, they are limited by their spherical geometry (see \citealt{Ivanova2016} for a discussion on the difficulties of connecting 1D and 3D simulations). At the same time, 3D calculations including recombination assume the envelope to be fully opaque, resulting in the entire amount of recombination energy injected into the envelope to be delivered as kinetic energy, while some may possibly leak out on account of the low opacities of recombined hydrogen and provide the observational features of transients such as the \virg Red Novae'' (\citealt{Ivanova2013b}).
Without understanding the fraction of recombination energy that can be used to do work, it is likely that using the entire recombination energy budget overestimates the amount of unbound gas, though not including any of it likely underestimates it (detailed discussions on the distribution, time-scale and amount of energy that can be deployed into different envelopes by recombination can be found in \citealt{Ivanova2015} and \citealt{Clayton2017}).

In this paper we carry out a range of simulations similar to those carried out by P12, but with a more bound envelope. The giant is this time a 2.0~\ms\ RGB star with the same core mass as the 0.88~\ms\ giant used by P12. The envelope is therefore more massive and more compact. While such an envelope could be more difficult to eject, it may also induce the orbit to shrink farther, thereby mining more orbital energy and ejecting more mass. Hence the fraction of unbound envelope for a more bound envelope cannot be easily predicted. We also carry out tests with additional resolutions to perform a resolution study.

This paper is structured as follows: in Section~\ref{sec:hydrodynamics_simulations} we describe the numerical setup. In Sections~\ref{sec:energy_conservation} and \ref{sec:resolution_study} we analyse the conservation properties of our numerical algorithms and the effect of increasing resolution on our simulations, respectively (with further numerical considerations in Appendix~\ref{sec:numerical_caveats}). In Section~\ref{sec:results} we analyse our results, focusing on the evolution of the orbital separation (Section~\ref{ssec:final_separation}) and on the dynamics of the envelope ejection (Section~\ref{ssec:unbound_mass}). In Section~\ref{sec:gravodrag} we discuss the strength of the gravitational friction or drag in our simulations.
Our discussion is presented in Section~\ref{sec:discussion}, where we consider the frequency of mergers inside the common envelope (Section~\ref{ssec:roche_lobe_radii_and_roche_radii}), the binding energy of the envelope at the end of the in-spiral (Section~\ref{ssec:bound_mass}) and the efficiency of the unbinding process (Section~\ref{ssec:unbinding_inefficiency}). Finally, we summarise and conclude in Section~\ref{sec:conclusions}.
\section{Hydrodynamic simulations}
\label{sec:hydrodynamics_simulations}
In this section we will describe how the hydrodynamic simulations of this work are setup and some caveats related to the numerics involved.

\subsection{Setup}
\label{ssec:simulations_setup}
All simulations are carried out with our modified version of the multi-dimensional hydrodynamics and $N$-body grid-based code {\sc Enzo} (\citealt{OShea2004}, \citealt{Bryan2014}). For details on the numerical technique and the equations solved in our particular version of the code, see P12 and \citet{Passy2014}. We use adaptive mesh refinement (AMR) with different levels of refinement; cells are split by a factor two along each dimension, according to the density in the cell.
  
We use two different 1D stellar models. The first is exactly the same as the one used by P12 and \citet{Iaconi2017}: a star with a zero-age main sequence mass of 1~\ms, evolved to the red giant branch (RGB) with the stellar evolution code {\sc evol} (\citealt{Herwig2000}) until its core mass reached 0.39~\ms, with a total mass of 0.88~\ms \ and a radius of 83~\rs. The second model is a 2~\ms \ zero-age main sequence star, modelled using the stellar evolution code {\sc mesa}  (\citealt{Paxton2011}, \citealt{Paxton2013}, \citealt{Paxton2015}). We stop the evolution on the RGB once the core has reached a mass of 0.40~\ms, an almost identical core mass as for the first model. At this point, the RGB star has a total mass of 2.0~\ms \ and a radius of 66~\rs. 
\begin{figure*}
\centering     
\includegraphics[scale=0.35, trim=0.5cm 0.9cm 0.0cm 0.0cm, clip]{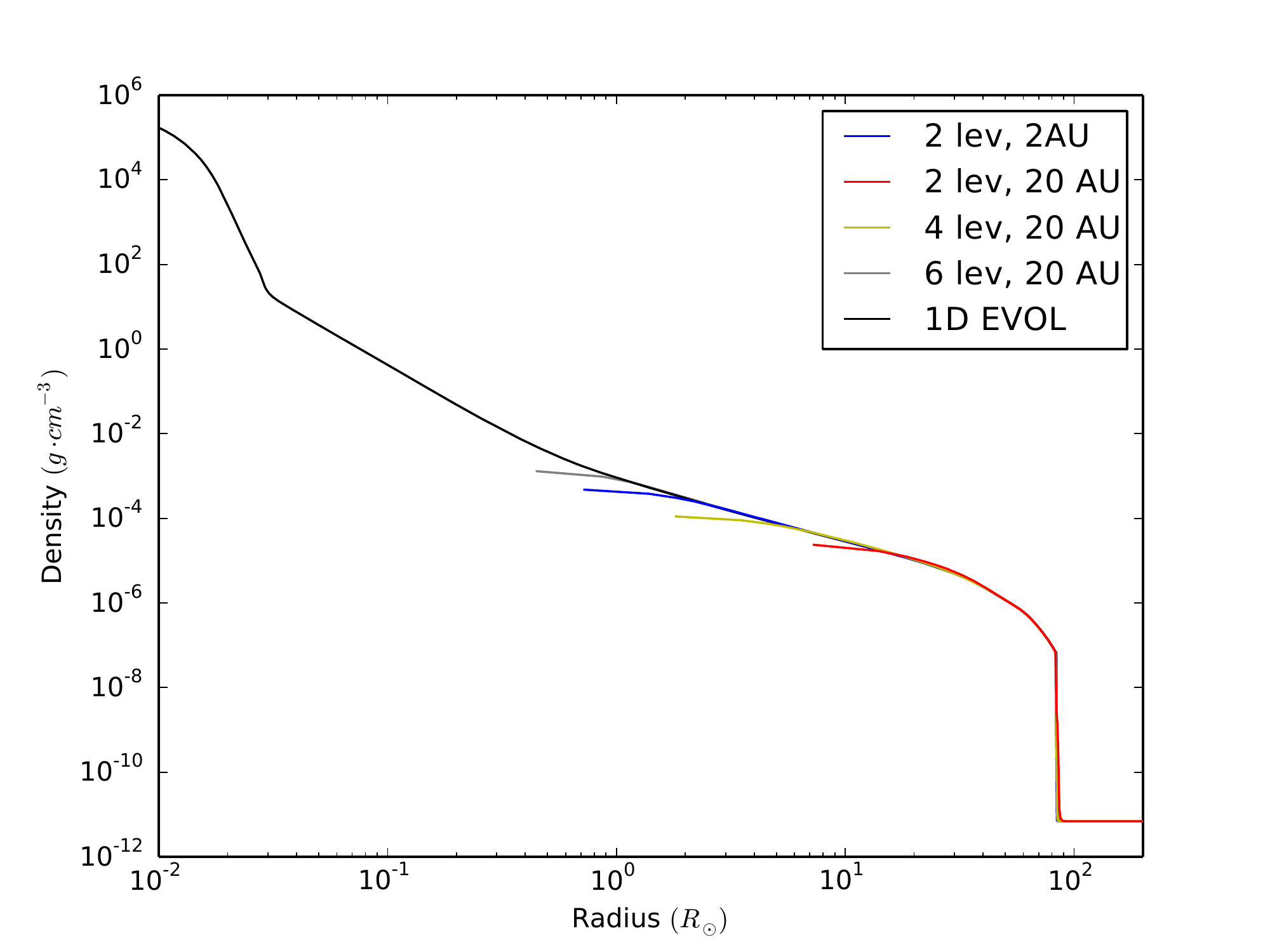}
\includegraphics[scale=0.35, trim=0.5cm 0.9cm 0.0cm 0.0cm, clip]{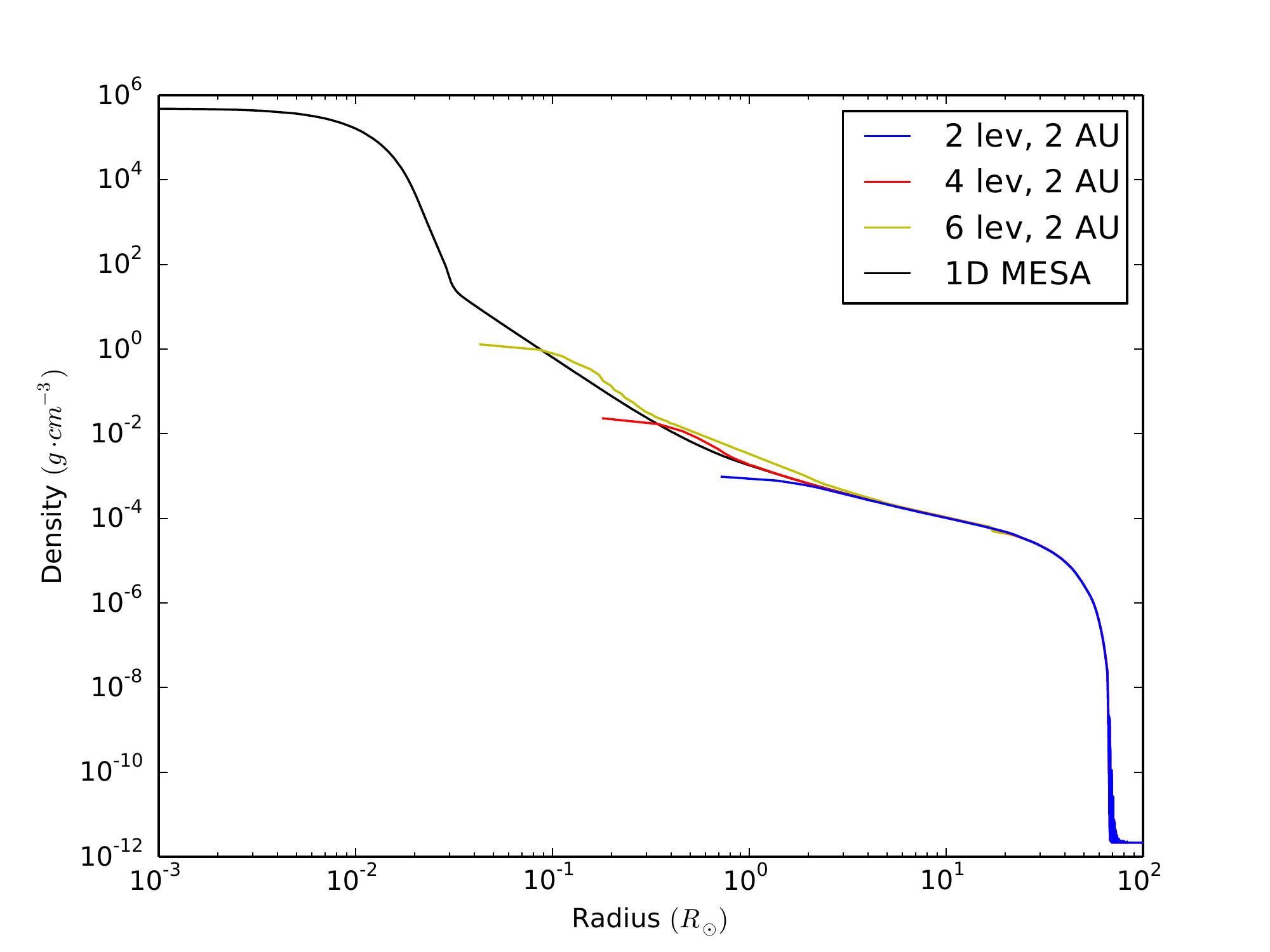}
\includegraphics[scale=0.35, trim=0.5cm 0.4cm 0.0cm 0.0cm, clip]{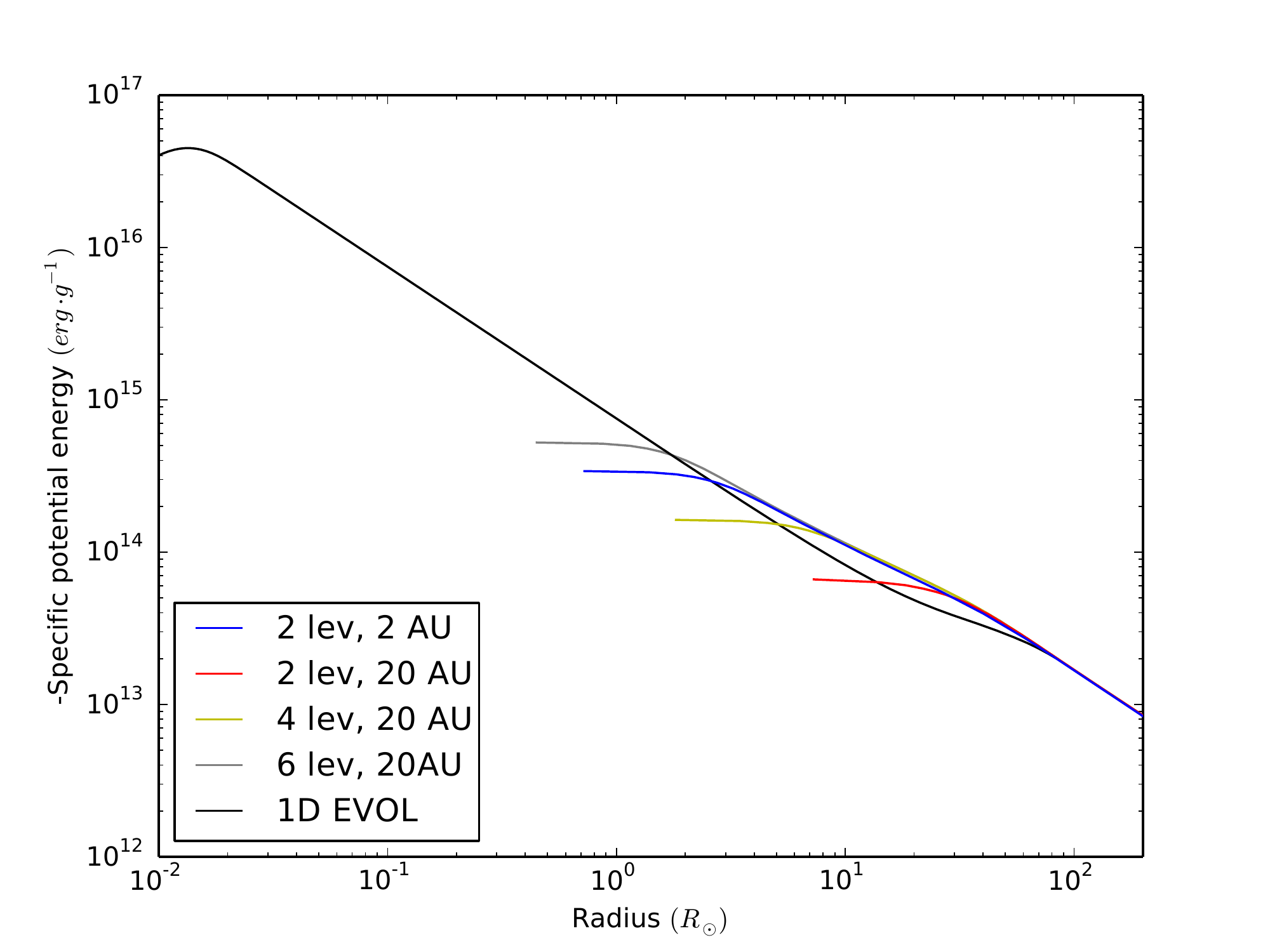}
\includegraphics[scale=0.35, trim=0.5cm 0.4cm 0.0cm 0.0cm, clip]{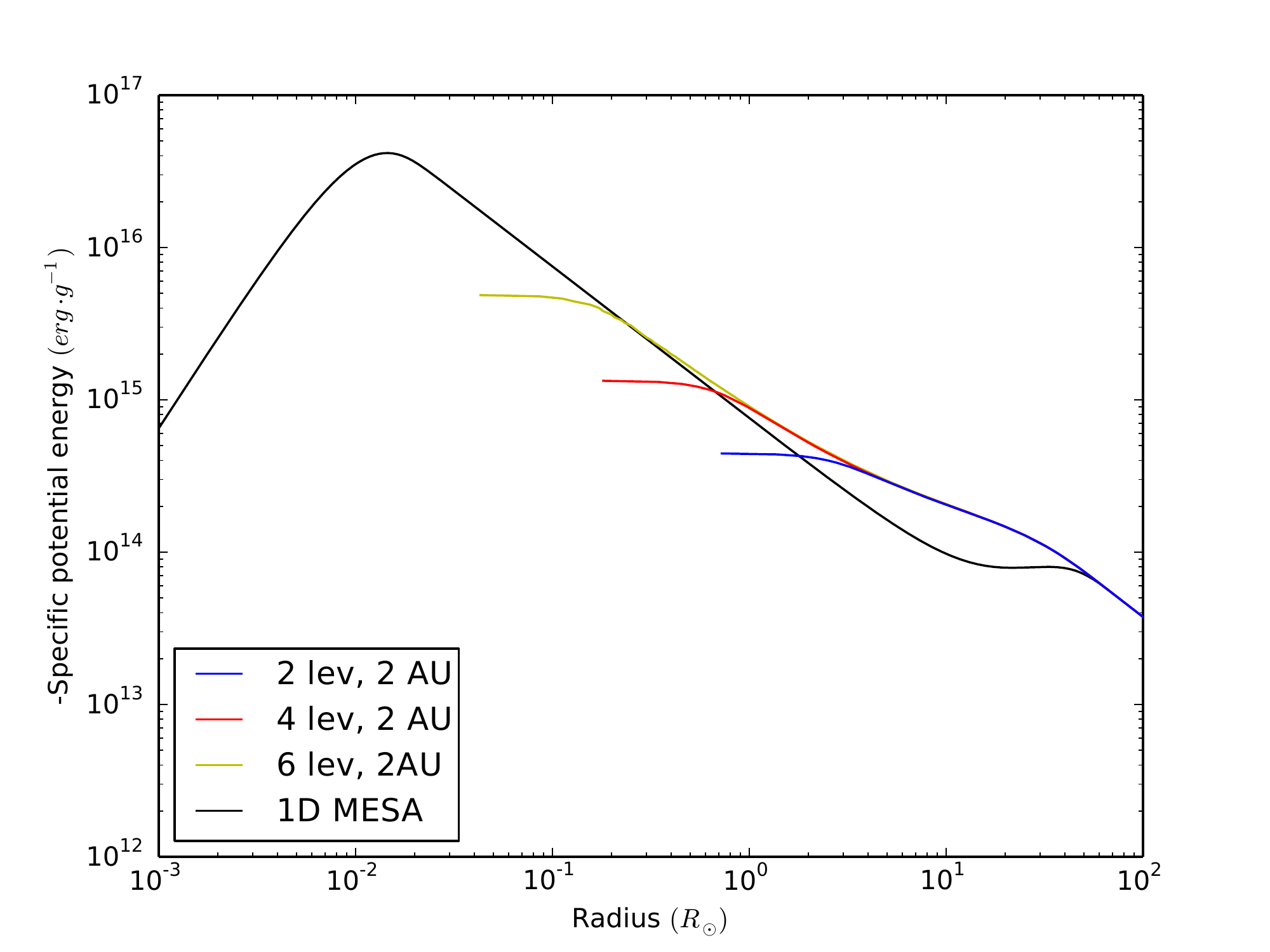}
\includegraphics[scale=0.35, trim=0.5cm 0.0cm 0.0cm 0.0cm, clip]{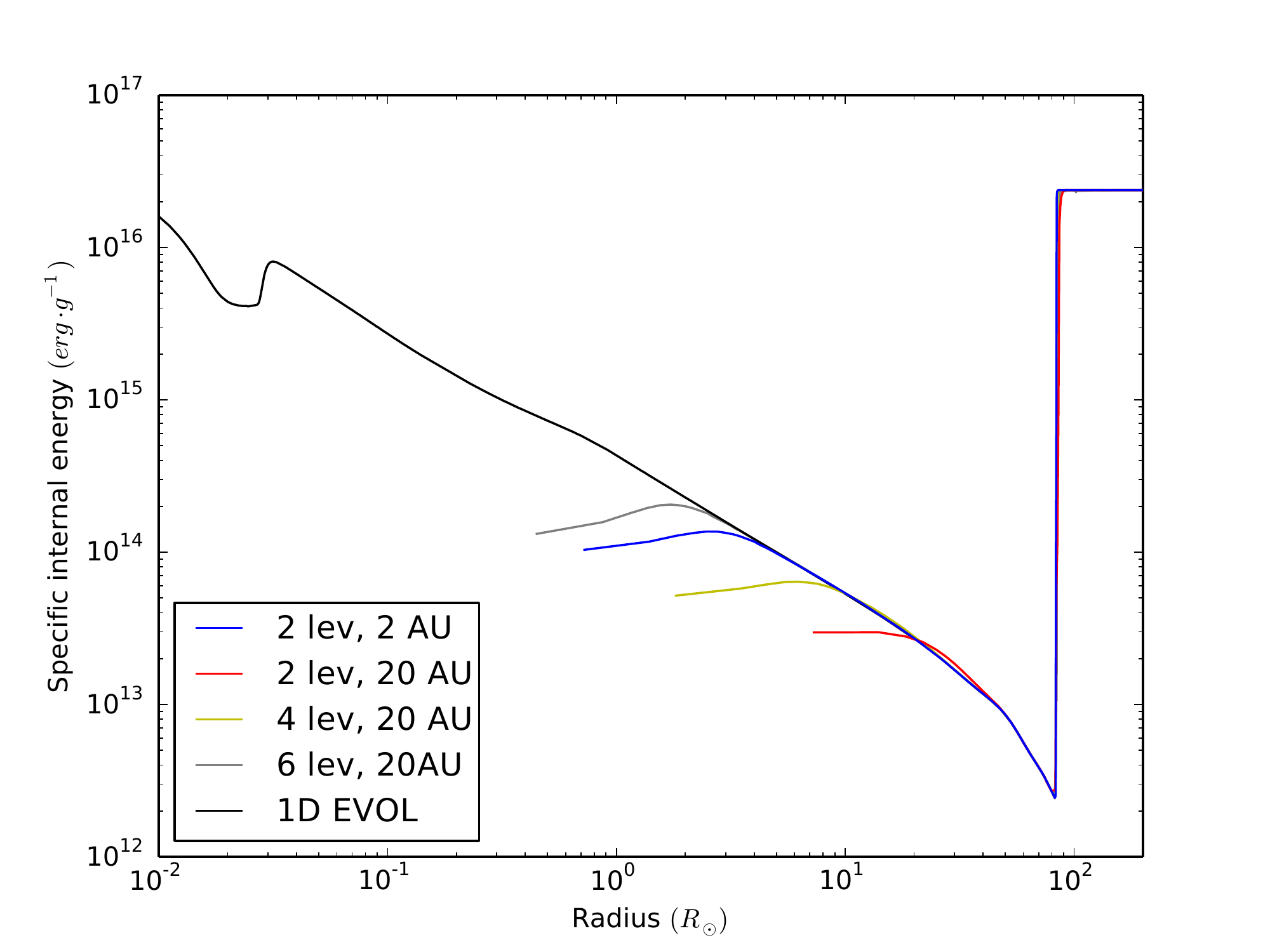}
\includegraphics[scale=0.35, trim=0.5cm 0.0cm 0.0cm 0.0cm, clip]{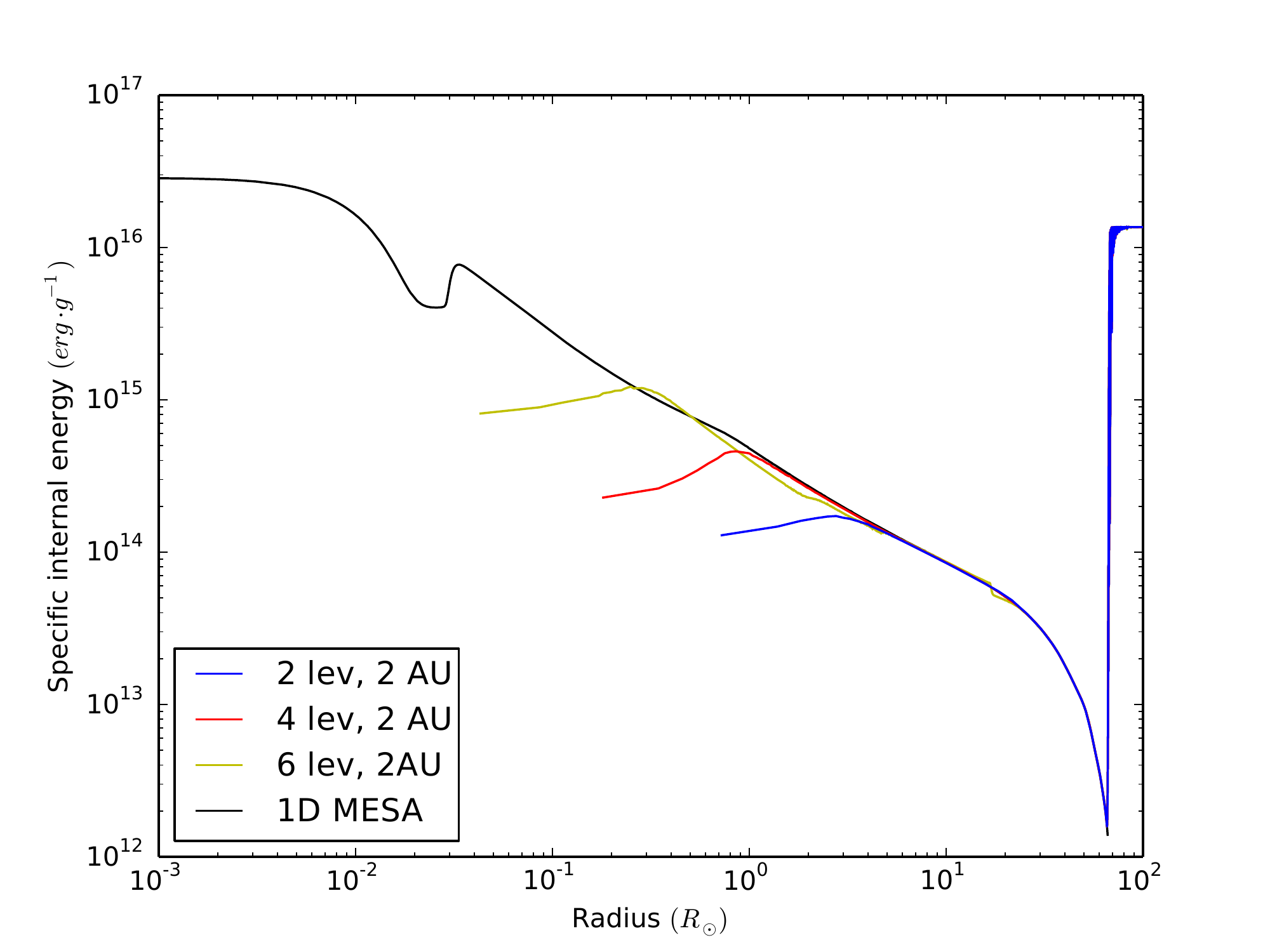}
\centering
\caption{\protect\footnotesize{{Upper left panel:} density profiles for the 1D {\sc evol} model of the 0.88~\ms \ giant, SIM1-SIM5, SIM13, SIM14 and SIM15. {Upper right panel:} density profiles for the {\sc mesa} model of the 2.0~\ms \ giant, SIM6-SIM10, SIM11 and SIM12. Middle left panel: modulus of the specific potential energy for the 0.88~\ms\ simulations. Middle right panel: modulus of the specific potential energy for the 2.0~\ms\ simulations. Lower left panel: specific internal energy for the 0.88~\ms\ simulations. Lower right panel: specific internal energy for the 2.0~\ms\ simulations.}}
\label{fig:star_profiles}
\end{figure*}

The primary stars have been modelled in 1D, mapped and stabilised in the {\sc Enzo} grid following the same procedure used by P12 and \citet{Iaconi2017}. In Figure~\ref{fig:star_profiles} we show the result of the mapping. More discussion on the mis-match between 1D and 3D resolution is carried out in Appendix~\ref{sec:numerical_caveats}.
To keep our models stable in the 3D computational domain, we fill the computational domain outside the primary star with a high-pressure (to keep the outer layers of the giant from outflowing), low-density (not to affect the gravitational layout and to minimise its effect on the dynamics) gas, which we will refer to as {\it vacuum}. This {\it vacuum} is set up to have the same pressure as the outer layers of the star and a constant density equal to $10^{-4}$ times the density of the outermost layer of the primary and the same pressure. Since the code uses an ideal gas adiabatic equation of state, this results in the gas having a very high temperature. While this is a common practice of grid simulations of the CE interaction \citep[e.g.,][]{Sandquist1998}, it is fair to question the validity of this setup. To first approximation we have already evaluated this setup not to have a substantial effect on the results by carrying out comparison simulations with a smooth particle hydrodynamics code \citep[P12 and][]{Iaconi2017}, which does not have the {\it vacuum}. On the other hand the {\it vacuum} is far more problematic when evaluating the thermodynamic and emissivity properties of the gas \citep[for more details see][]{Galaviz2017}. For additional details on this setup in Appendix~\ref{sec:numerical_caveats}.

For all our adopted resolutions we ensure that the primary is stable for at least 100 dynamical times inside our computational domain, where the dynamical times of the two envelopes are $\simeq 21$~days and $\simeq 7$~days, for the 0.88~\ms \ and the 2.0~\ms \ models, respectively. Since the CE simulations carried out here have a maximum duration of $\simeq 300$~days, this guarantees that our CE simulations are not affected by numerical instabilities.
Additionally, as suggested by \citet{Staff2016a}, we use a smoothing length equal to three times the smallest cell length in all our simulations, to achieve acceptable levels of energy conservation.

In Table~\ref{tab:simulation_parameters} we summarise the most relevant initial and final parameters of the simulations, where $M_1$ is the mass of the primary, $M_\mathrm{c}$ is the mass of the primary's core, $R_1$ is the radius of the primary, $M_2$ is the mass of the companion, $a_\mathrm{i}$ is the initial separation between the core of the primary and the companion and $a_\mathrm{f}$ is the final separation. We have carried out 4 sets of simulations:

\begin{enumerate}
\item  SIM1 to SIM5 reproduce those of P12, except for the fact that in this case we use AMR and smoothing length of 3 times the smallest cell length (see above) instead of 1.5. In this way the smoothing length has the same physical size in both simulations, namely 2.53~\rs. This set of simulations is used both to test the effect of AMR on the results of P12 and to provide a direct comparison with our set of simulations carried out with a more massive primary. 
\item SIM6 to SIM10 aim to study the effect of a more compact and massive envelope on the outcome of the CE interaction. We use the 2.0~\ms \ star in an initial setup similar to SIM1-SIM5, with the same five companion masses, placed on the primary's surface and in circular orbit at the beginning of the simulations. No rotation has been given to the primary. 
\item SIM11 and SIM12 are similar in all aspects to SIM9, except that they are carried out with additional AMR levels: 4 and 6, respectively. These simulations serve as a resolution test. We note that the minimum cell size of 0.05~\rs\ in SIM12 is just larger than the core size, though it is smaller than the presumed companion radius if the companion is a main sequence star ($R_{\rm 2}=0.66$~\ms\ for $M_{\rm 2}=0.6$~\ms). Even the smoothed potential radius, at 0.15~\rs, is smaller than the main sequence radius of the companion. This means that the flow around the companion is relatively well reproduced.
\item Finally, for SIM13, SIM14 and SIM15 we use the same physical setup of SIM4, but a larger simulation domain of 20~AU per side to contain all the expanding envelope for the entire time of the CE interaction. 
We also use periodic boundary conditions, so that no gas can escape the computational domain. 
Due to the presence of a large amount {\it vacuum}, these simulations are {\it exclusively} used to consider global energy and angular momentum conservation properties (see Appendix~\ref{sec:numerical_caveats}).
\end{enumerate}

\begin{table*}
\begin{center}
\begin{adjustbox}{max width=\textwidth}
\begin{tabular}{ccccccccccccccc}
\hline
ID & $M_1$ & $M_\mathrm{c}$ & $R_1$ & $M_2$ & $a_\mathrm{i}$ & Domain size & Top grid & Levels of & Min. cell & Boundary & $a_\mathrm{f}$ & Mass inside the  & Mass in the & Unbound \\
&&&&&&& resolution & refinement & size & conditions & & computational & original volume & mass \\
&&&&&&&&&&&& domain & of the primary &\\
& (\ms) & (\ms) & (\rs) & (\ms) & (\rs) & (\rs) & (cells/side) & (\#) & (\rs) &  & (\rs) & (\%) & (\%) & (\%) \\
\hline
SIM1 & 0.88 & 0.39 & 83 & 0.1 & 83  & 431 & 128 & 2 & 0.84 & outflow & 3.1 & 94 & 55 & 3 \\
SIM2 & 0.88 & 0.39 & 83 & 0.15 & 83 & 431 & 128 & 2 & 0.84 & outflow & 3.4 & 86 & 35 & 4  \\
SIM3 & 0.88 & 0.39 & 83 & 0.3 & 83  & 431 & 128 & 2 & 0.84 & outflow & 7.3 & 72 & 18 & 7 \\
SIM4 & 0.88 & 0.39 & 83 & 0.6 & 83  & 431 & 128 & 2 & 0.84 & outflow & 12 & 59 & 12 & 10 \\
SIM5 & 0.88 & 0.39 & 83 & 0.9 & 83  & 431 & 128 & 2 & 0.84 & outflow & 17 & 49 & 6 & 9 \\
\hline                                
SIM6 & 2.0 & 0.39 & 66 & 0.1 & 66  & 431 & 128 & 2 & 0.84 & outflow & 2.2 & 98 & 88 & 1 \\
SIM7 & 2.0 & 0.39 & 66 & 0.15 & 66 & 431 & 128 & 2 & 0.84 & outflow & 1.8 & 97 & 87 & 2 \\
SIM8 & 2.0 & 0.39 & 66 & 0.3 & 66  & 431 & 128 & 2 & 0.84 & outflow & 1.6 & 94 & 70 & 4 \\
SIM9 & 2.0 & 0.39 & 66 & 0.6 & 66  & 431 & 128 & 2 & 0.84 & outflow & 1.8 & 81 & 47 & 10 \\
SIM10 & 2.0 & 0.39 & 66 & 0.9 & 66 & 431 & 128 & 2 & 0.84 & outflow & 2.8 & 78 & 13 & 10 \\
\hline
SIM11 & 2.0 & 0.39 & 66 & 0.6 & 66 & 431 & 128 & 4 & 0.21 & outflow & 0.67 & 52 & 4 & 15 \\
SIM12$^1$ & 2.0 & 0.39 & 66 & 0.6  & 66 & 431 & 128 & 6 & 0.05 & outflow & 0.87 & 84 & 28 & 13 \\
\hline
SIM13 & 0.88 & 0.39 & 83 & 0.6 & 83 & 4310 & 128 & 2 & 8.4 & periodic & 21 & 100 & 6 & N/A$^2$\\
SIM14 & 0.88 & 0.39 & 83 & 0.6 & 83 & 4310 & 128 & 4 & 2.1 & periodic & 11 & 100 & 10 & N/A$^2$ \\
SIM15 & 0.88 & 0.39 & 83 & 0.6 & 83 & 4310 & 128 & 6 & 0.5 & periodic & 10 & 100 & 12 & N/A$^2$ \\
\hline
\end{tabular}
\end{adjustbox}
\vspace{-0.3cm}
\flushleft \scriptsize{
$^1$ The simulation has been run for a shorter time with respect to the others due to its computational cost.\\
$^2$ Due to the larger computational domain, the expanding layers of the envelope, with a decreasing density, are in contact with the {\it vacuum} for a much longer time and heated up, something that unbinds more gas. Therefore, to avoid misleading numbers, we do not report the amount of mass unbound.\\}
\end{center}
 \begin{quote}
  \caption{\protect\footnotesize{Initial parameters and final results of the simulations performed for this publication. Masses at the end of the simulations are expressed as a percentage of the initial envelope mass.}} \label{tab:simulation_parameters}
\end{quote}
\end{table*} 
\section{Energy and angular momentum conservation}
\label{sec:energy_conservation}

With mass flowing out of the computational domain it is difficult to check the conservation properties of our simulations. Various expedients have been devised by \citet{Iaconi2017}, from using a larger domain to accounting approximately for lost mass and energy by measuring the flow of mass and energy through the domain boundary.

\begin{figure}
\centering     
\includegraphics[scale=0.4, trim=0.5cm 0.0cm 0.0cm 0.0cm, clip]{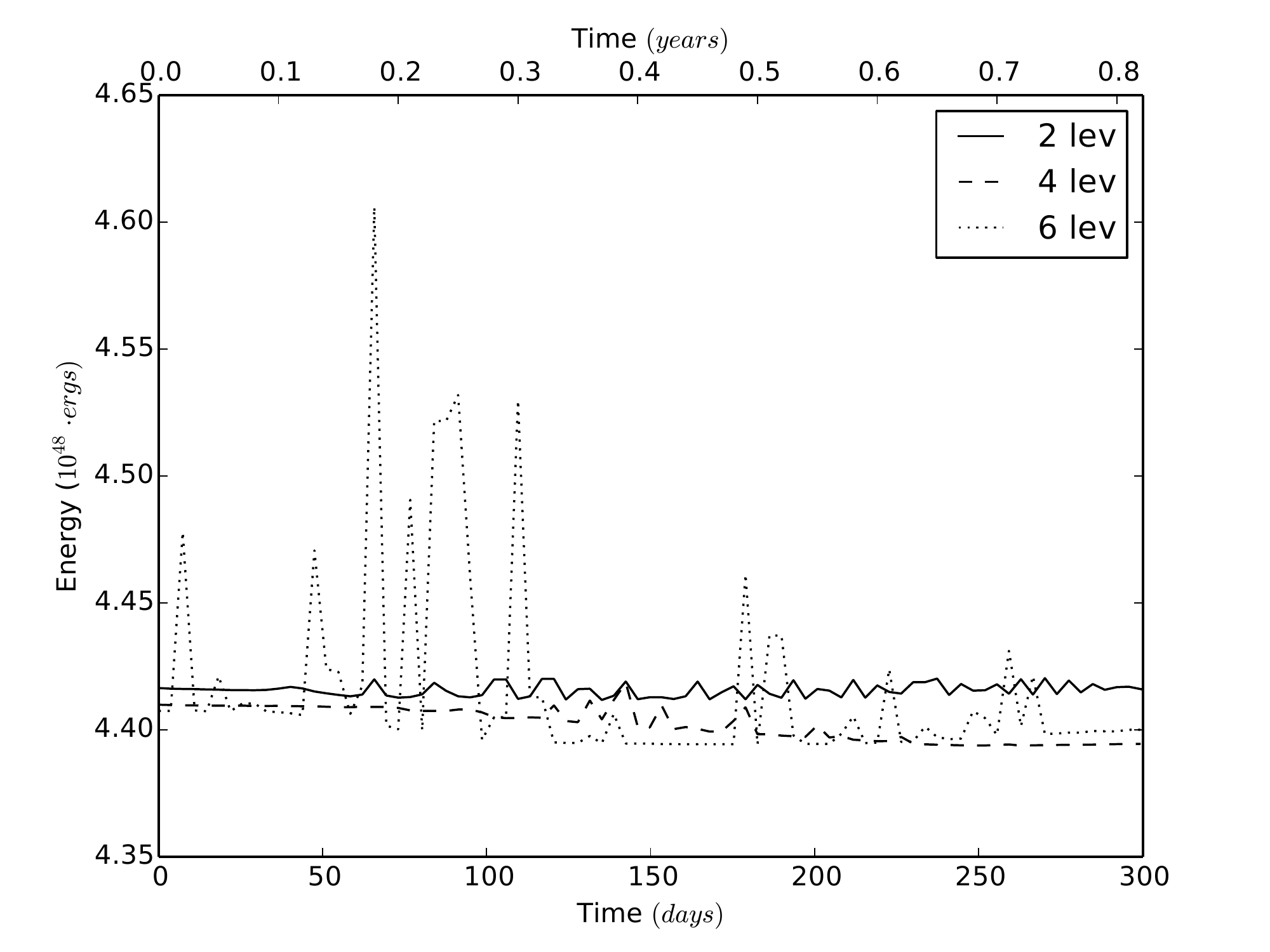}
\centering
\includegraphics[scale=0.4, trim=0.5cm 0.0cm 0.0cm 0.0cm, clip]{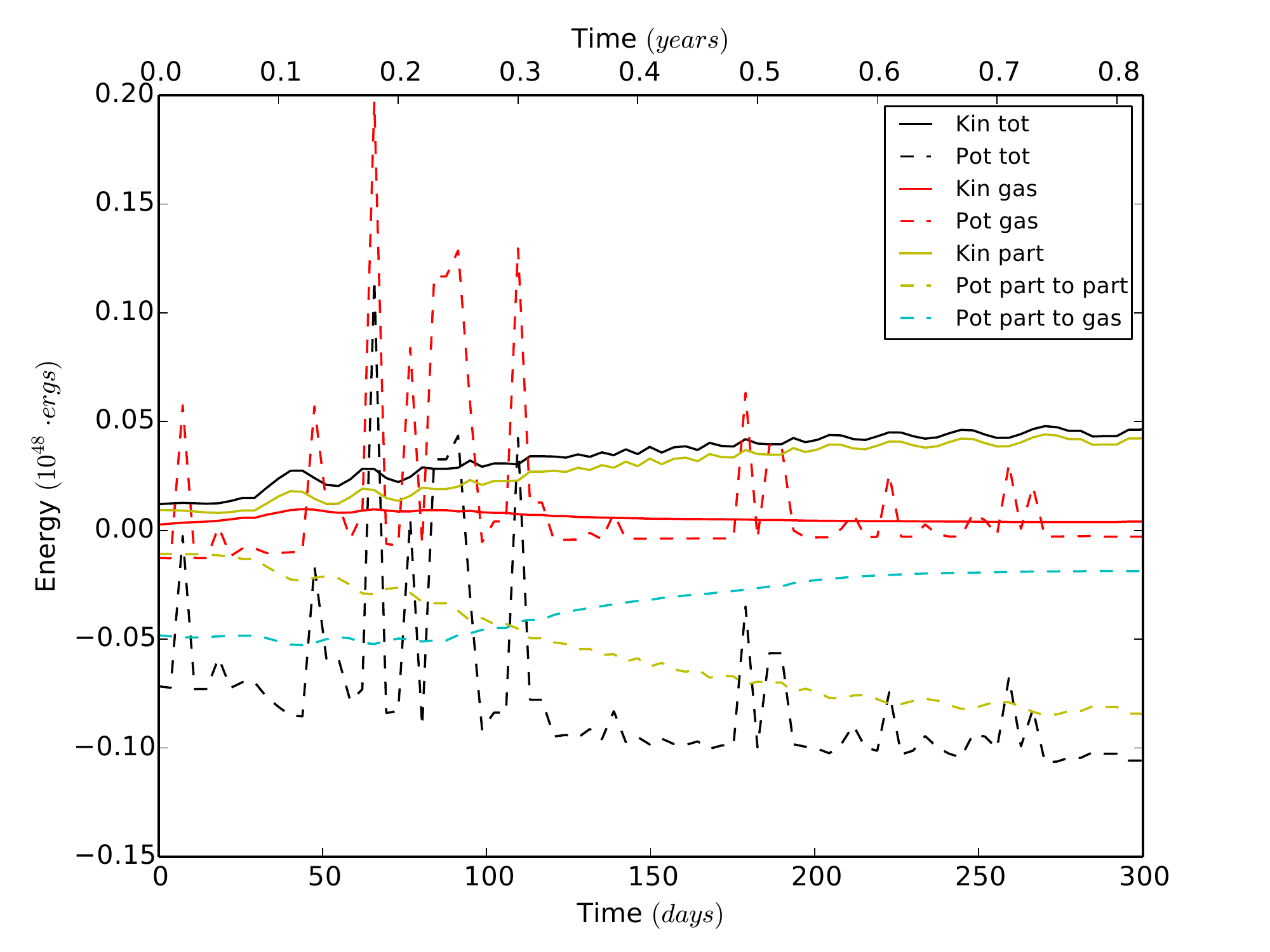}
\caption{\protect\footnotesize{{Upper panel:} Total energy as a function of time for SIM13-SIM15. {Lower panel:} components of the energy as a function of time for SIM15: \virg Kin tot'' is the total kinetic energy, \virg Pot tot'' is the total potential energy, \virg Kin gas'' is the kinetic energy of the gas, \virg Pot gas'' is the potential energy of the gas , \virg Kin part'' is the kinetic energy of the point-mass particles representing primary's core and companion, \virg Pot part to part'' is the point-mass to point-mass potential energy, \virg Pot part to gas'' is the point-masses to gas potential energy. Note that total energy, total thermal energy and gas thermal energy are not plotted as the gas thermal energy dominates over all the other components due to the presence of a large amount of {\it vacuum}. Its value is constant at $\sim 4.47 \times 10^{48}$~erg for the entire simulation. The ``spikes" in the potential energy of SIM15 are discussed in Section~3.}}
\label{fig:energy_conservation}
\end{figure}

In this work, as explained in Section~\ref{sec:hydrodynamics_simulations}, we carried out SIM13-SIM15, using the ``light" primary star, with a simulation domain able to contain the envelope for the entire time of the rapid in-spiral and with periodic boundary conditions (see Table~\ref{tab:simulation_parameters}), so that even small outflows of {\it vacuum} gas (low mass but high thermal energy) can be accounted for. While, as we discuss in Appendix~\ref{sec:numerical_caveats} these simulations may have their dynamics somewhat affected by the large amount of {\it vacuum}, we use them here exclusively to evaluate energy conservation. The results for the conservation of energy are shown in Figure~\ref{fig:energy_conservation}, where in the upper panel we compare the total energy as a function of time for the three simulations, while in the lower panel we show the various components of the energy for SIM15, that has a resolution similar to that of the two main sets of simulations (SIM1-SIM5 and SIM6-SIM10). 

We find that energy is reasonably conserved on the time-scale of the interaction for all the refinement levels used. 
The three simulations show on average a slow decrease of the total energy over time, the values of the total energy at 300~days differ by 0.1\%, 0.4\% and 0.2\%, for SIM13, SIM14 and SIM15, respectively, from the initial values of the total energy ($4.4 \times 10^{48}$~erg). If we consider instead the initial binding energy of the star as a representative reference energy, then the decline is  $3.8\%$, $15\%$ and $7.5\%$ from the initial values of the binding energies of the primaries for SIM13, SIM14 and SIM15, respectively (the initial binding energies of the primaries, as reported in Table~\ref{tab:energy_initial_parameters} - sum of the first two columns, are $-1.59 \times 10^{46}$~erg, $-2.24 \times 10^{46}$~erg and $-2.49 \times 10^{46}$~erg for SIM13, SIM14 and SIM15, respectively).

We also observe fluctuations in the total energy during the three simulations, significant in the case of SIM15. 
From the lower panel of Figure~\ref{fig:energy_conservation} we observe that these fluctuations are driven by the potential energy of the gas.
These fluctuations are the result of approximations when AMR grids are rearranged between top-grid time-steps and are always contained inside a single cell. 
We are confident that these single cell excursions do not affect the overall behaviour of the simulation because we repeated part of SIM15 using slightly different setups and while the number and location of the cells with energy excursions changed, the overall behaviour of the simulation were identical.

\begin{figure}
\centering     
\includegraphics[scale=0.4, trim=0.5cm 0.0cm 0.0cm 0.0cm, clip]{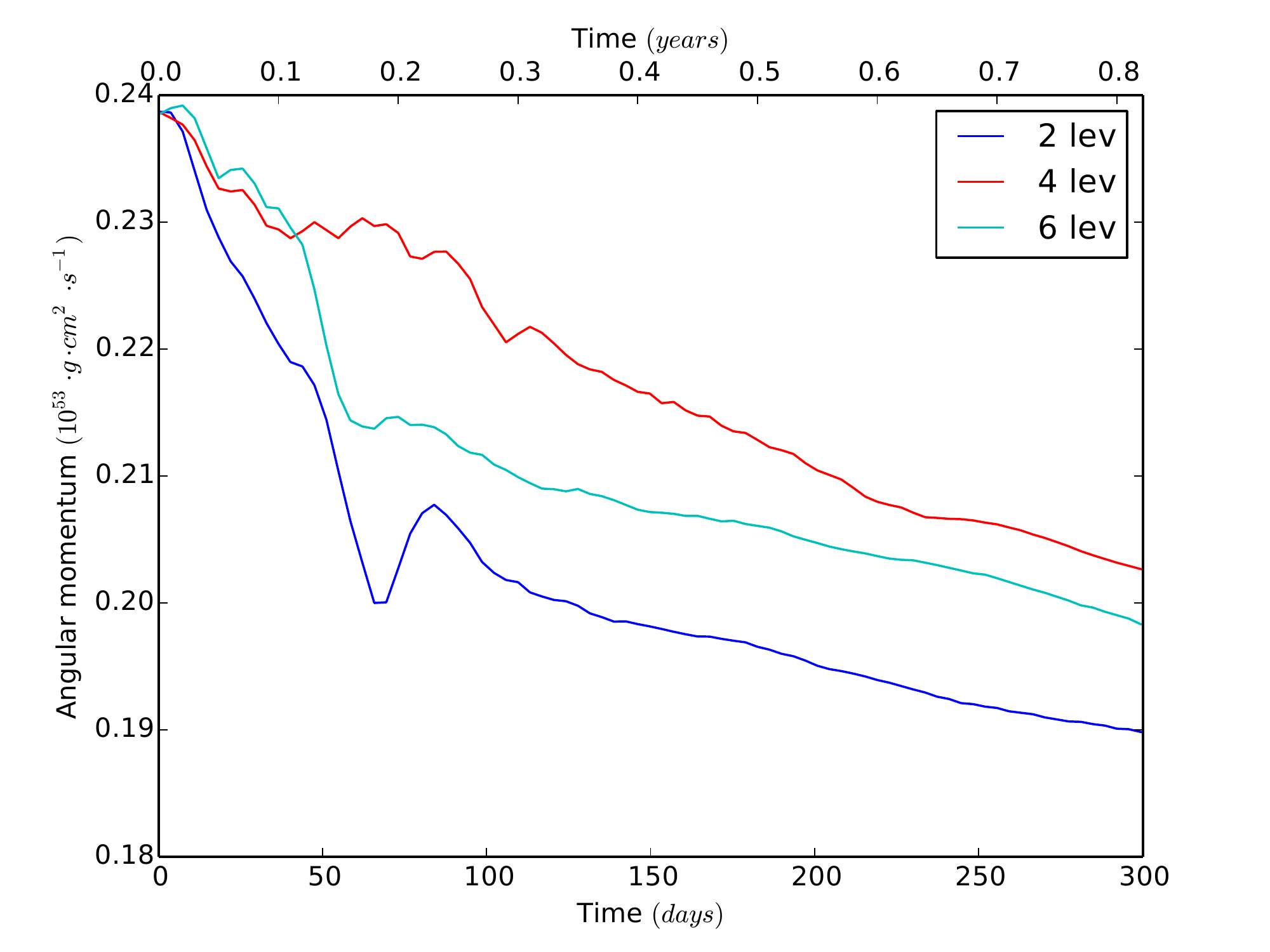}
\centering     
\includegraphics[scale=0.4, trim=0.5cm 0.0cm 0.0cm 0.0cm, clip]{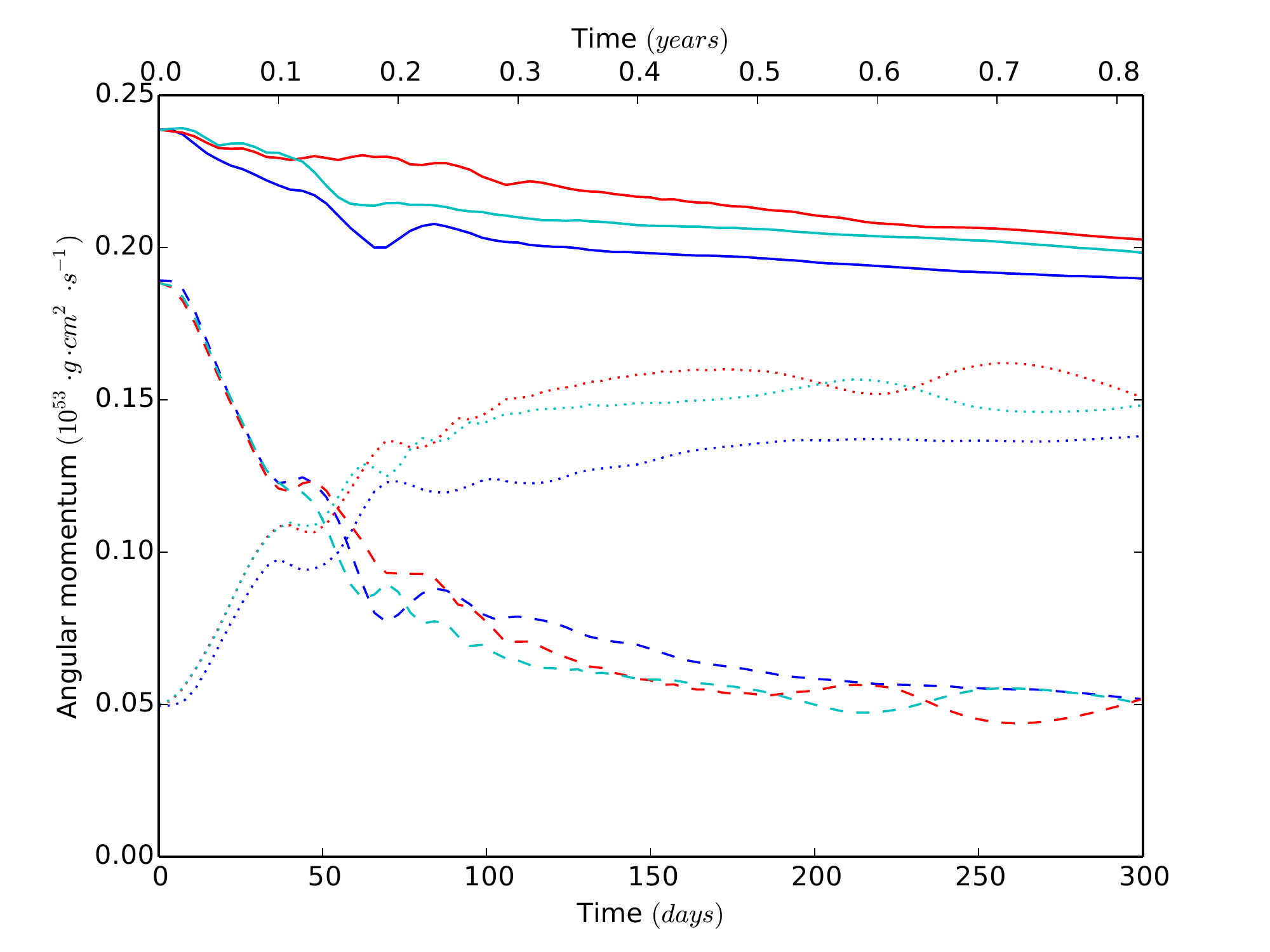}
\caption{\protect\footnotesize{{Upper panel:} total angular momentum along $z$ with respect to the centre of mass of the system as a function of time for SIM13 (blue), SIM14 (red) and SIM15 (cyan). {Lower panel:} components of the angular momentum as a function of time for SIM13, SIM14 and SIM15: gas (dotted line), particles (dashed line), total (solid line). Colours correspond to those used in the upper panel.}}
\label{fig:angular_momentum_conservation}
\end{figure}

In Figure~\ref{fig:angular_momentum_conservation} we plot the the angular momentum along the $z$ axis for SIM13-SIM15 (upper panel) and the various components of the $z$ angular momentum for SIM15 (lower panel). Angular momentum components along the $x$ and $y$ axes have negligible values. Note that adopting periodic boundary conditions does not ensure angular momentum conservation: gas exiting the computational domain re-enters it from the opposite side, but with the same velocity vector, therefore with a different angle between radius and velocity vectors. Nevertheless, both mass and velocities of the gas leaving and re-entering the domain are negligible. Therefore, this does not affect the estimate of the total angular momentum we carry out below.

We find that, on the time-scale of the interaction angular momentum is conserved to $21\%$, $15\%$ and $17\%$ for SIM13, SIM14 and SIM15, respectively. Where the initial value for the $z$ angular momentum measured in the computational domain is $\simeq 2.4 \times 10^{52}$~g~cm$^2$~s$^{-1}$.
This level of angular momentum conservation is not satisfactory and may impact the results to an extent. A loss of angular momentum may lead to more compact orbits or more bound gas and it may also cause second order effects like a stronger gravitational drag. The issue of angular momentum conservation in grid codes remains a pressing one. To control angular momentum non-conservation effects we have run SPH simulations alongside grid simulations (see for example P12 and \citealt{Iaconi2017}). SPH simulations conserve angular momentum to a far greater level of precision, usually below 1\%. We have never observed large differences in simulation outputs between identical grid and SPH simulations that could be ascribed to a lesser angular momentum conservation observed in grid simulations (usually at the 10\% level). Therefore, we doubt that the conclusions of this paper would be significantly altered by the issue of the angular momentum non-conservation.

The more resolved simulation (SIM15) conserves angular momentum slightly less than the intermediate resolution one (SIM14), breaking the expected monotonic behaviour whereby more resolved simulations conserve angularfmomentum better. The behaviour of these simulations is complex and just {\it where} in the simulation domain the non conservation happens is difficult to track. Looking at Figure~\ref{fig:angular_momentum_conservation}, it is clear that the least resolved simulation (SIM13) starts to lose angular momentum right away, and far faster than the other two (SIM14 and SIM15), which are instead similar to one other. Then SIM15 has its closest periastron passage yet, with a subsequent dip in the total angular momentum possibly due to a locally stronger interaction. SIM14 has a periastron passage slightly later and its total angular momentum dips, too at that time. On the other hand, the intermediate resolution simulation (SIM14) has a more circular orbit, with smaller changes in the environment experienced by the companion and seems, as a result, to avoid conservation dips. The dips scramble the expected order of overall conservation (least resolved should conserve worse) and make it a non monotonic progression. Ultimately a parallel SPH simulation should be carried out to control for non-conservation of angular momentum.

The degree of angular momentum non-conservation in the current simulations is only slightly worse than what is observed in previous simulations, including with other codes: \citet{Sandquist1998} measured a non-conservation of angular momentum at the 10\% level over the 800 days of their simulation. Interestingly most of the non-conservation happened for them over the early part of their simulation, where the in-spiral takes place.  In our simulations instead, the total angular momentum has a steady decline. We ascribe this worse-than-usual behaviour to the AMR aspect of the simulation, but we defer an investigation to future work.
\section{Resolution study}
\label{sec:resolution_study}
A common problem of CE numerical simulations is the lack of convincing resolution tests. This is brought about by the enormous computational expense of these simulations. Ideally we would carry out at least three simulations with spatial resolution increasing by approximately a factor of two in each of three dimensions. This is not a hard factor, but with smaller increases may fail to notice a difference in the output.
 
A second difficulty is the fact that resolution is not the only aspect that can alter the final outputs of the simulations. For example, the smoothing length\footnote{The smoothing length refers to the size scale by which we smooth the point mass potential to prevent it from reaching arbitrarily large values. This is different from the smoothing length in SPH simulations. When SPH simulations utilise point masses, as those carried out by Iaconi et al. (2017) do, the potential smoothing length is often called the ``softening length".} of our point masses can also alter the outcome of a simulation. A third issue is that sometimes one of the simulation's outputs (e.g., final separations) may be unchanged by increasing resolution, while another (e.g., unbound mass) may vary. 

Finally, as discussed further in Appendix \ref{ssec:resolution_testing}, the initial binding energy of the primaries increases with increasing resolution.
As a result our resolution test results will depend on more than just resolution. This means that our resolution test is not a convergence test. Nevertheless, it is essential to evaluate the behaviour of the code at increasing resolution. Therefore, we perform a resolution study by carrying out two additional simulation each with one additional level of AMR refinement. Each successive level has an improvement of a factor of 2 in linear resolution.

We start by noticing that the final separation in SIM13-15, which have progressively better resolution, seems to level off at 10~\rs. SIM4 is not identical to SIM13-15 because of the different boundary conditions, but it is similar. Its final separation (12~\rs) is only slightly larger than for SIM14 (11~\rs) and SIM15 (10~\rs), while its resolution is intermediate (0.84~\rs vs. 2.1 and 0.5~\rs, respectively). We therefore conclude that, even increasing the resolution, for the lighter of the two primaries (0.88~\ms) the final separation would not decrease.

\begin{figure}
\centering     
\includegraphics[scale=0.4, trim=0.5cm 0.0cm 0.0cm 0.0cm, clip]{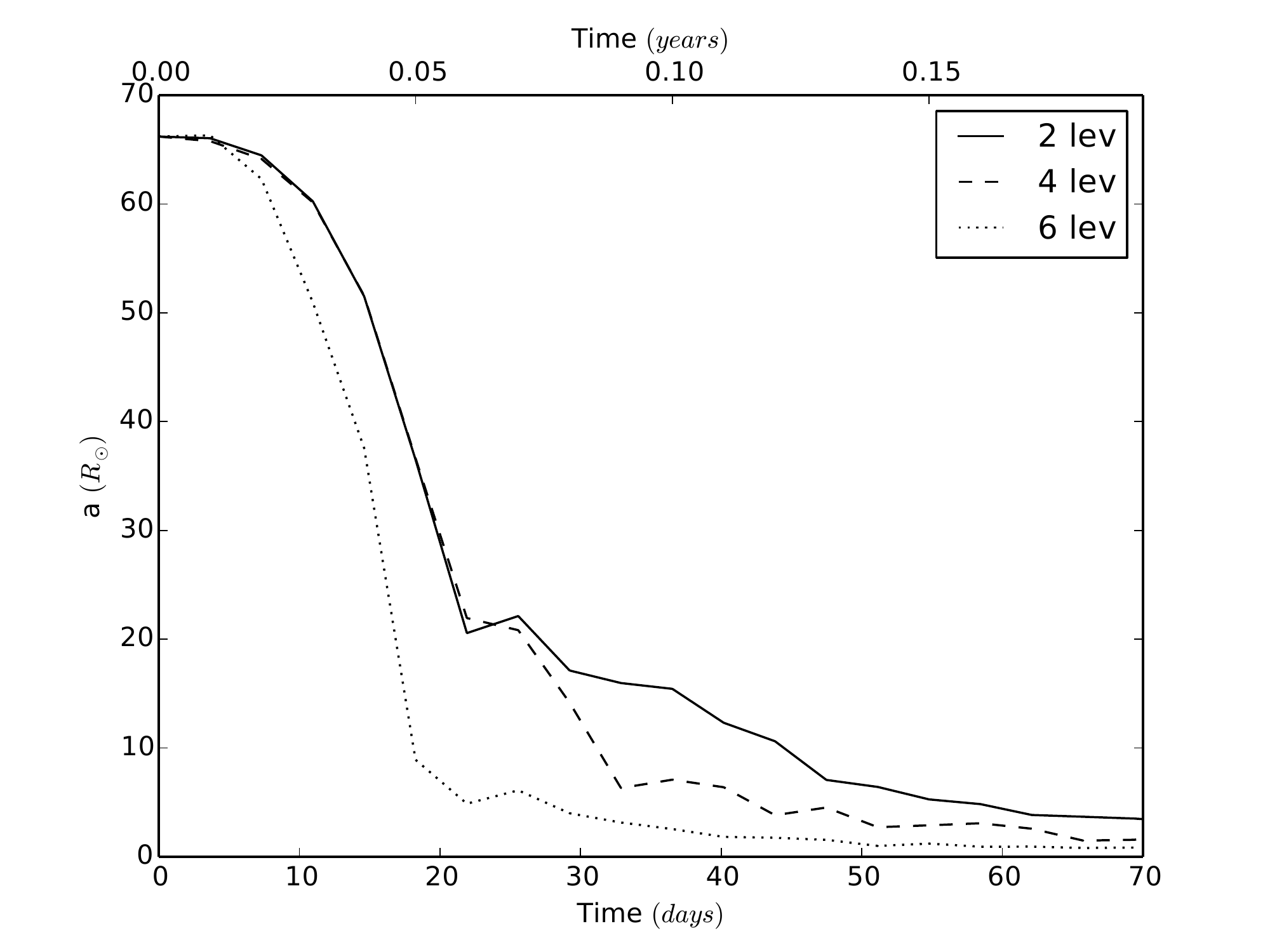}
\centering
\includegraphics[scale=0.4, trim=0.5cm 0.0cm 0.0cm 0.0cm, clip]{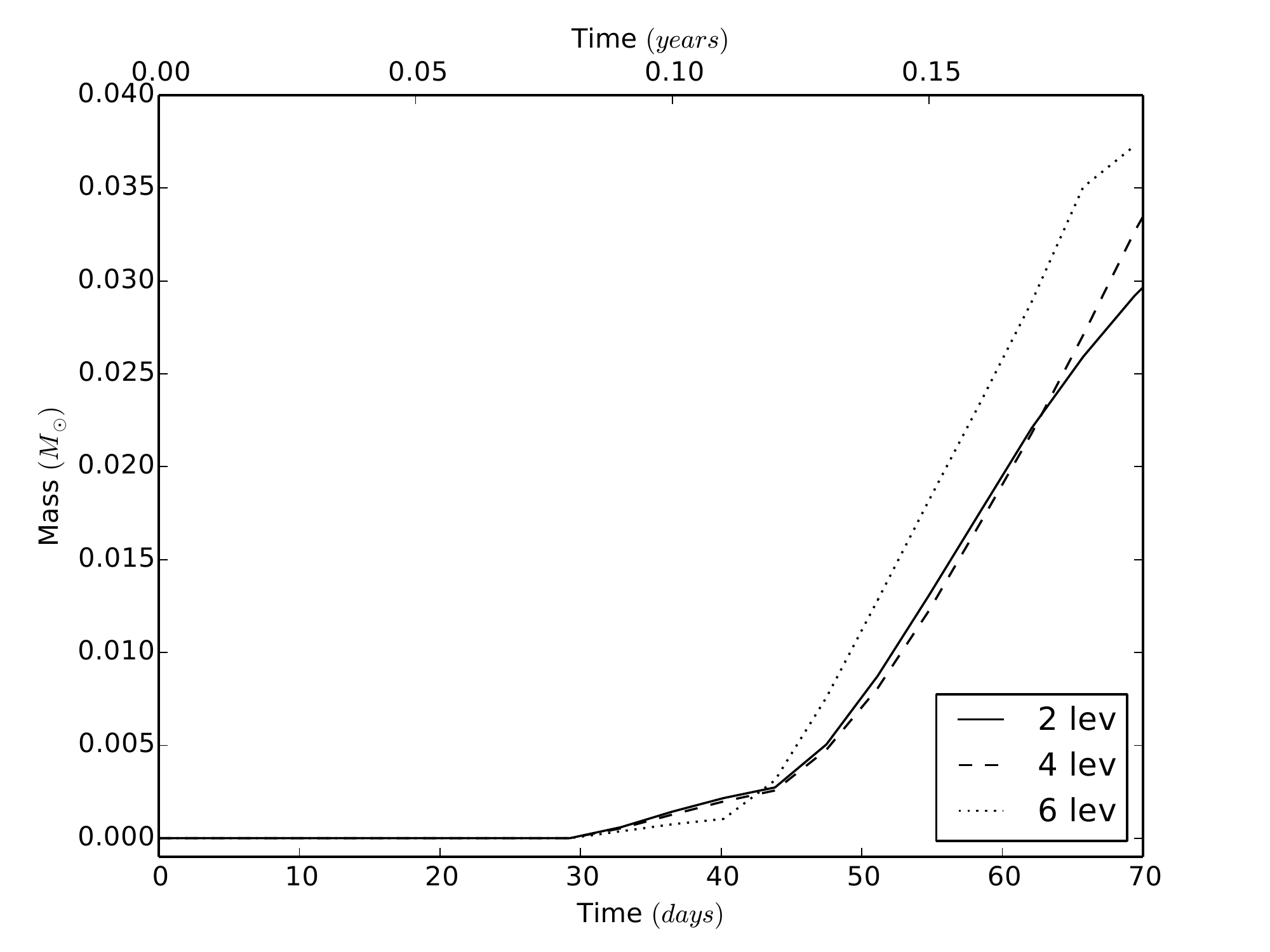}
\caption{\protect\footnotesize{{Upper panel:} evolution of the separation, $a$, between the two particles representing the core of the primary and the companion for SIM9, SIM11 and SIM12. {Lower panel:} cumulative unbound mass for SIM9, SIM11 and SIM12, estimated with the same method used for the bottom panels of Figure~\ref{fig:mass_time_1MsunPrimary_128_2lev_2AUbox} and \ref{fig:mass_time_2MsunPrimary_128_2lev_2AUbox}. In both the panels we limit the abscissa range to 70~days, the maximum time reached by SIM12, and the line corresponding to the maximum refinement level used in the simulations is shown in the legend.}}
\label{fig:resolution_study}
\end{figure}

\begin{figure*}
\centering     
\includegraphics[scale=0.23, trim=0.4cm 0.0cm 1.2cm 0.0cm, clip]{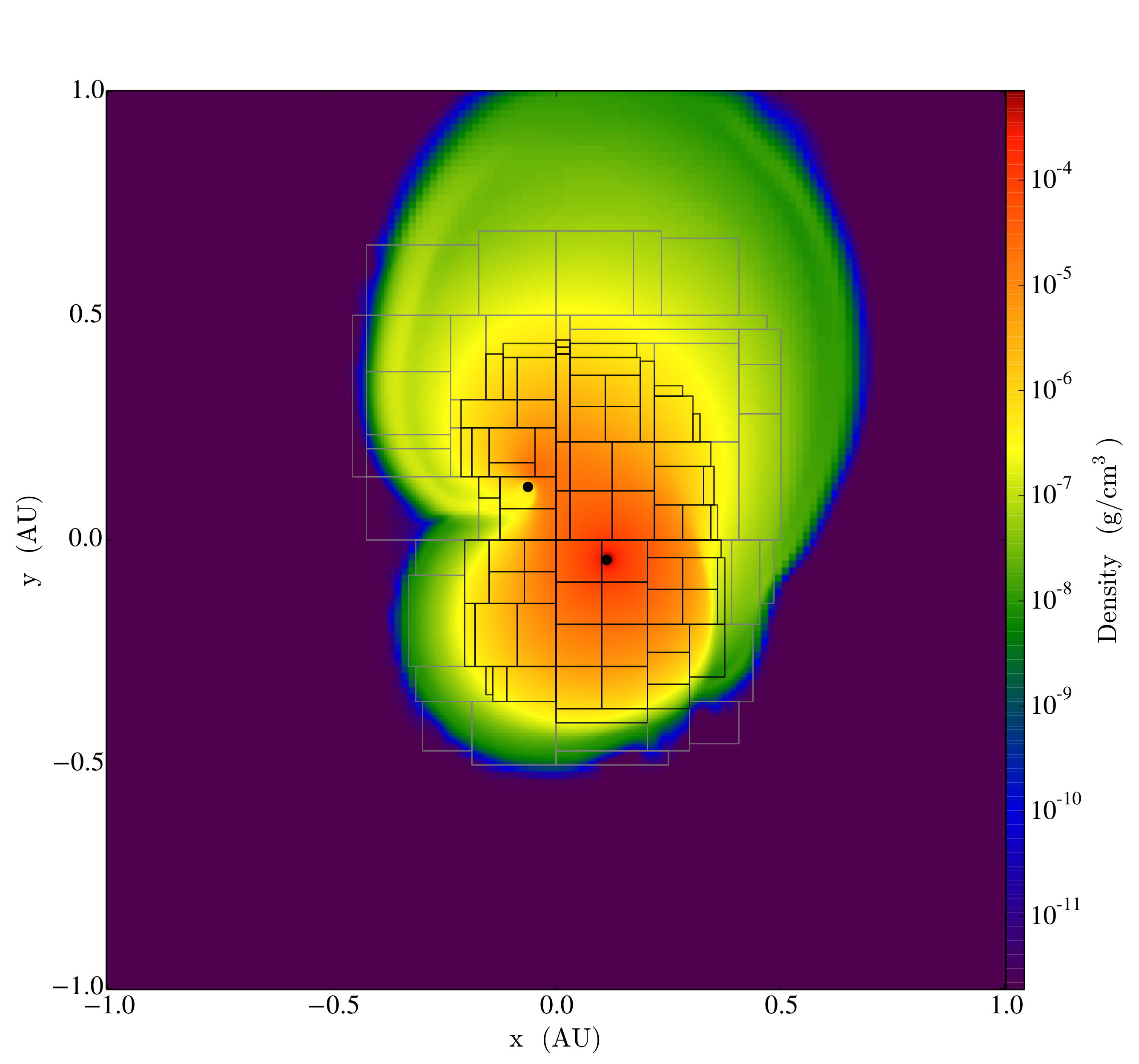}
\includegraphics[scale=0.23, trim=1.3cm 0.0cm 1.2cm 0.0cm, clip]{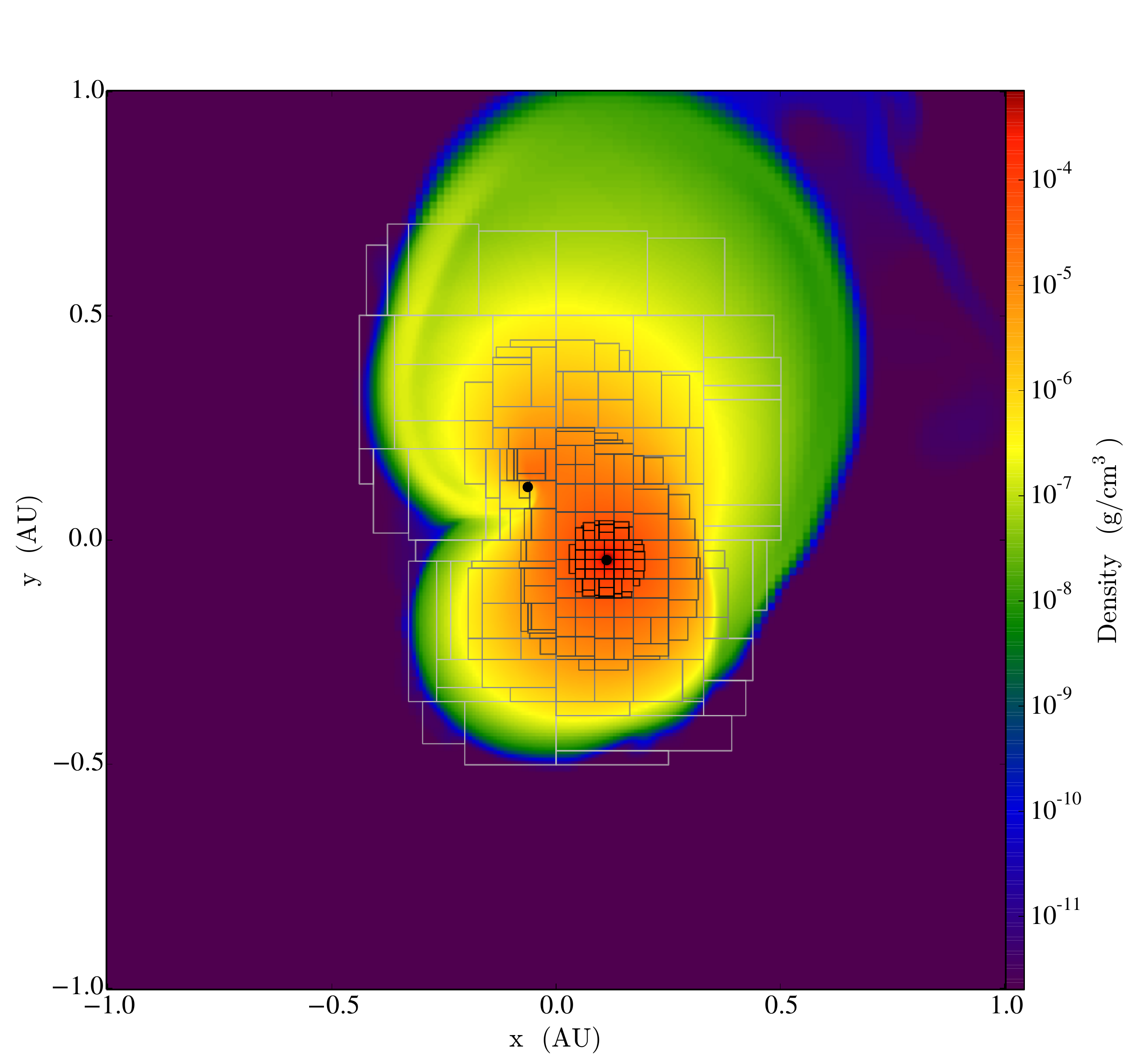}
\includegraphics[scale=0.23, trim=1.3cm 0.0cm 0.0cm 0.0cm, clip]{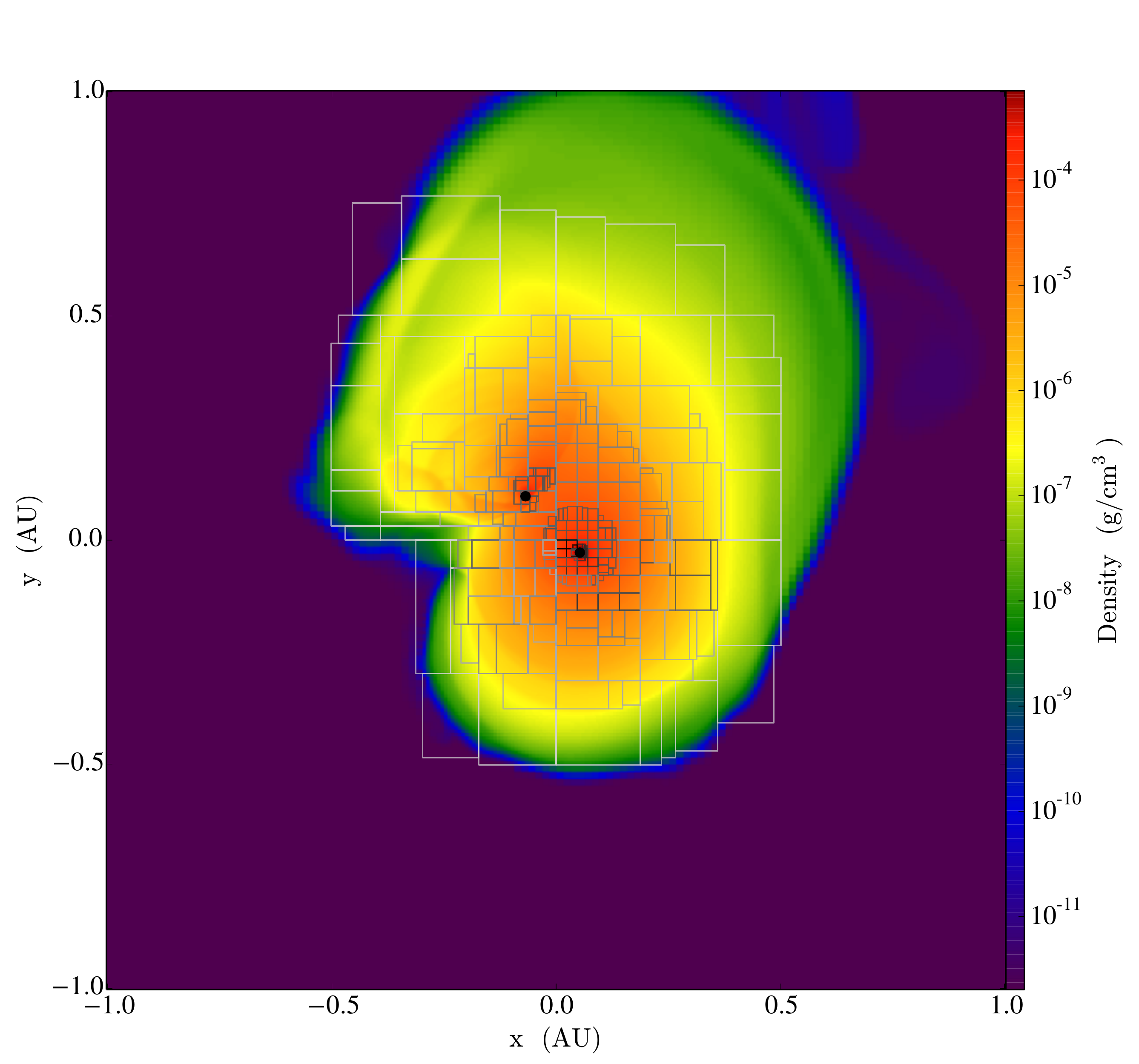}
\caption{\protect\footnotesize{Density slices perpendicular to the $z$ axis in the orbital plane at 15~days for SIM9 { (left panel)}, SIM11 { (centre)} and SIM12 { (right panel)}. The borders of the different refinement levels are over-plotted.}}
\label{fig:grid_slices}
\end{figure*}

This may not be so for the case of the 2.0-\ms, heavier primary. In this case for the resolution study we compare SIM9, SIM11 and SIM12, carried out with 2, 4 and 6 levels of refinement (Table~\ref{tab:simulation_parameters}) and the same base grid resolution. The evolution of the orbital separation and the cumulative unbound mass as a function of time for the three simulations are shown in Figure~\ref{fig:resolution_study}. There (Figure~\ref{fig:resolution_study} -- upper panel) we see that, in the first part of the rapid in-spiral, SIM9 and SIM11 have a similar behaviour, while SIM12's orbit decays faster. This is due in part to differences in the distribution of the initial AMR sub-grids. For larger maximum levels of refinement the central regions of the star near the point-mass particle and around the companion (once mass starts accumulating in its potential well) are resolved on smaller length-scales. The presence of smaller cells around the companion in the first part of the in-spiral is in particular relevant to its outcome, in fact a better sampling generates a more realistic drag. The clear difference between SIM9 and SIM11 on the one hand and SIM12 on the other is that two extra grid levels provide the extra refinement around the companion, which prompts a faster in-spiral early in the simulation. Both the different distribution of the grids and the increased maximum density around the point-masses can be appreciated in Figure~\ref{fig:grid_slices}.

At the end of the simulations the orbital separation is smaller for more resolved simulations, a behaviour similar to that observed for the SPH code {\sc phantom} by \citet{Iaconi2017}, but the difference is smaller between the two more resolved simulations. We only show the behaviour for 70 days because SIM12, the most resolved of the three could not be evolved for as long as the other two due to the fact that the time-step needed to satisfy the Courant condition became impossibly small. Additionally, at higher resolution, also the cumulative unbound mass recorded during the simulations is larger, displaying a similar behaviour to the orbital separation evolution (Figure~\ref{fig:resolution_study}, lower panel). Here too the high resolution simulation, unbinds more mass earlier, in line with its in-spiral being steeper, and likely caused by a strong gravitational drag at the hand of a more resolved companion-gas interaction. For now we can say that in more resolved simulations we unbind slightly more mass. This behaviour is the same for our {\sc phantom} simulations in \citep{Iaconi2017}, which allow a precise computation of the unbound mass value and show convergent behaviour for the unbound mass at increasing number of particles.

In conclusion, what we observe gives us a quantitative idea of the effect of increasing resolution on the in-spiral. There are still questions about the precise effects of resolution, e.g., on the evolution of the separation, but our capability of investigating more resolved setups is hampered by the long run times these simulations.
\section{Simulations' results}
\label{sec:results}
Having assessed the conservation and the impact of increasing resolution in our simulations, we now review the results.

\subsection{Orbital separation}
\label{ssec:final_separation}
The main physical effect driving the in-spiral is the gravitational drag exerted by the envelope of the primary on the primary's core and companion (\citealt{Ricker2012}). Gravitational drag depends mainly on three factors \citep{Iben1993}: the density of the gas surrounding the companion (and the core), the velocity of the companion with respect to the envelope gas and whether the companion speed is subsonic or supersonic \citep[e.g.,][]{Ostriker1999}. Determining the factors that play a major role during the simulations' in-spiral is a difficult task, because it is difficult to calculate the magnitude of the different effects in the proximity of the companion (\citealt{Staff2016b,Iaconi2017}). The expectation is that a star that is both more massive and more compact should generate a larger gravitational drag compared to a lighter and larger one, therefore yielding a faster in-spiral and a smaller final separation.

\begin{figure}
\centering     
\includegraphics[scale=0.4, trim=0.5cm 0.0cm 0.0cm 0.0cm, clip]{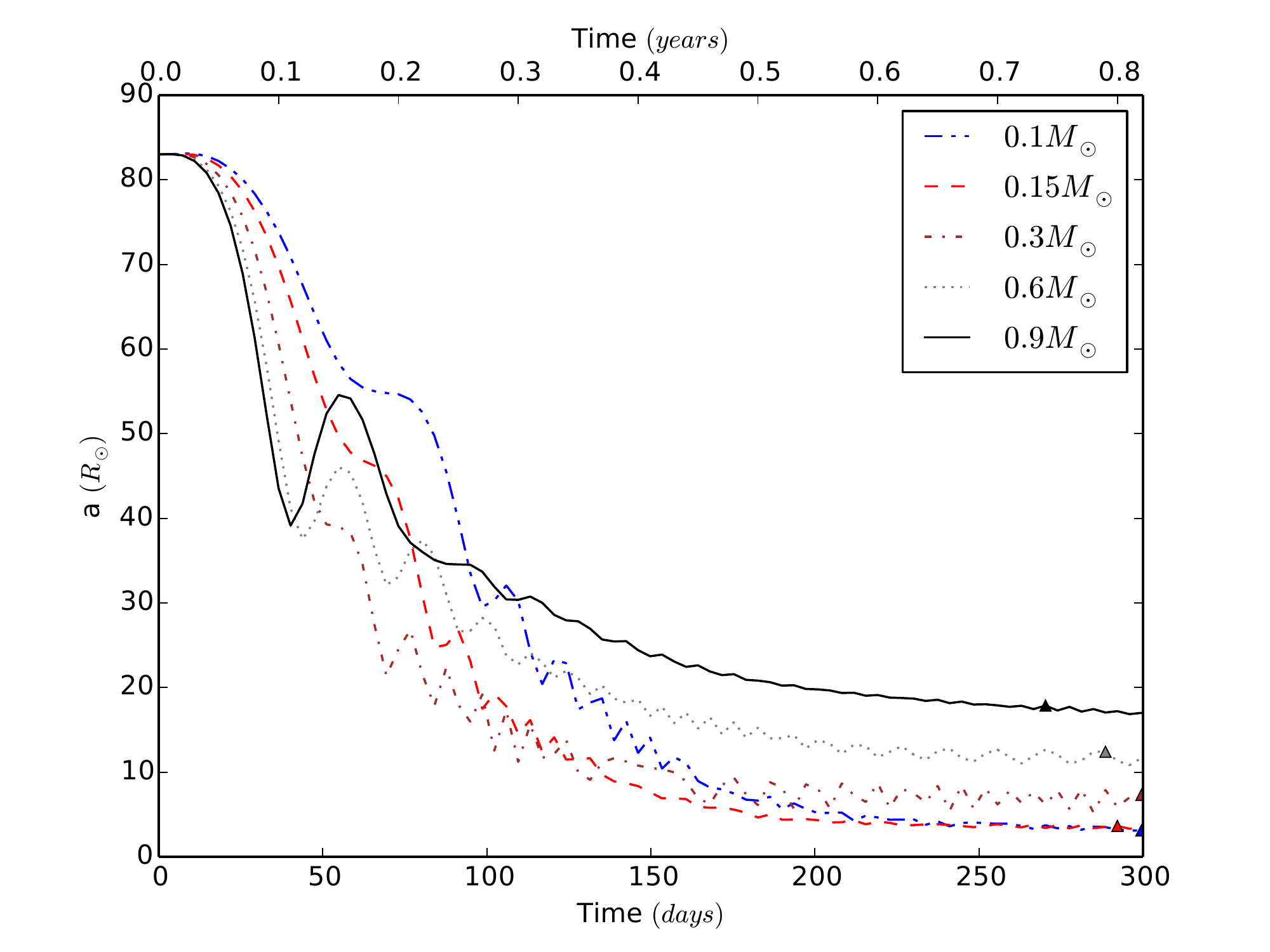}
\centering
\includegraphics[scale=0.4, trim=0.5cm 0.0cm 0.0cm 0.0cm, clip]{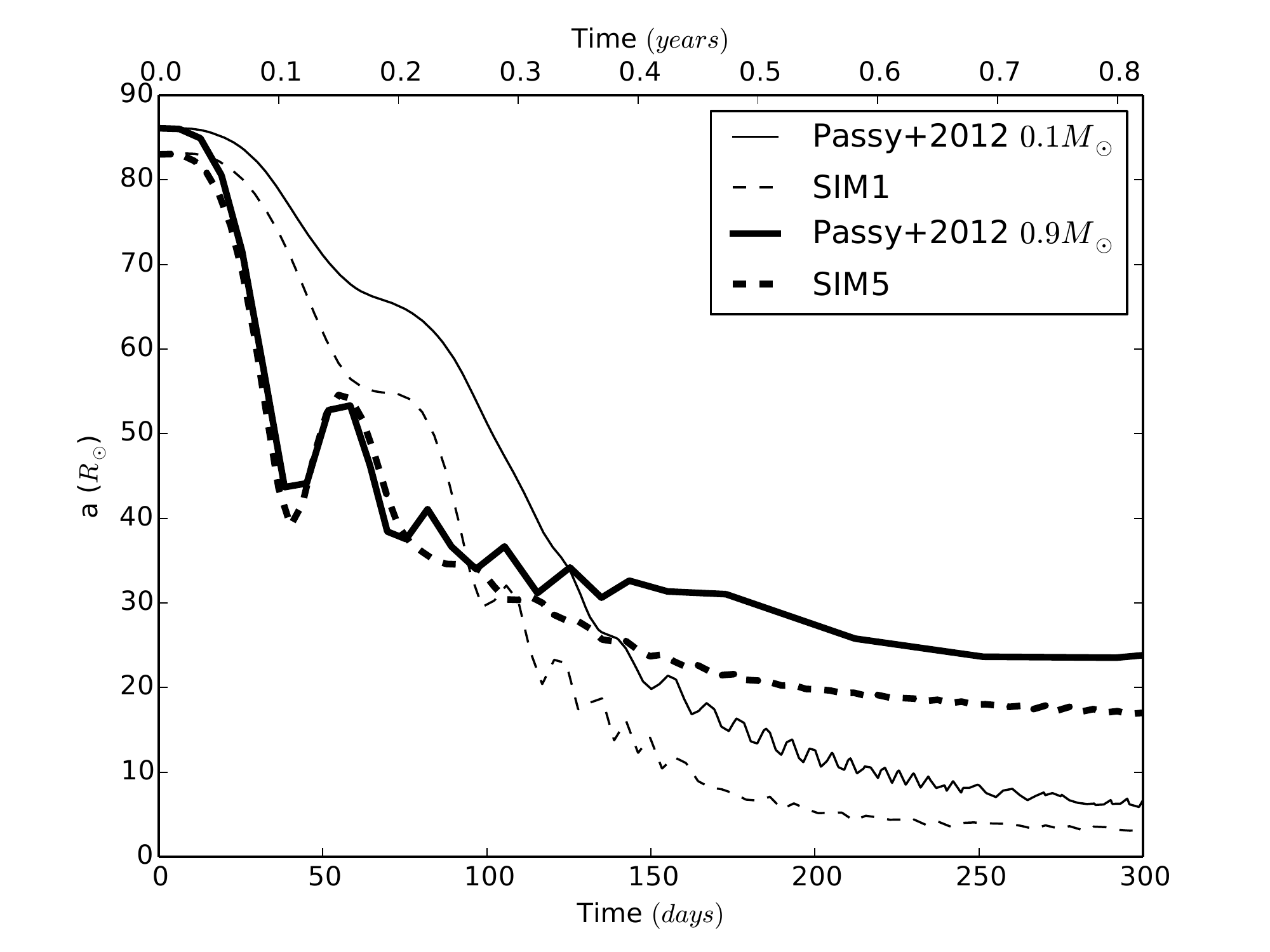}
\caption{\protect\footnotesize{{Upper panel:} evolution of the separation, $a$, between the two particles representing the core of the primary and the companion for SIM1-SIM5. The coloured triangles mark the point where we determined the end of the rapid in-spiral phase. { Lower panel:} comparison of the evolution of the separations between this work and P12 for the least and most massive companions.}}
\label{fig:separations_time_1MsunPrimary_128_2lev_2AUbox}
\end{figure}

In Figure~\ref{fig:separations_time_1MsunPrimary_128_2lev_2AUbox} (upper panel) we show the evolution of the separation in SIM1-SIM5. We compare these to those of P12 for the $256^3$ {\sc Enzo} simulations (their figure~4; which had a cell size of 1.7~\rs, compared to our smallest cell size of 0.84~\rs). We observe that the trend of the curves for different companion masses is the same, with the most massive companions in-spiraling faster at the beginning of the interaction but reaching a larger stable separation. 

\begin{figure}
\includegraphics[scale=0.4, trim=0.0cm 0.0cm 0.0cm 0.0cm]{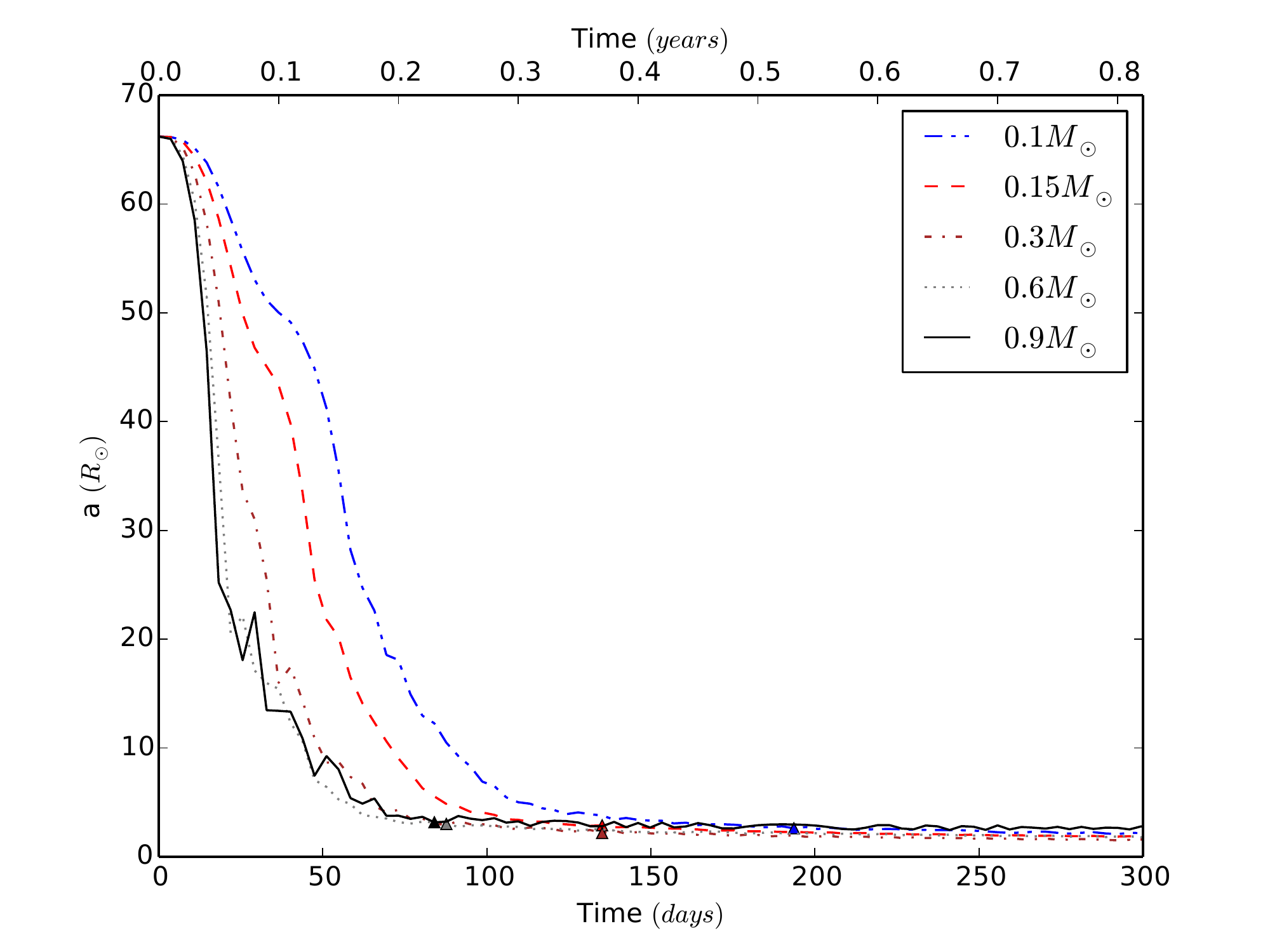}
\caption{\protect\footnotesize{Evolution of the separation, $a$, between the two particles representing the core of the primary and the companion for SIM6-SIM10.  The coloured triangles mark the point where we determined the end of the rapid in-spiral phase.}}
\label{fig:separations_time_2MsunPrimary_128_2lev_2AUbox}
\end{figure}

In our simulations we start with an initial separation slightly smaller than P12, 83~\rs \ vs. 85~\rs, due to small differences in how the initial model becomes stable in different grids. However, even accounting for this initial offset, we find that the separation during the initial part of the in-spiral tends to evolve differently with decreasing companion's mass. At lower companion masses the uniform grid separation is larger than for the AMR grid (Figure~\ref{fig:separations_time_1MsunPrimary_128_2lev_2AUbox}, lower panel). This could be an effect of the different numerics (we solve the particles' gravity using the new solver introduced by \citealt{Passy2014}), but more likely of the different resolutions, which result in a slower in-spiral and a weaker interaction between companion and primary's envelope/core for the P12 simulations. 
The different resolutions also results in different final separations, with our more resolved, new simulations having smaller final separations (lower panel of Figure~\ref{fig:separations_time_1MsunPrimary_128_2lev_2AUbox} and Figure~\ref{fig:separations_time_2MsunPrimary_128_2lev_2AUbox}).
We also remark here that we adopt a larger smoothing length with respect to the grid simulations of P12 (see Section~\ref{sec:hydrodynamics_simulations}). This may partially affect the final separation for SIM1 and SIM2, where the companions in-spiral deeper, closer to the region where the gravity of the point-mass particle is smoothed. However, the final separations obtained for SIM1 and SIM2 are larger than the smoothing length value ($\sim2.53$~\rs). Therefore we believe that the effect of the smoothing length on the simulation outcome is minimal (this is not the case for SIM6-SIM10, see below). 

Comparing SIM1-SIM5 to SIM6-SIM10 (upper panel of Figure~\ref{fig:separations_time_1MsunPrimary_128_2lev_2AUbox} and Figure~\ref{fig:separations_time_2MsunPrimary_128_2lev_2AUbox}) we see how the increased gravitational drag in the case of the more massive and compact envelope generates faster in-spirals in for the entire range of companion masses (SIM6-SIM10). To determine the end of the rapid in-spiral phase we utilise the same criterion used by \citet{Sandquist1998} and P12, that is when the orbital shrinkage parameter reaches 10 per cent of its maximum value:
\begin{equation}
 \frac{1}{a} \frac{da}{dt} \simeq 0.1 \Big( \frac{1}{a} \frac{da}{dt} \Big)\Big|_{\mathrm{max}} \ ,
 \label{eq:final_separation_criterion}
\end{equation}
where $a$ is the separation between the primary's core and the companion. We report the values of the final separation obtained following this criterion, the maximum value of the orbital shrinkage parameter and the time at which we record this value in Table~\ref{tab:shrinking_paramter}. We also report the final separations in Table~\ref{tab:simulation_parameters} and lable the time when they have been determined with a triangular symbol in Figure~\ref{fig:separations_time_1MsunPrimary_128_2lev_2AUbox} and \ref{fig:separations_time_2MsunPrimary_128_2lev_2AUbox}.

\begin{table}
\begin{center}
\begin{adjustbox}{max width=\textwidth}
\begin{tabular}{cccc}
\hline
ID & $a_{\mathrm{f}}$  &  $-(\dot{a}/a)$ (Max) & $t(a_{\mathrm{f}})$  \\
 & (\rs) & (days$^{-1}$) & (days) \\
\hline
SIM1 & 3.1 & 0.028 & 303  \\
SIM2 & 3.4 & 0.037 & 292  \\
SIM3 & 7.3 & 0.032 & 299  \\
SIM4 & 12 & 0.029 & 288  \\
SIM5 & 17 & 0.027 & 270  \\
\hline
SIM6 & 2.2 & 0.044 & 193  \\
SIM7 & 1.8 & 0.052 & 135  \\
SIM8 & 1.6 & 0.060 & 135  \\
SIM9 & 1.8 & 0.077 & 87  \\
SIM10 & 2.8 & 0.076 & 84  \\
\hline
SIM11 & 0.67 & 0.111 & 80  \\
SIM12 & 0.87 & 0.311 & 62  \\
\hline
SIM13 & 21 & 0.028 & 547  \\
SIM14 & 11 & 0.029 & 332  \\
SIM15 & 10 & 0.028 & 219  \\
\hline
\end{tabular}
\end{adjustbox}
\end{center}
 \begin{quote}
  \caption{\protect\footnotesize{Final separations, maximum values of the orbital shrinkage parameter, time corresponding to the final separations in our sets of simulations.}} \label{tab:shrinking_paramter}
 \end{quote}
\end{table}

Being the rapid in-spiral quicker and steeper in SIM6-SIM10, with respect to SIM1-SIM5, the values for $a_{\mathrm{f}}$ are naturally located earlier in time. All the values take place at a moment of the separation evolution where the curve has levelled. If we compare the moment where the values are recorded for SIM1-SIM5 with those of P12 (last two columns in Table~\ref{tab:shrinking_paramter}), we can see that there are deviations between the corresponding simulations, likely due to the numerical differences discussed above, however the values fall in similar time zones of the orbital evolution.

In SIM6-SIM10 the lack of an obvious $q$-$a_{\mathrm{f}}$ correlation, observed instead in SIM1-SIM5 is likely due to the separation becoming similar to the smoothing length. We used a smoothing length of 3 cells or 2.53~\rs, similar to the final separations obtained. At this distance the gravitational force between the two point-masses is therefore weakened, forcing the particles to orbits larger than those they would reach in non-smoothed gravity conditions. The final separation of the higher resolution SIM11 (see Table~\ref{tab:simulation_parameters}), which has a smoothing length of $0.21 \times 3 = 0.63$~\rs, is 0.67~\rs. SIM12, with a smoothing length $0.05 \times 3 = 0.15$~\rs, has instead a final separation 0.87~\rs. Thus we can conclude that the final separation tends to decrease with increasing resolution, and it is greatly impacted by the smoothing length.

Irrespective of numerical limitations, the final separations obtained in SIM6-SIM10 is much smaller and comparable to those of other simulations \citep[see table~1 of][]{Iaconi2017}, with binding energies comparable to those of SIM6-10\footnote{The only simulation that has a substantially larger binding energy is that of Rasio and Livio (1996) with a massive, compact RGB star.}. This was also concluded by \citet{Iaconi2017} using the heterogeneous set of  simulations from the literature.

\subsection{Envelope ejection}
\label{ssec:unbound_mass}
\begin{figure}
\centering     
\includegraphics[scale=0.4, trim=0.0cm 0.0cm 0.0cm 0.0cm]{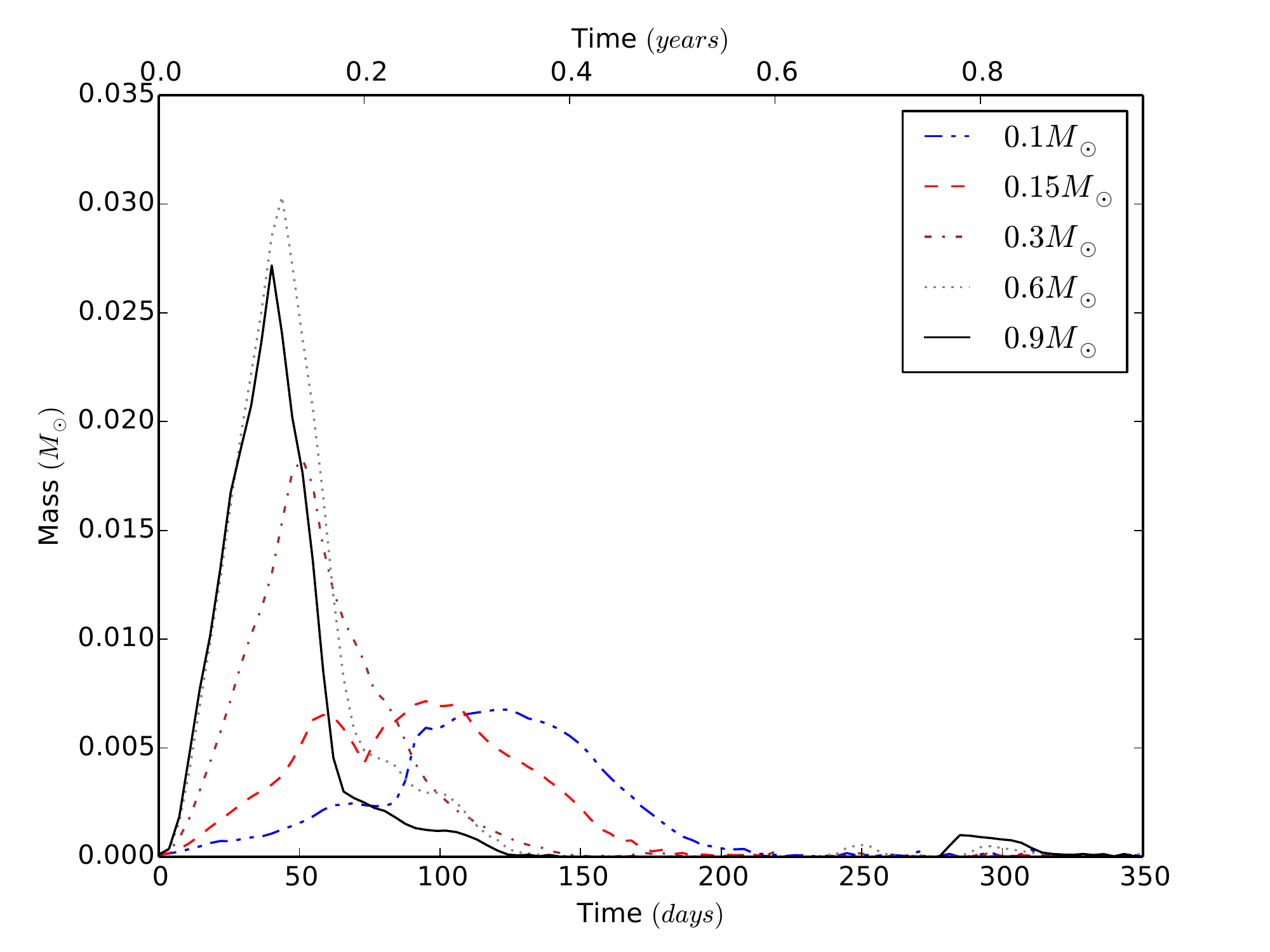}
\includegraphics[scale=0.4, trim=0.0cm 0.0cm 0.0cm 0.0cm]{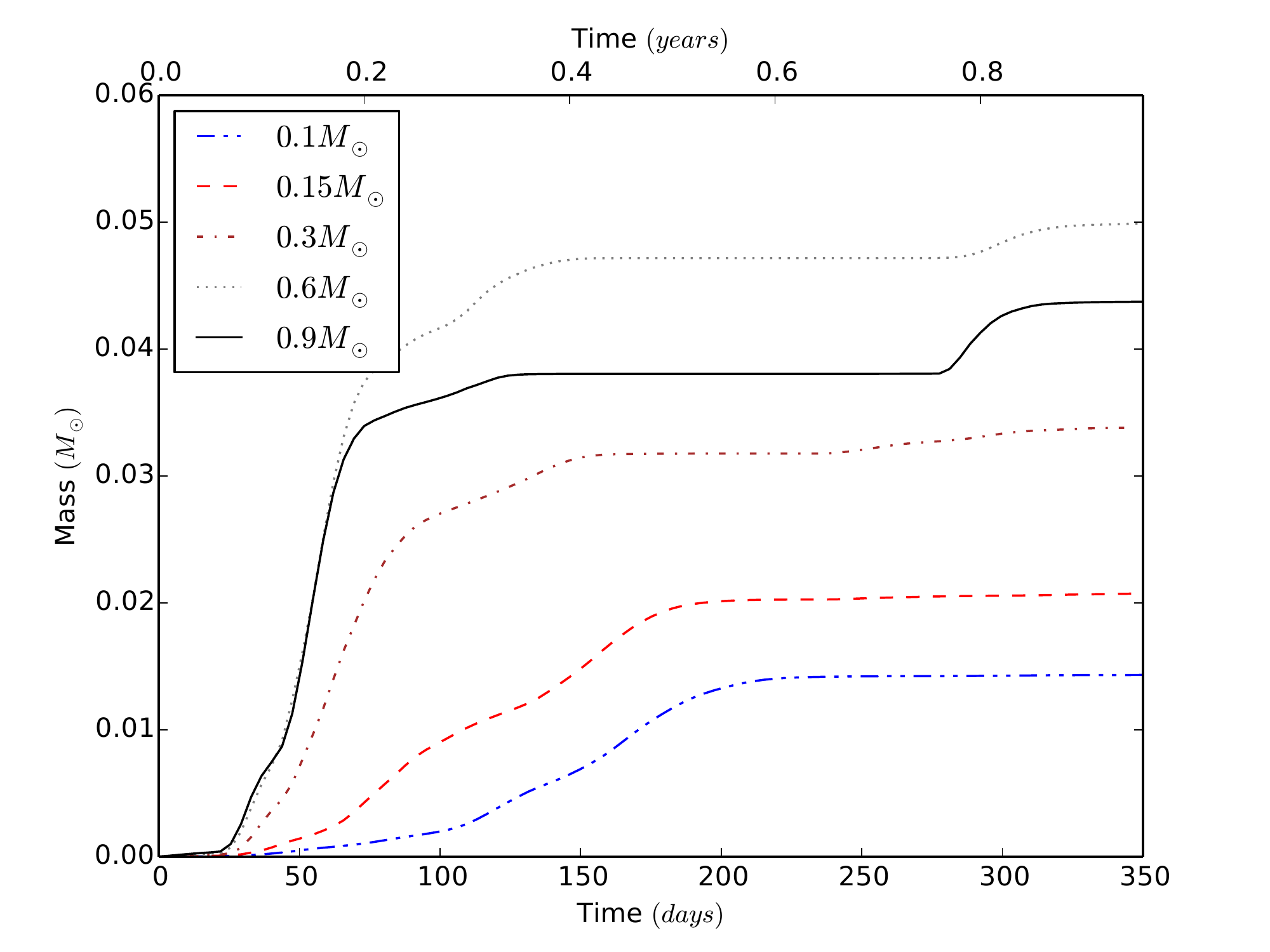}
\caption{\protect\footnotesize{{ Upper panel:} unbound mass inside the simulation domain for SIM1-SIM5. { Lower panel:} cumulative unbound mass outside the simulation domain for SIM1-SIM5.
Note that the range of the abscissa as been slightly increased to 350~days to accommodate the mass that was unbound late in the simulation and therefore left the domain after the 300~days mark.}}
\label{fig:mass_time_1MsunPrimary_128_2lev_2AUbox}
\end{figure}

\begin{figure}
\centering     
\includegraphics[scale=0.4, trim=0.0cm 0.0cm 0.0cm 0.0cm]{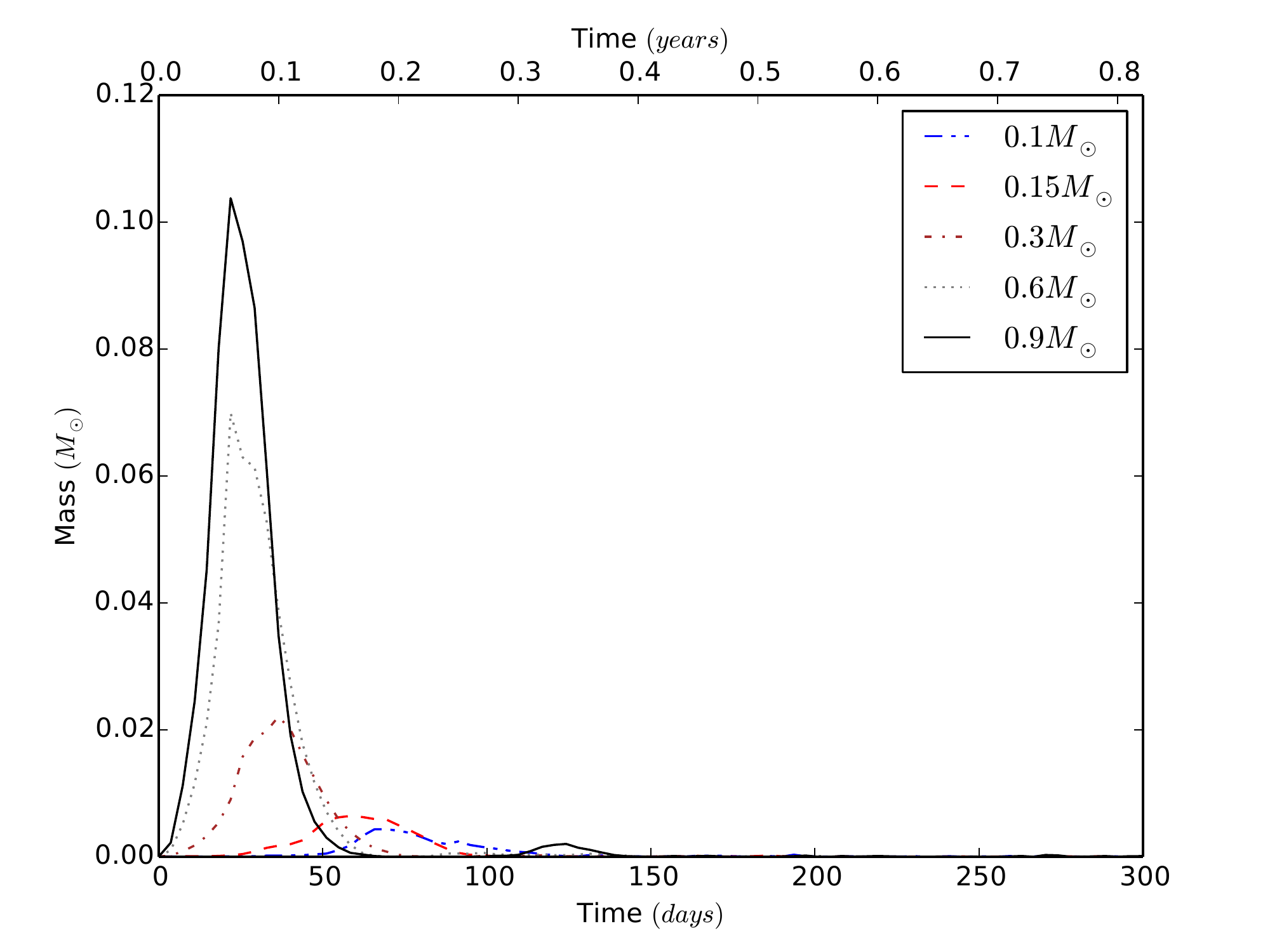}
\includegraphics[scale=0.4, trim=0.0cm 0.0cm 0.0cm 0.0cm]{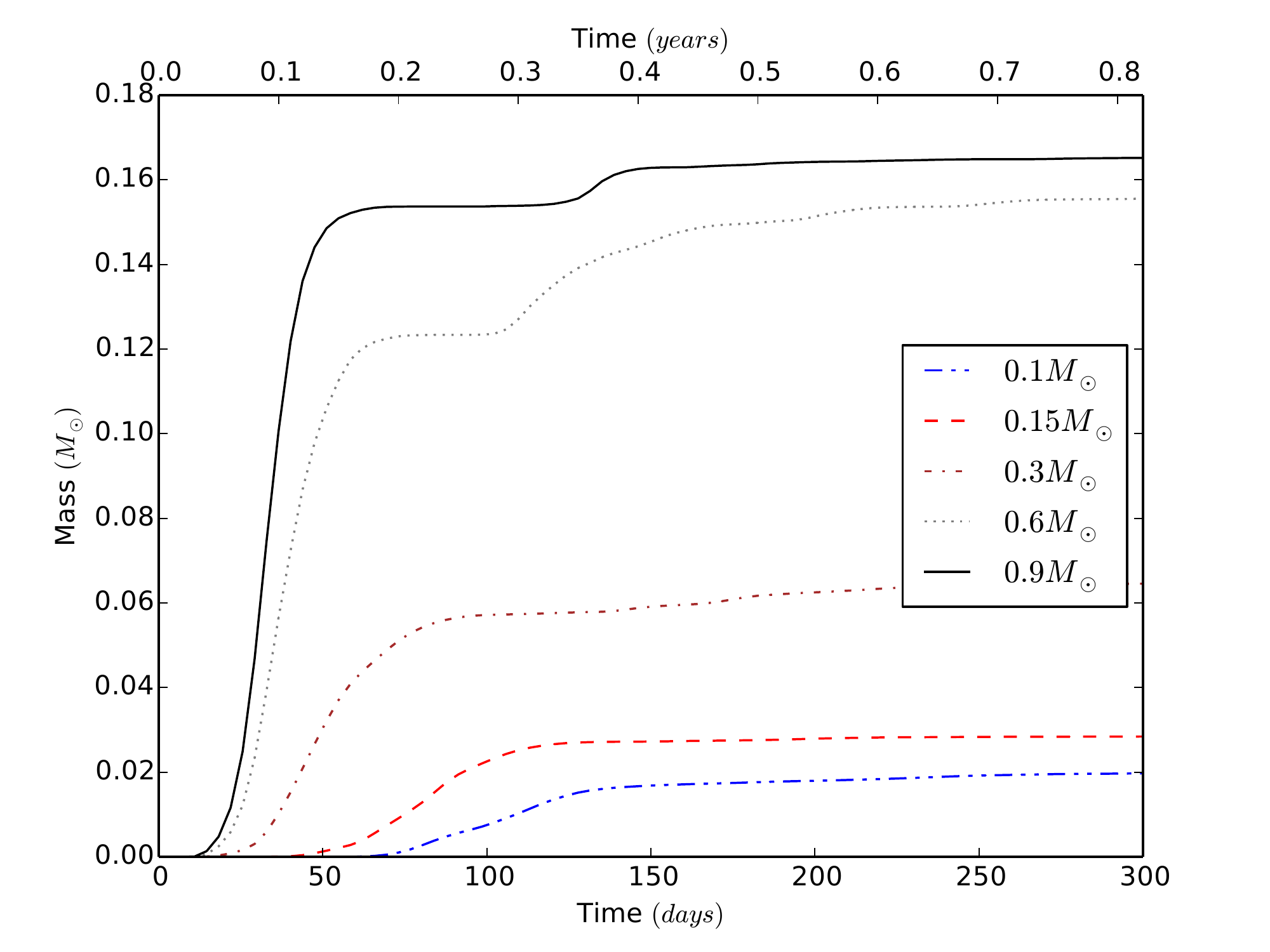}
\caption{\protect\footnotesize{{Upper panel:} unbound mass inside the simulation domain for SIM6-SIM10. { Lower panel:} cumulative unbound mass outside the simulation domain for SIM6-SIM10.}}
\label{fig:mass_time_2MsunPrimary_128_2lev_2AUbox}
\end{figure}

From their comparison of all past simulations, \citet{Iaconi2017} concluded that more compact envelopes do not necessarily result in more unbound mass: despite the fact that more orbital energy is deposited in the envelope, the envelope extra ``weight" leads to a similar level of unbinding. This conclusion, however was hampered by the inherent diversity of the literature simulations that were being compared and there was also a suspicion that resolution effects  were adding noise to the results.
Here we provide a more solid analysis on the effect of binding energy by comparing SIM1-SIM5 with a less bound envelope to SIM6-SIM10 with a more bound one. 

Gas is considered unbound if the sum of its potential, kinetic and thermal energies is greater than zero. 
To estimate the total amount of unbound mass, including the gas leaving the computational domain, we carried out the same interpolation used in \citet[their section~3.2]{Iaconi2017}.

In Table~\ref{tab:simulation_parameters} we list the fraction of unbound mass in the simulations. SIM1-SIM5 unbind gas masses not dissimilar to those obtained by P12 in their SPH simulations (their table~2). Since the grid simulations performed by P12 mimicked the behaviour of SPH, we can also expect that similar amounts of unbound mass were achieved by their {\sc Enzo} simulations. We therefore conclude that there are only minor differences in the unbinding of the envelope between our AMR simulations and those carried out with static, uniform grids by P12.

Comparing the results from SIM1-SIM5 and SIM6-SIM10 we can see that for companion masses $\leq 0.3$~\ms \ SIM1-SIM3 unbind approximately twice as much as SIM6-SIM8, while for companion masses $\geq 0.6$~\ms \ the unbound mass fractions are similar. Similar results were obtained by \citet{Sandquist2000}, who analysed the effect of increasing the envelope mass by a factor two while keeping the core mass constant in their \virg Simulation 5''. They used two giants with same core mass of 0.45~\ms, total masses of 1~\ms \ and 2~\ms\ and radii of 243~\rs \ and 177~\rs, respectively. With a relatively light companion they observed an unbound mass of $11\%$ of the envelope for the lighter envelope and $6\%$ for the heavier one. 

SIM11 and SIM12, with a higher resolution can be compared to SIM9. A larger fraction of unbound mass seems to be promoted by higher resolution, likely, in this case, because the orbit shrinks farther, with additional release of energy. The unbound mass in SIM11 is 15 \% larger than the 10\% in SIM9. SIM12 has even higher resolution but it was not carried out as far as the others. It does seem therefore likely that the fraction of unbound mass in the simulations is resolution-dependent, though we think it unlikely that the value would increase significantly enough to solve the envelope unbinding problem (e.g., Passy et al. 2012).

\begin{figure*}
\centering     
\includegraphics[scale=0.71, trim=7cm 0.0cm 7cm 0.0cm, clip]{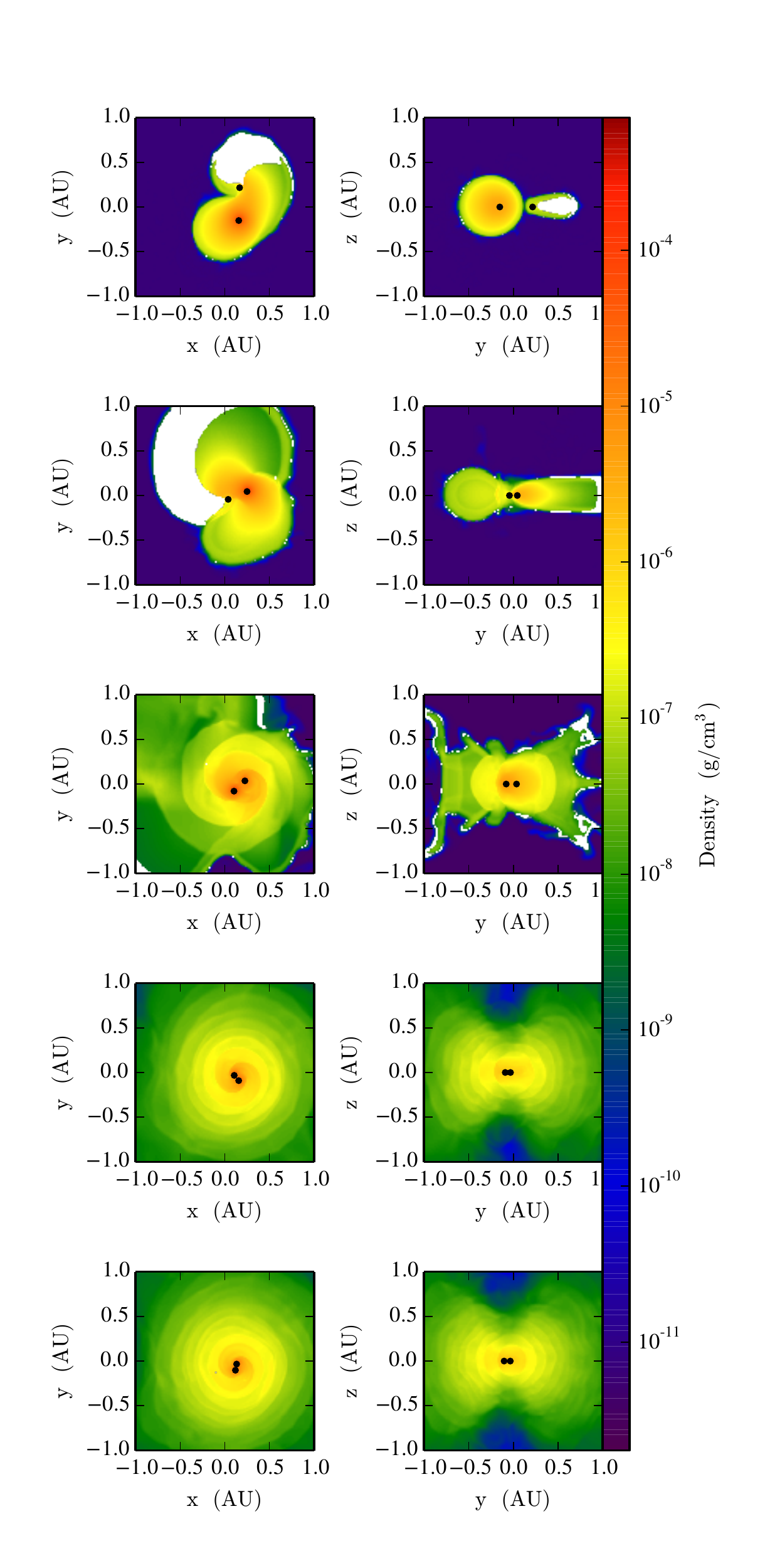}
\includegraphics[scale=0.71, trim=7cm 0.0cm 7cm 0.0cm, clip]{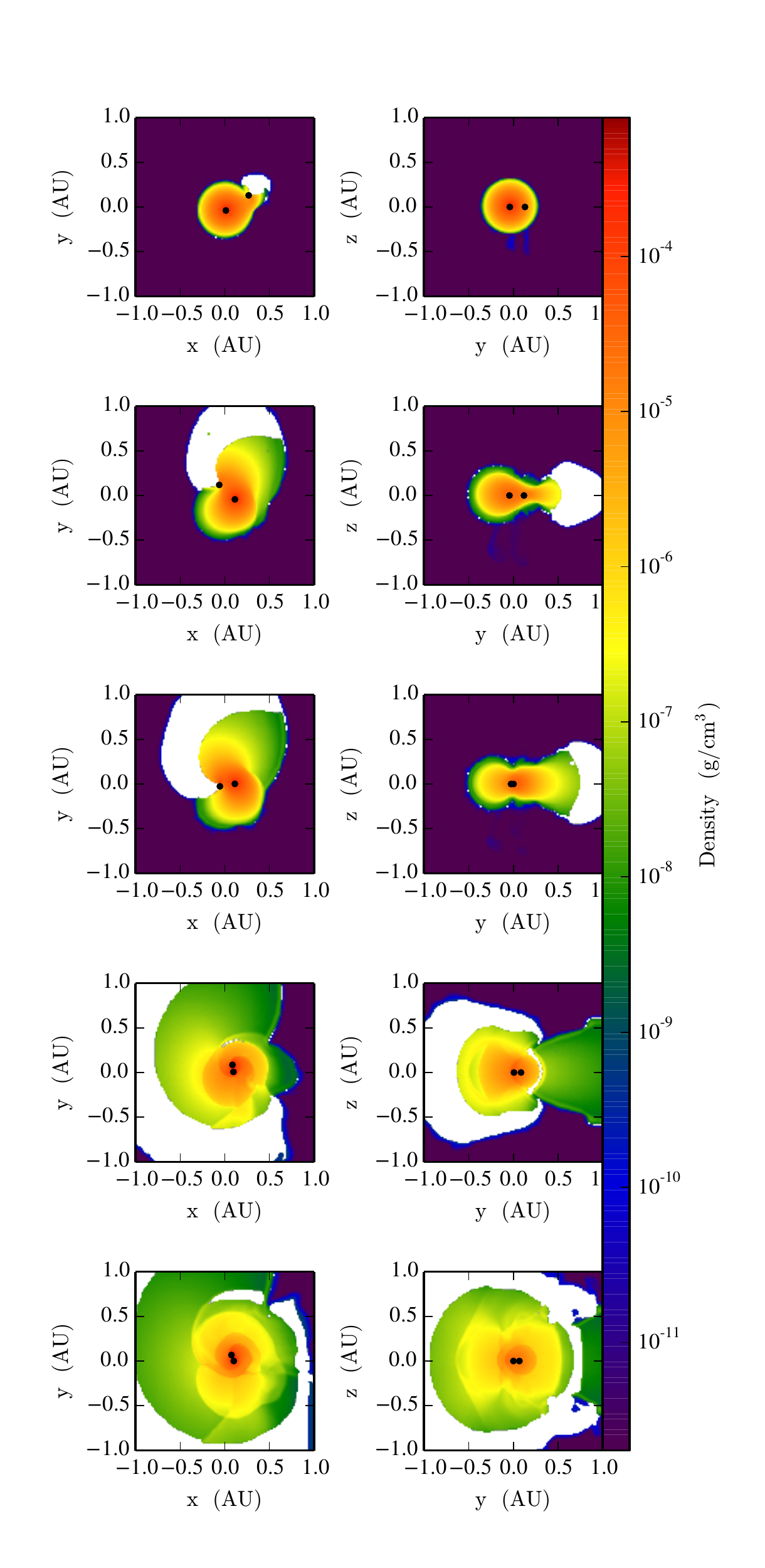}
\caption{\protect\footnotesize{Density slices perpendicular to the $z$ axis in the orbital plane for SIM4 {(left panel, left column)} and SIM9 {(right panel, left column)}. Right columns in both the panels are the same as the left columns, but the slicing is performed perpendicular to the $x$ axis at $x=0$. Slices are captured at times when the orbital separations are approximately comparable, these times correspond to $18$, $36$, $76$, $149$, $164$~days for SIM4 and $4$, $15$, $18$, $29$, $36$~days for SIM9.}}
\label{fig:bound_unbound_z_slices}
\end{figure*}

Figures~\ref{fig:mass_time_1MsunPrimary_128_2lev_2AUbox} and \ref{fig:mass_time_2MsunPrimary_128_2lev_2AUbox} show that the overall behaviour of the unbinding process is similar for both sets of simulations. The unbinding process is more gradual for less massive companions while it is more of a bursting event for larger mass companions. This can also be seen in Figure~\ref{fig:bound_unbound_z_slices}, where we show the geometry of the interaction and of the unbound portions of the envelope for the two comparable simulations SIM4 and SIM9 ($M_2=0.6$~\ms).
In all the simulations all the unbound mass is eventually pushed out of the simulation domain. The lower panels of Figures~\ref{fig:mass_time_1MsunPrimary_128_2lev_2AUbox} and \ref{fig:mass_time_2MsunPrimary_128_2lev_2AUbox} show an estimate of the cumulative unbound mass outside the simulation domain. The most prominent feature is that for the 0.6 and 0.9~\ms \ companions a smaller unbinding event takes place later in time and unbinds a further $1\%$ of the envelope gas. In all cases the bulk of the unbinding is due to envelope gas being accelerated to velocities greater than the escape velocity, with only a small percentage of the mass being unbound due to heating. This extra 1~\% unbinding is due to compression heating between ejected layers of the envelope.  

Strangely, SIM4 unbinds more mass than SIM5, against the trend observed in all other simulations, where the more massive companions unbind more mass. We attribute this to resolution effects: whether a simulation is more or less converged depends on the specifics of that simulation and it is possible that SIM5 is better converged than SIM4.

The slices of Figure~\ref{fig:bound_unbound_z_slices} show snapshots of the in-spiral for the two  primaries (SIM4 and SIM9, $M_2=0.6$~\ms) taken at similar orbital separations. The more massive primary leads to a much faster in-spiral. For the same separation SIM4 has completely pushed out of the simulation domain the first unbound layer and several orbits have been completed, while SIM9 has not yet pushed much mass outside the computational domain and has only completed a few orbits. A similar behaviour is observed for other companion masses. 

Similarly to what observed by \citet{Iaconi2017} and in most of the previous numerical work (see e.g., \citealt{Sandquist2000}, \citealt{Ricker2012}), we find the presence of mild bow shocks on the ejected layers of the envelope in all our simulations. Shocks form on the front of the expanding layers of the envelope that hit the layers lifted during previous orbits. Shocks are slightly stronger for our lighter primary at similar companion mass.

\section{Gravitational drag in simulations of the common envelope interaction}
\label{sec:gravodrag}
\begin{figure*}
\centering     
\includegraphics[scale=0.23, trim=0.5cm 0.cm 1.0cm 1.4cm, clip]{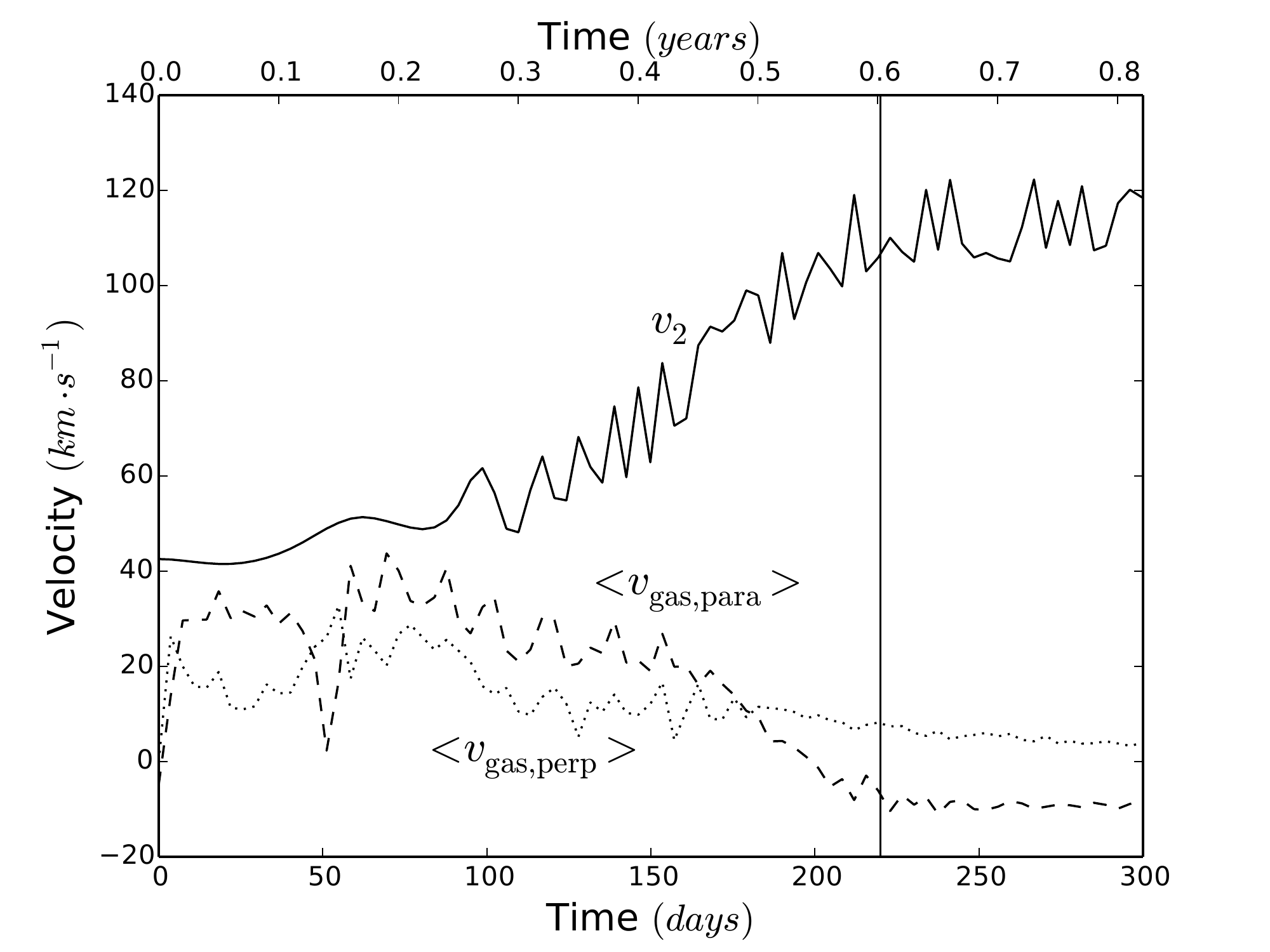}
\includegraphics[scale=0.23, trim=1.0cm 0.cm 1.0cm 1.4cm, clip]{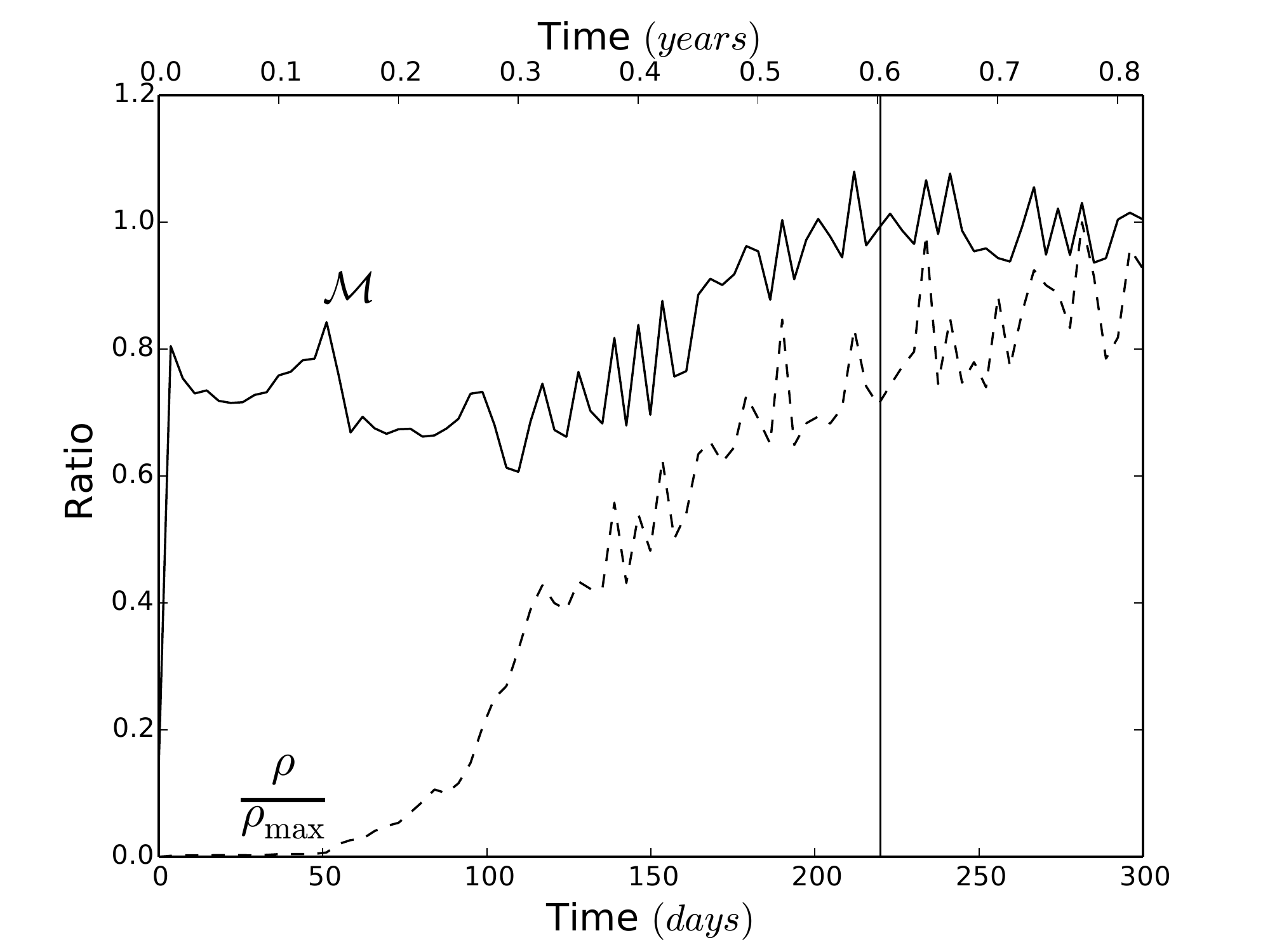}
\includegraphics[scale=0.23, trim=0.5cm 0.cm 1.0cm 1.4cm, clip]{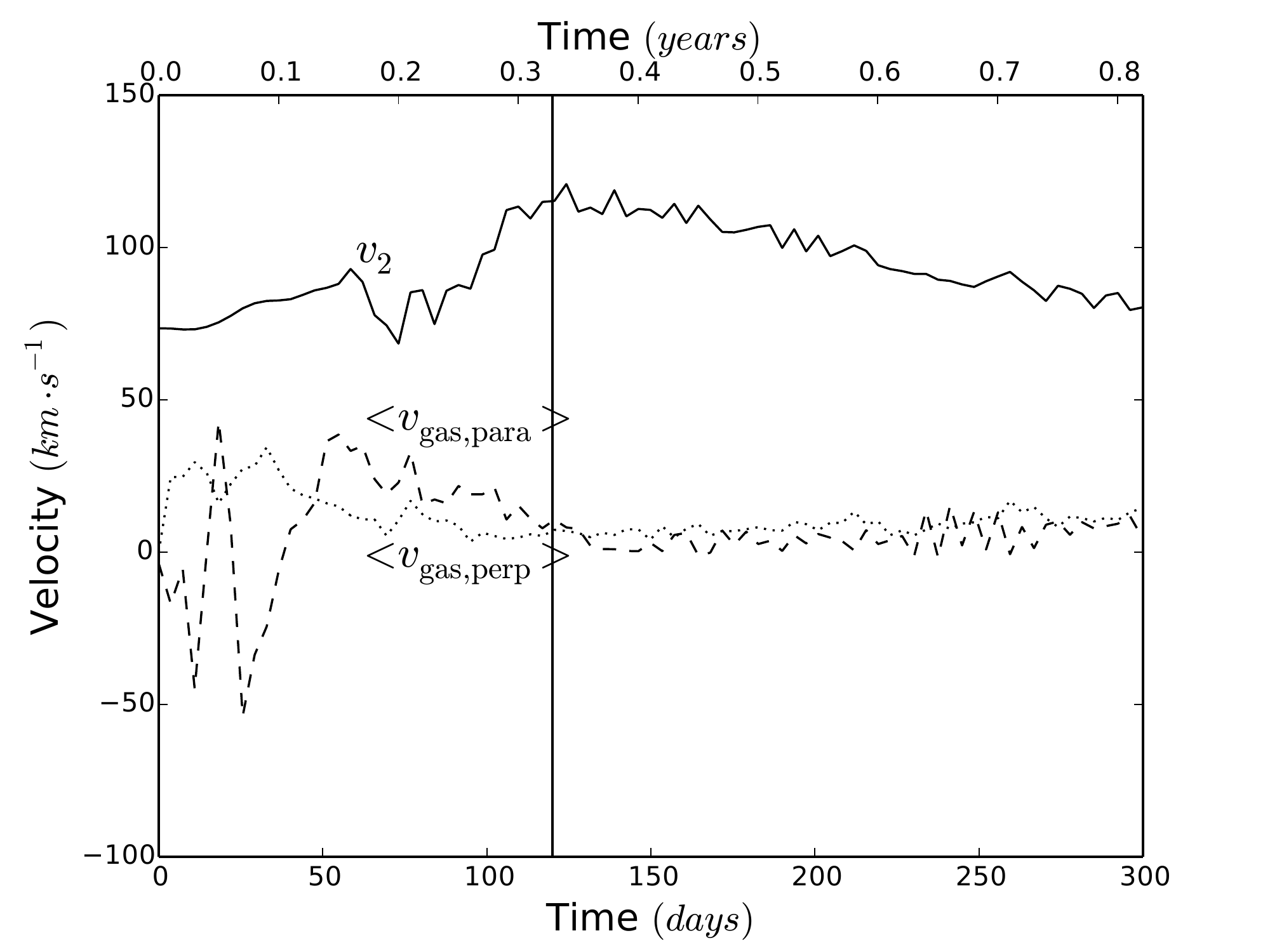}
\includegraphics[scale=0.23, trim=0.9cm 0.cm 1.0cm 1.4cm, clip]{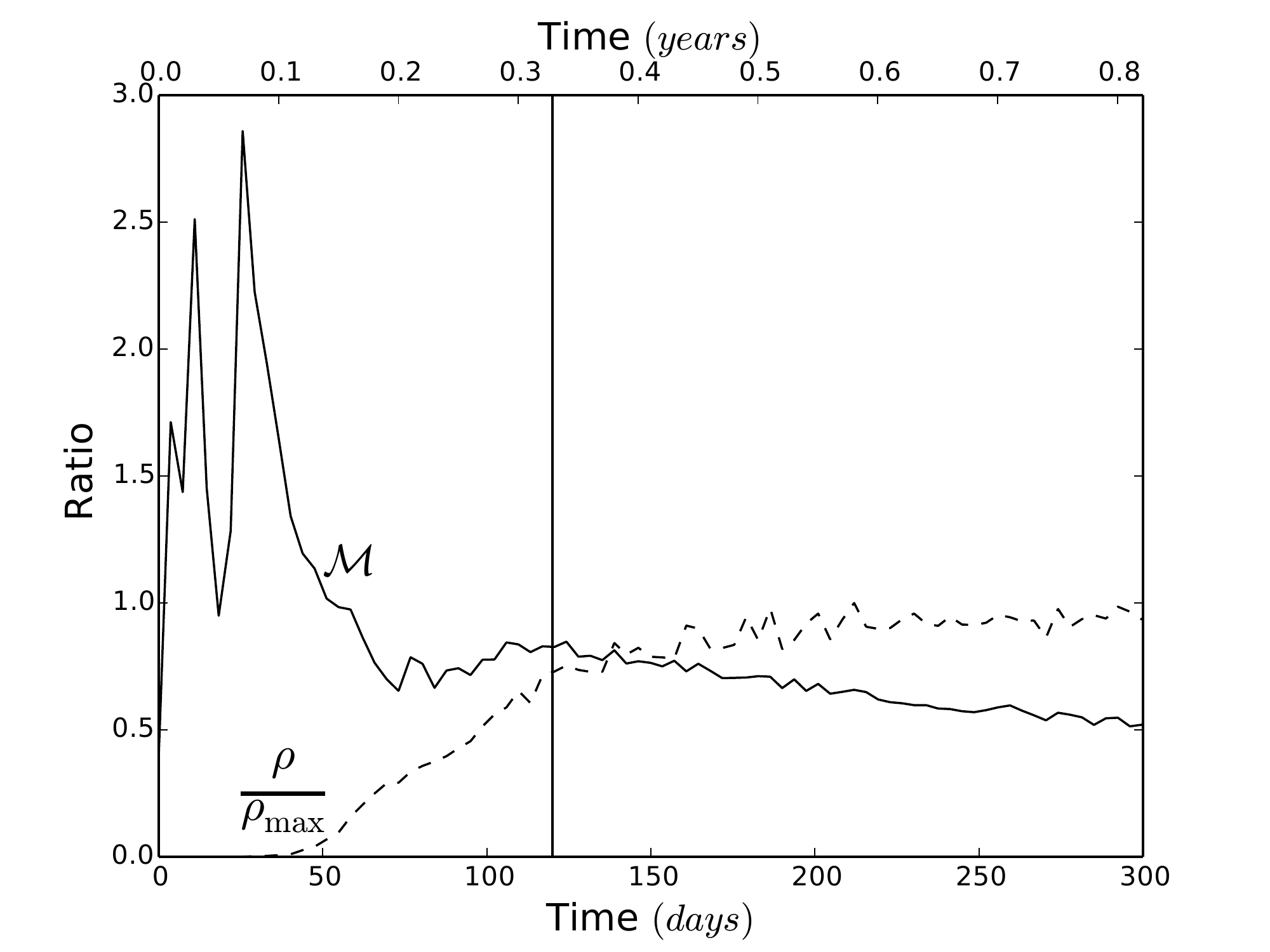}
\includegraphics[scale=0.23, trim=0.5cm 0.cm 1.0cm 1.4cm, clip]{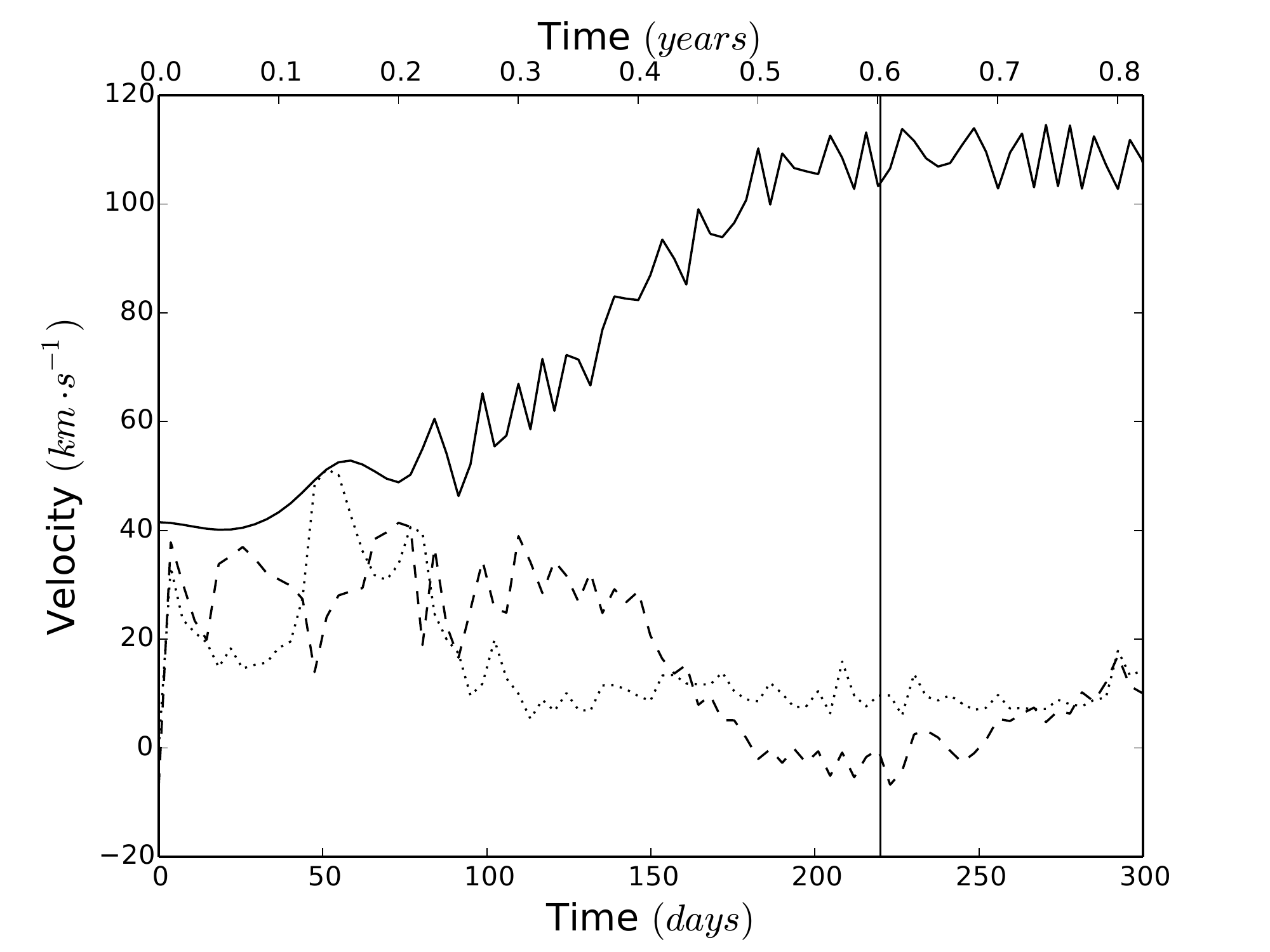}
\includegraphics[scale=0.23, trim=1.0cm 0.cm 1.0cm 1.4cm, clip]{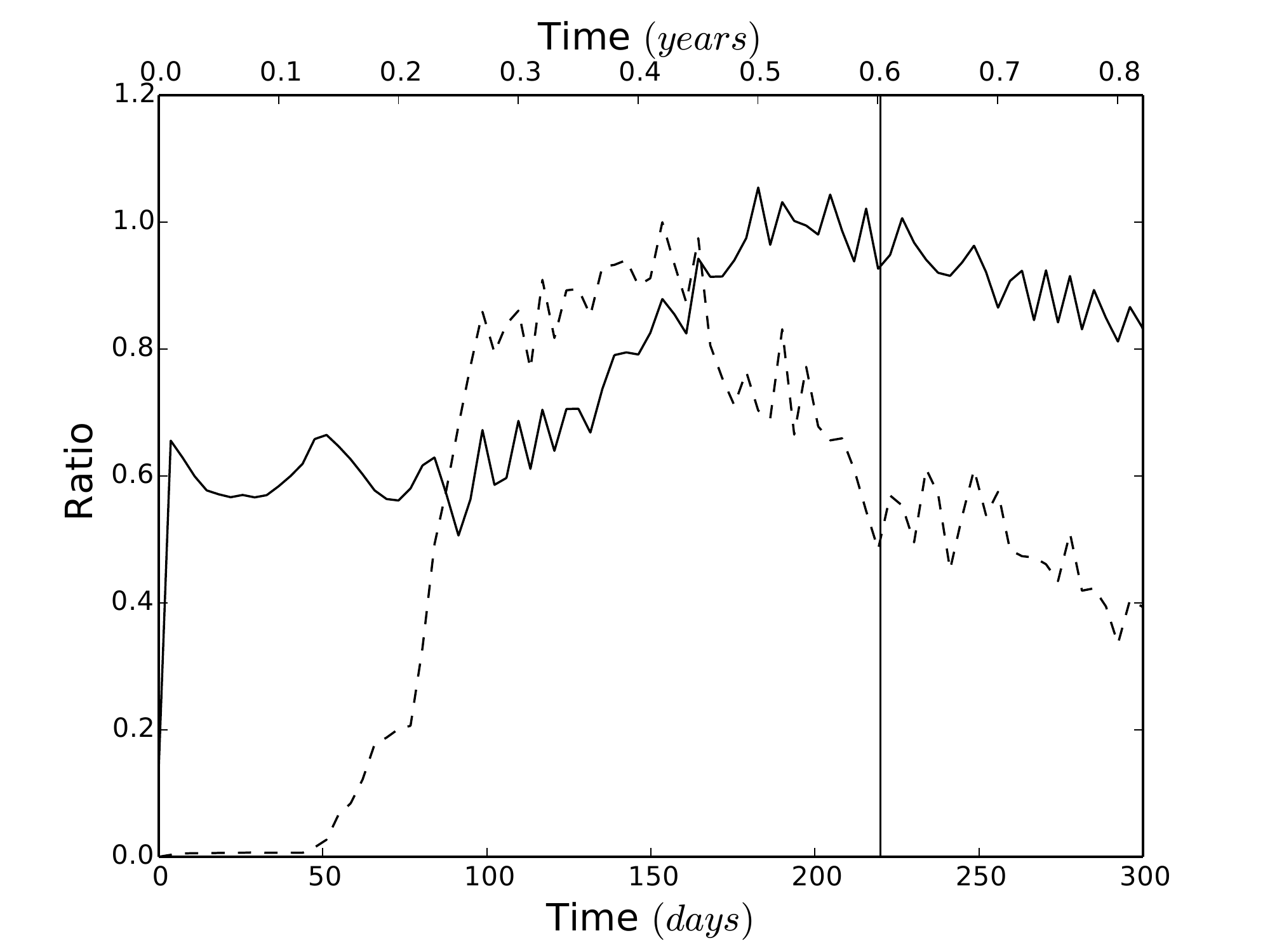}
\includegraphics[scale=0.23, trim=0.5cm 0.cm 1.0cm 1.4cm, clip]{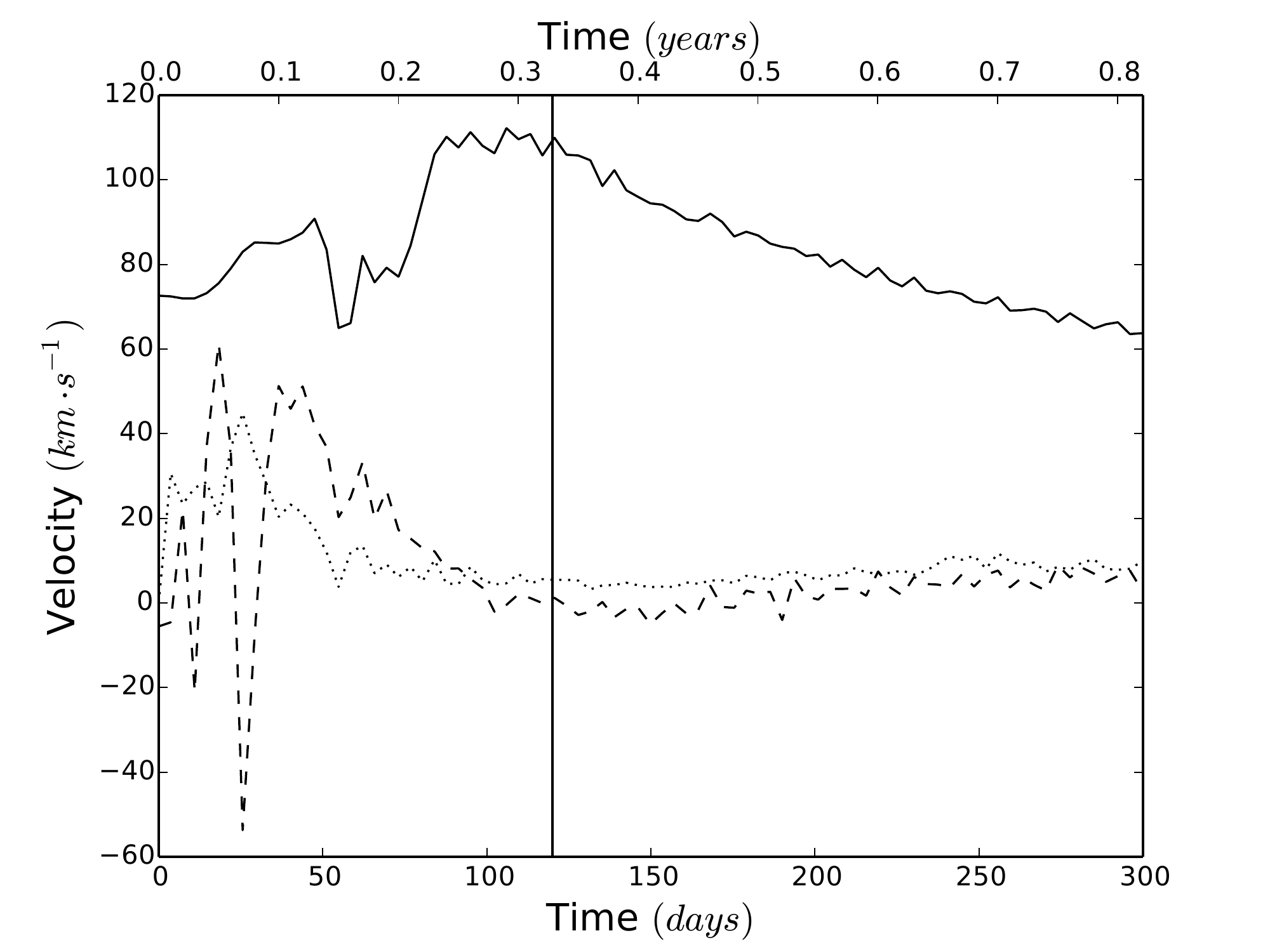}
\includegraphics[scale=0.23, trim=0.9cm 0.cm 1.0cm 1.4cm, clip]{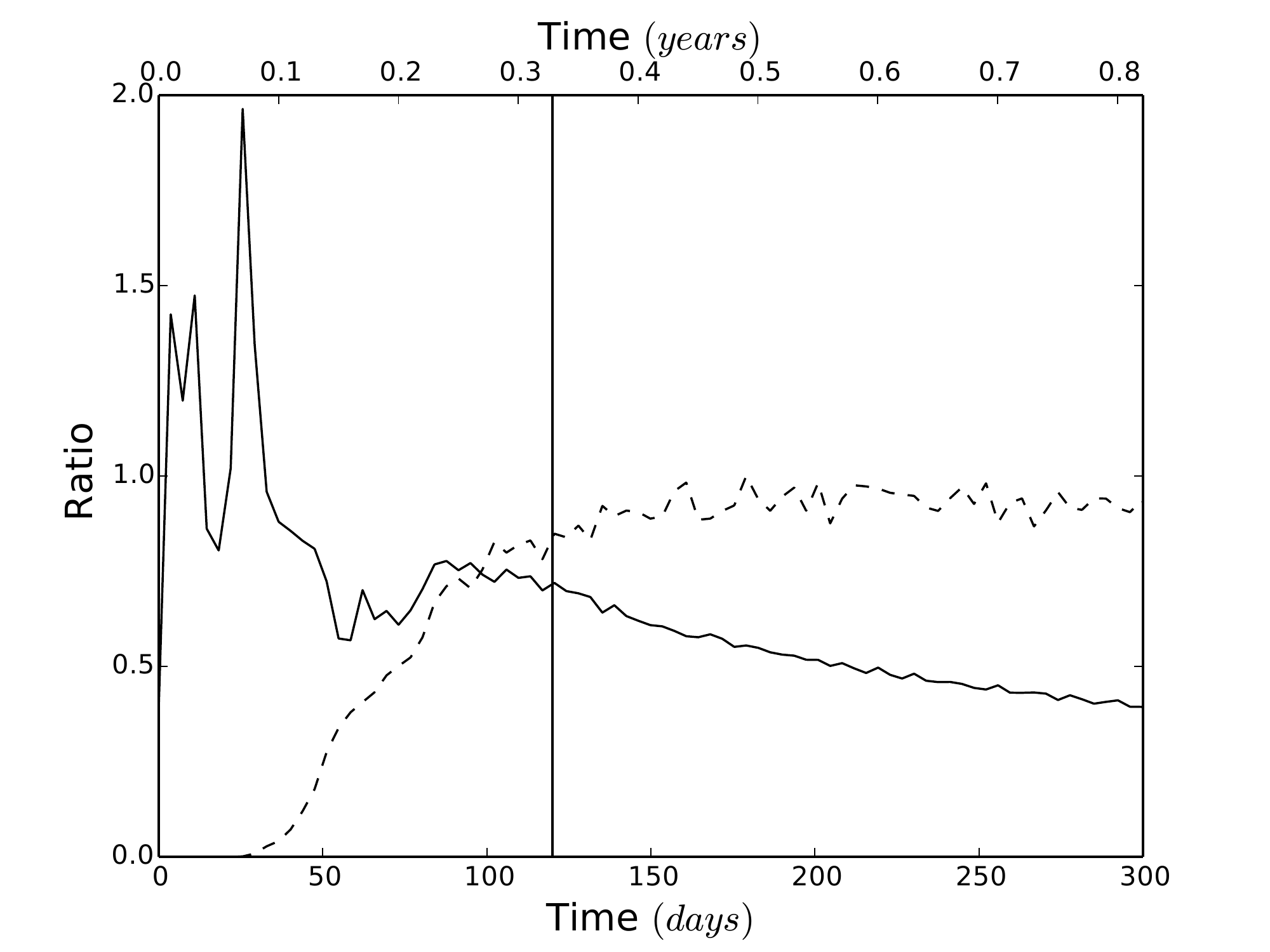}
\includegraphics[scale=0.23, trim=0.5cm 0.cm 1.0cm 1.4cm, clip]{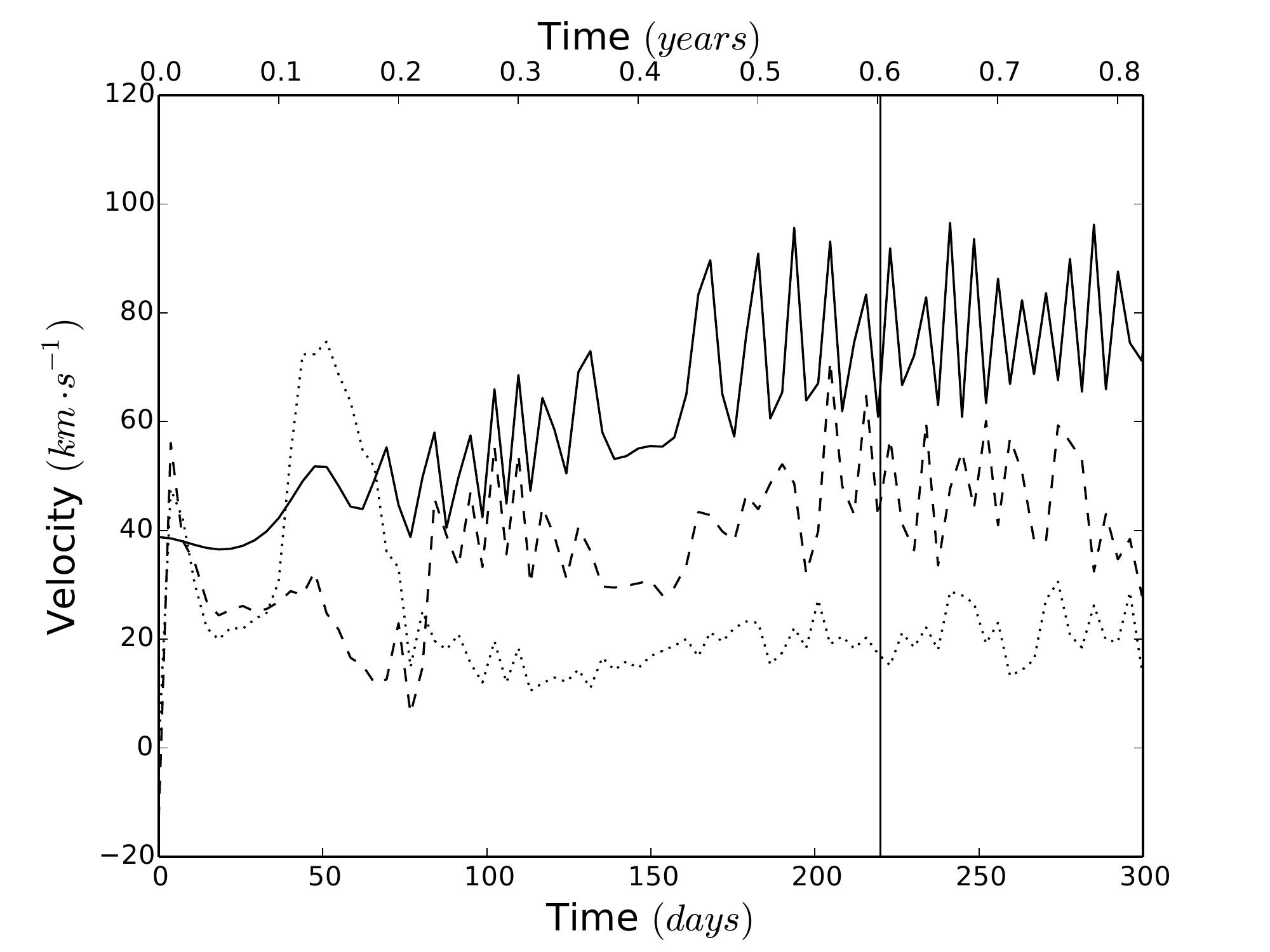}
\includegraphics[scale=0.23, trim=1.0cm 0.cm 1.0cm 1.4cm, clip]{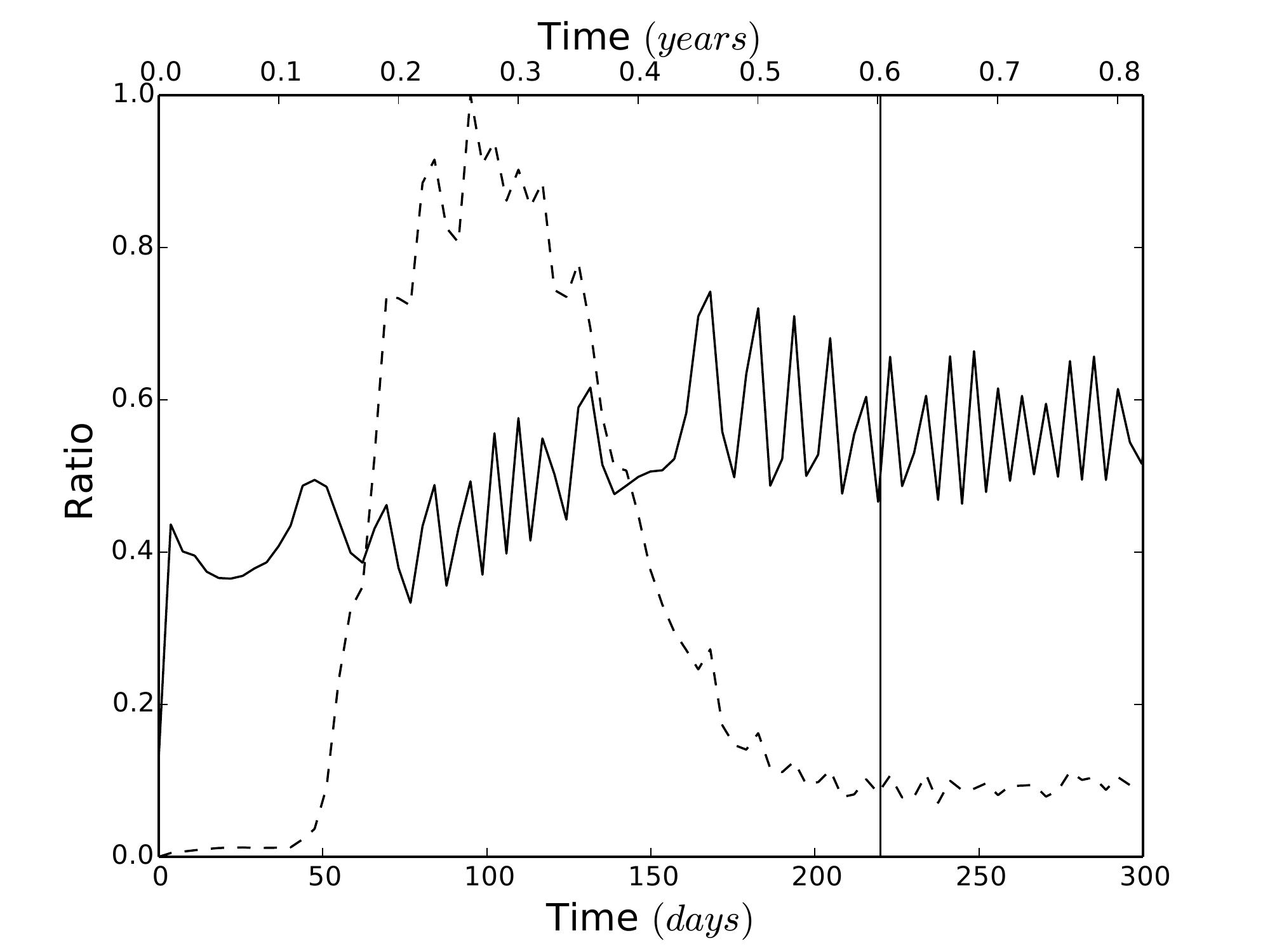}
\includegraphics[scale=0.23, trim=0.5cm 0.cm 1.0cm 1.4cm, clip]{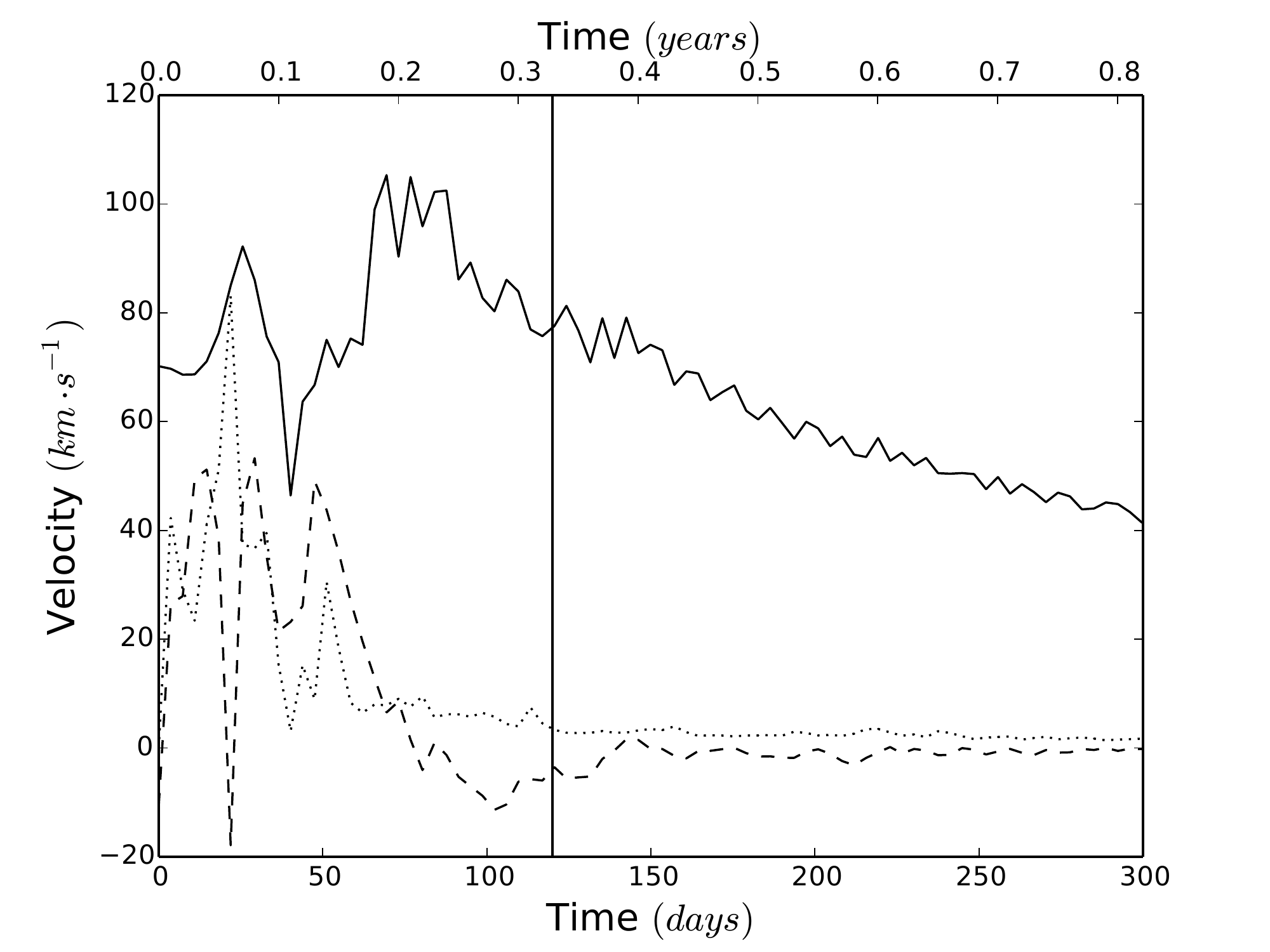}
\includegraphics[scale=0.23, trim=0.9cm 0.cm 1.0cm 1.4cm, clip]{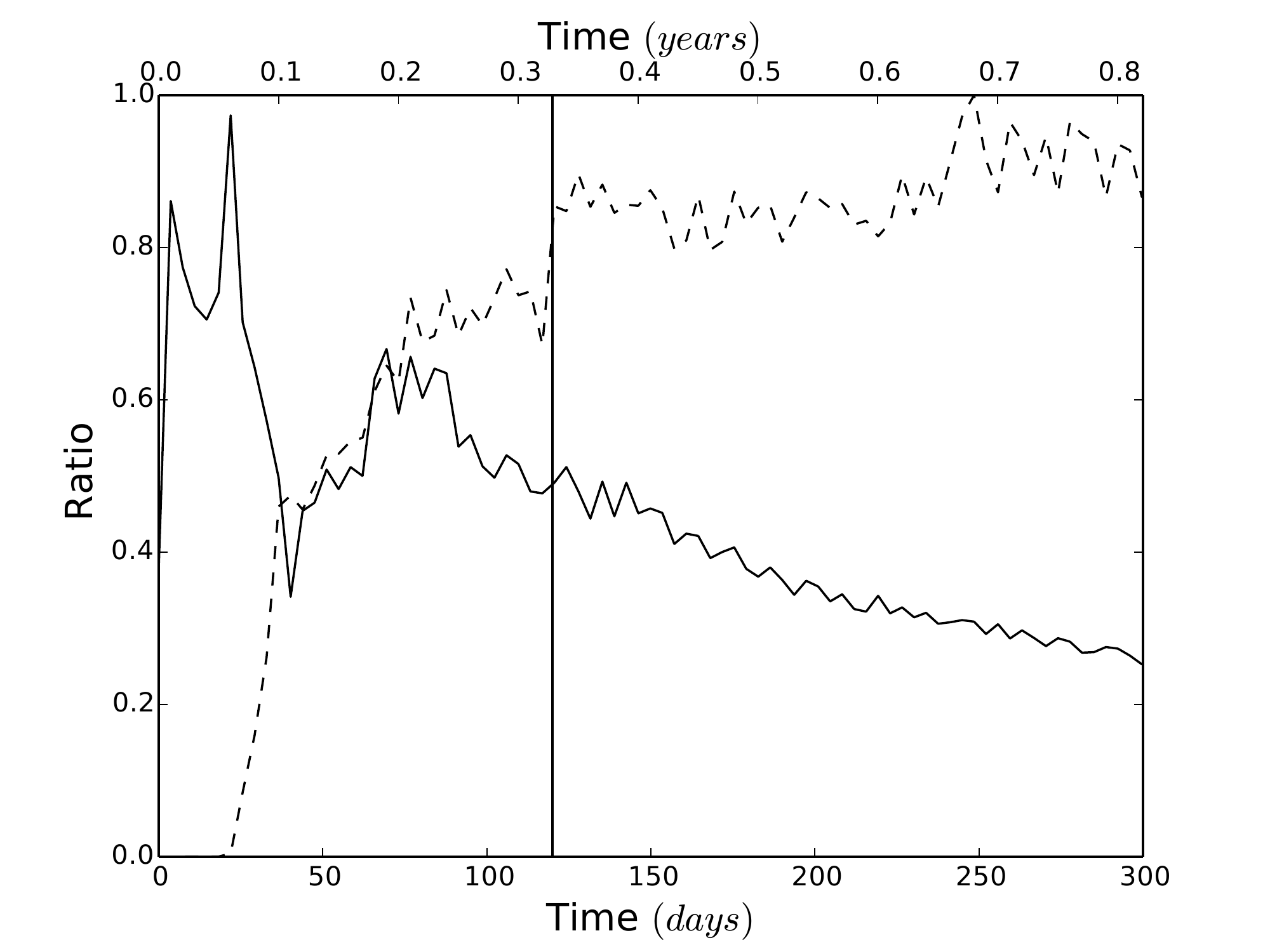}
\includegraphics[scale=0.23, trim=0.5cm 0.cm 1.0cm 1.4cm, clip]{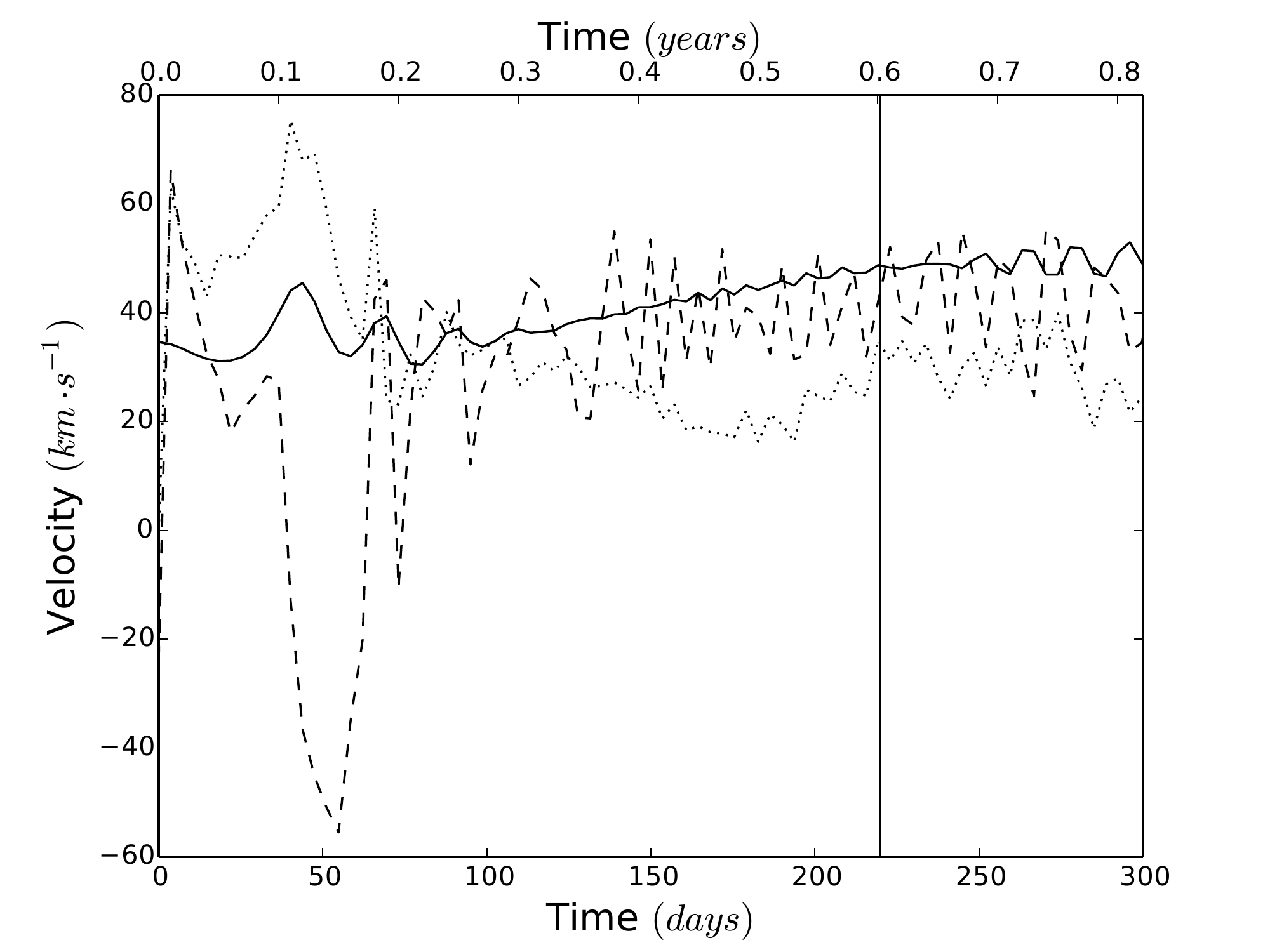}
\includegraphics[scale=0.23, trim=1.0cm 0.cm 1.0cm 1.4cm, clip]{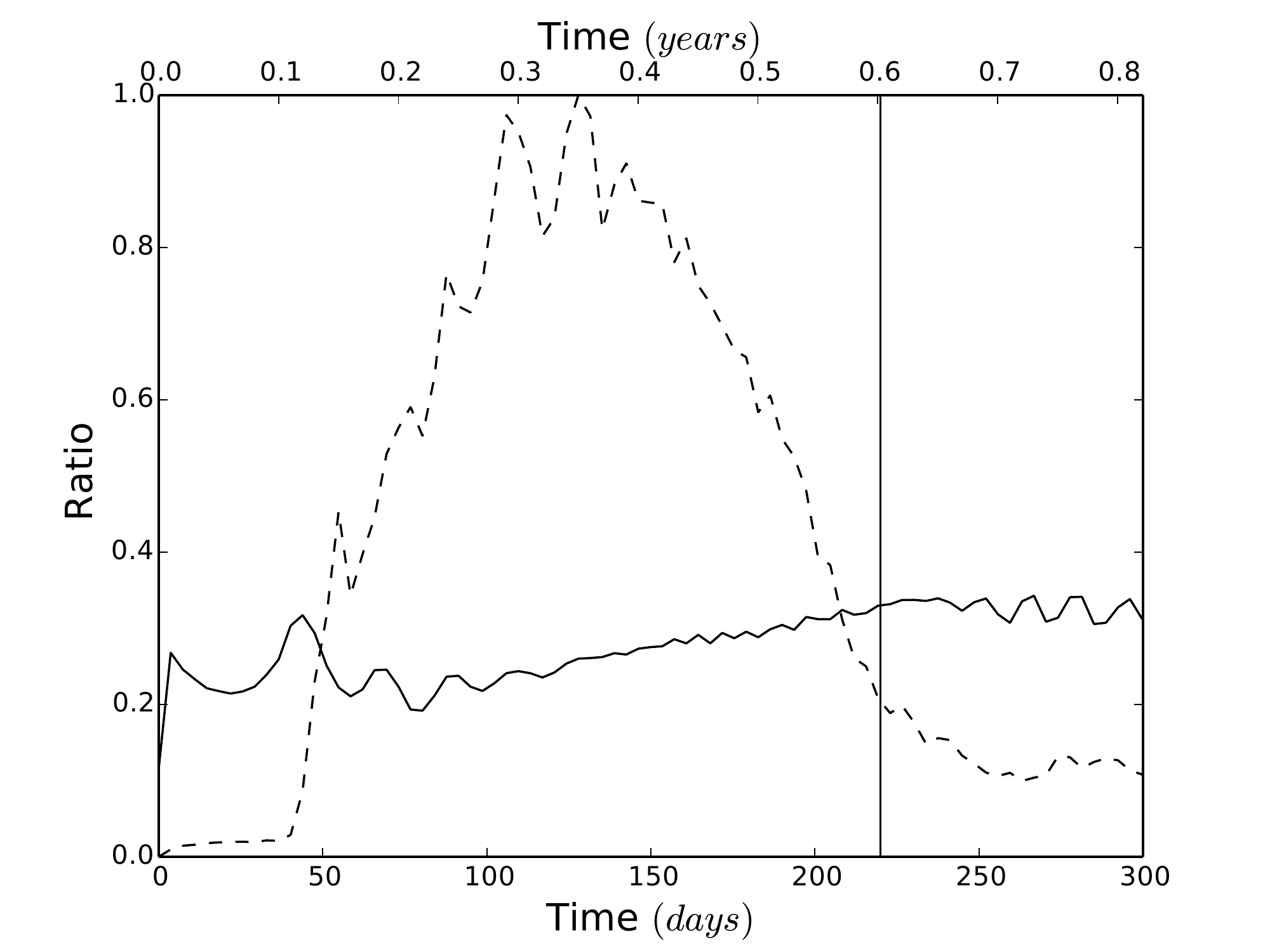}
\includegraphics[scale=0.23, trim=0.5cm 0.cm 1.0cm 1.4cm, clip]{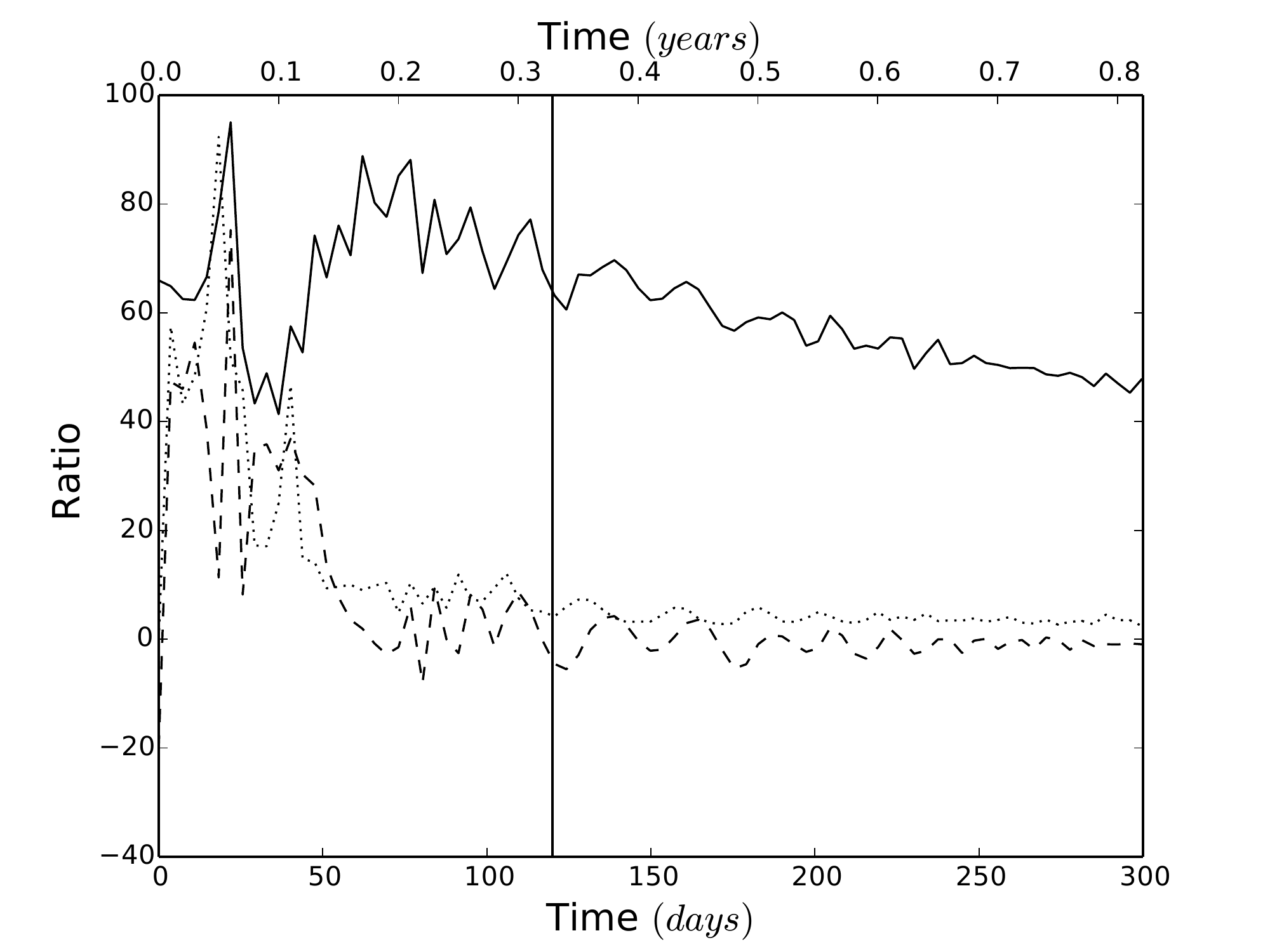}
\includegraphics[scale=0.23, trim=0.9cm 0.cm 1.0cm 1.4cm, clip]{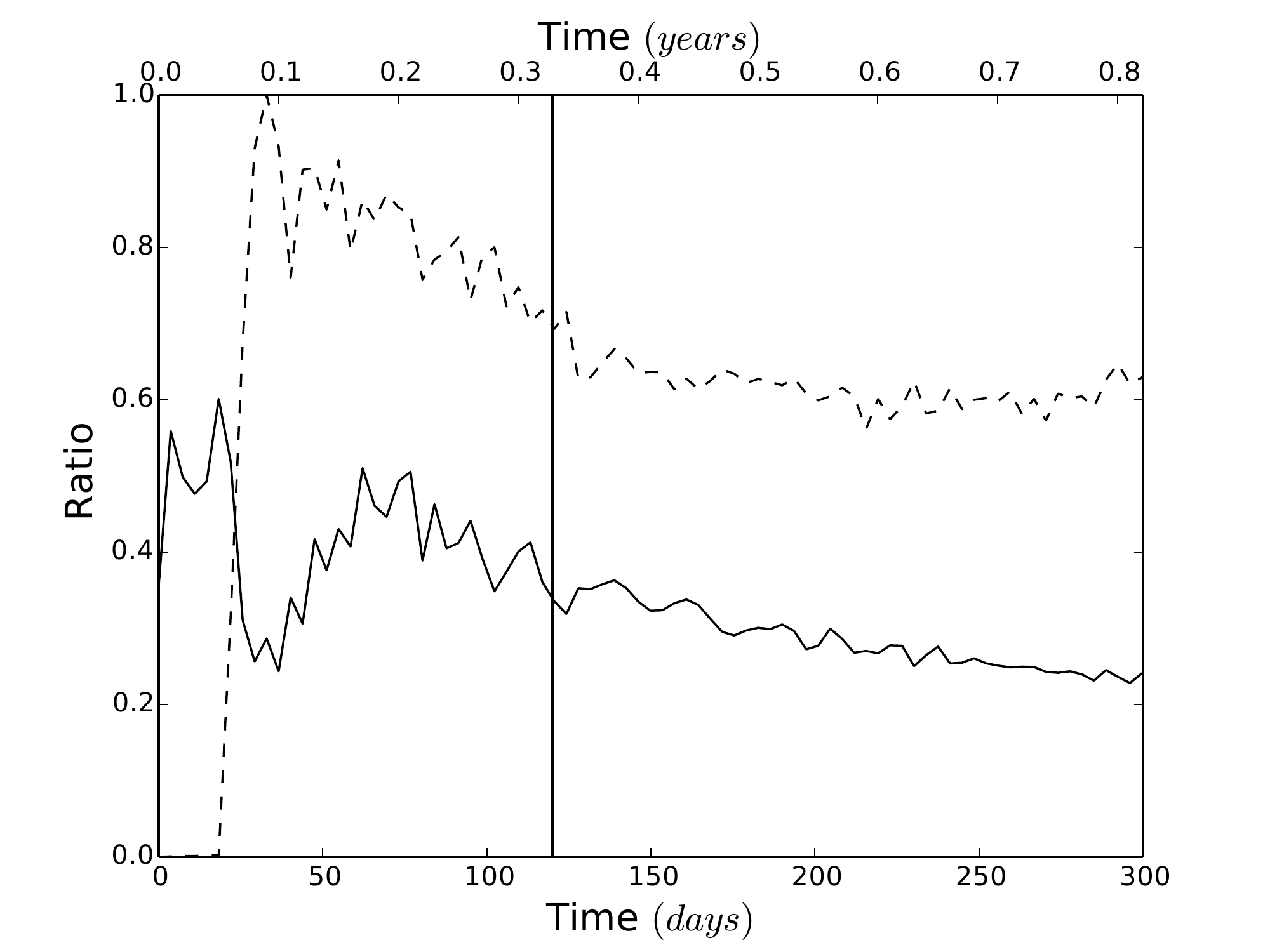}
\includegraphics[scale=0.23, trim=0.5cm 0.cm 1.0cm 1.4cm, clip]{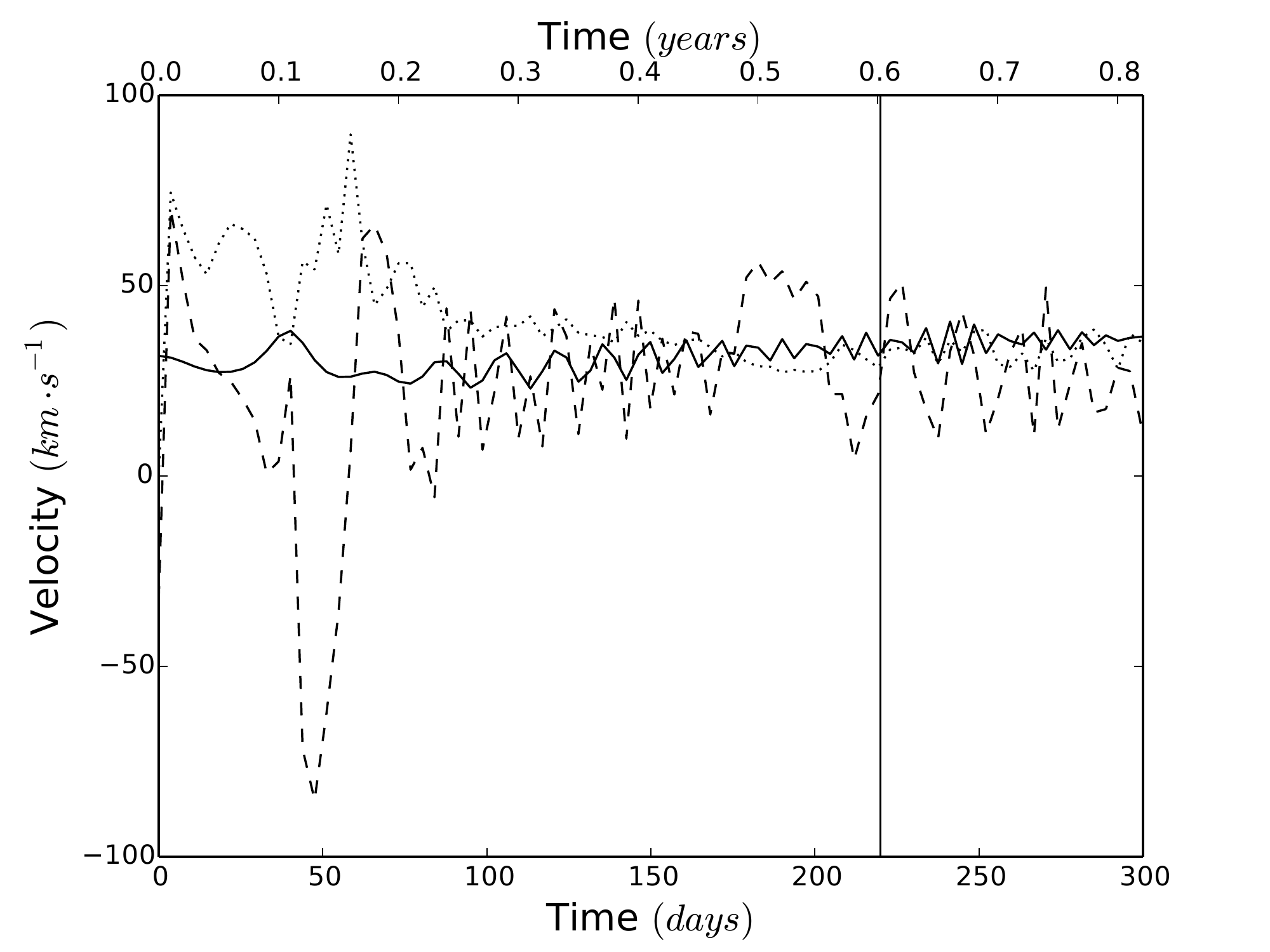}
\includegraphics[scale=0.23, trim=1.0cm 0.cm 1.0cm 1.4cm, clip]{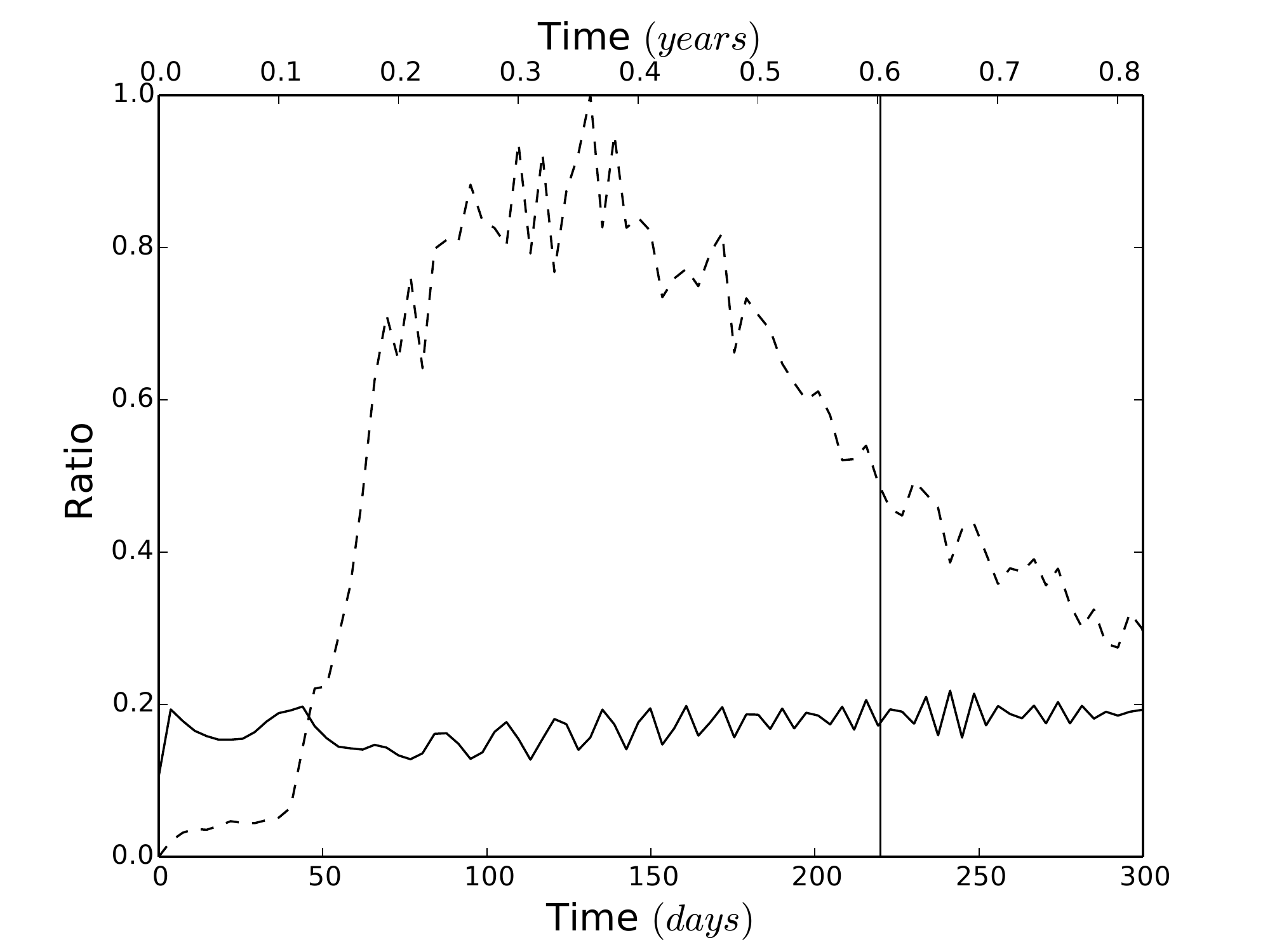}
\includegraphics[scale=0.23, trim=0.5cm 0.cm 1.0cm 1.4cm, clip]{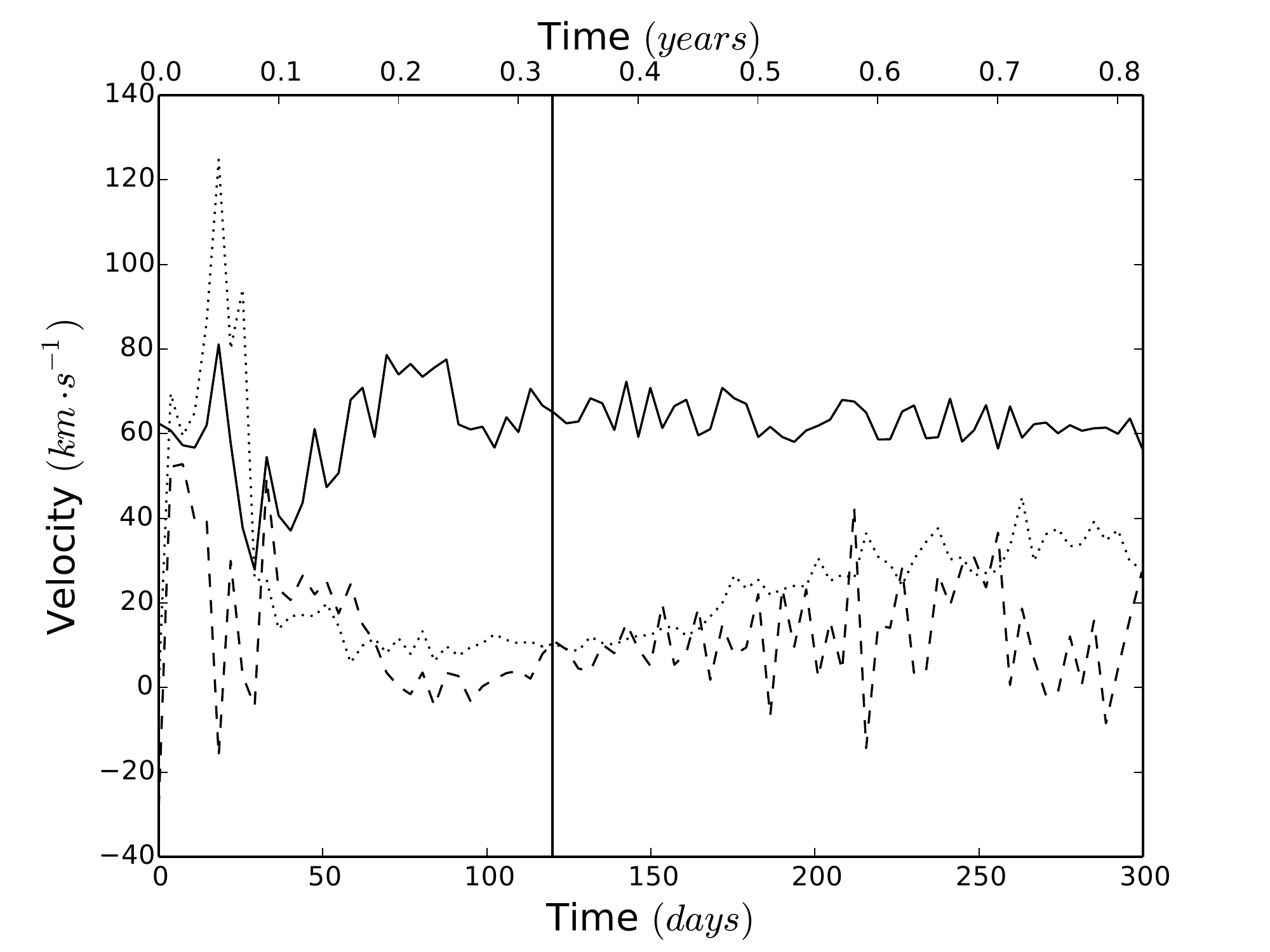}
\includegraphics[scale=0.23, trim=0.9cm 0.cm 1.0cm 1.4cm, clip]{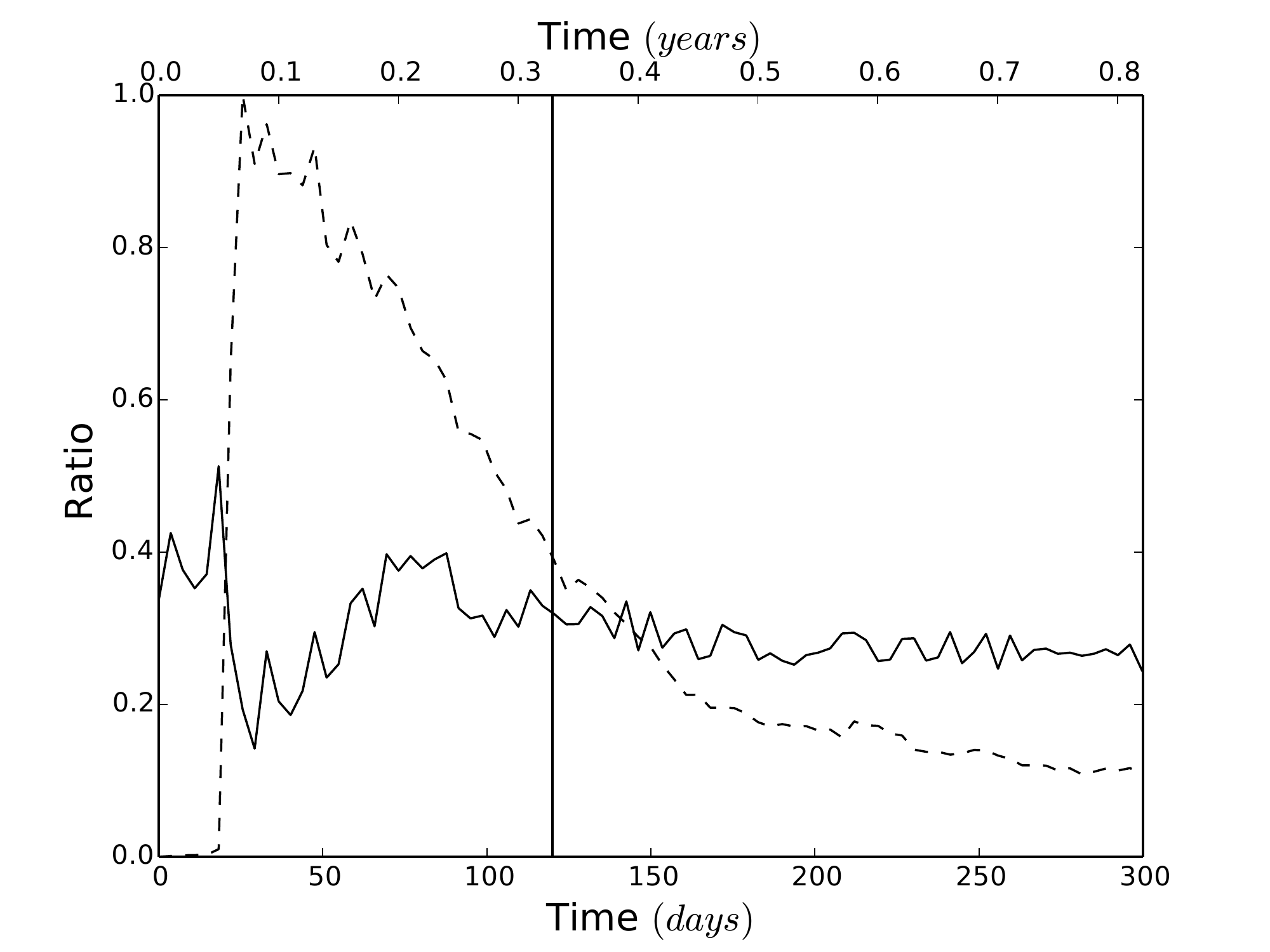}
\caption{\protect\footnotesize{{First column:} companion velocity (solid line), local average gas velocity projected on the direction of the companion velocity ($\left< v_{\mathrm{gas,\|}} \right>$, dashed line) and local average gas velocity perpendicular to the direction of the companion velocity ($\left< v_{\mathrm{gas,\bot}} \right>$, dotted line). From top to bottom we plot SIM1-SIM5.
{Second column:} companion Mach number (solid line) and normalised average gas density in the companion's proximity (dashed line). From top to bottom we plot SIM1-SIM5.
{Third column:} same as the first, but from top to bottom we plot SIM6-SIM10.
{Fourth column:} same as the second, but from top to bottom we plot SIM6-SIM10.
The vertical lines approximately represent the points where the rapid in-spiral terminates.}}
\label{fig:gravodrag}
\end{figure*}

Whether the simulations reproduce the gravitational drag correctly remains an open question.  The mechanism of orbital energy and angular momentum exchange in CE simulations has been scrutinised recently by \citet{Ricker2008}, \citet{Staff2016b} and \citet{Iaconi2017}. Particularly \citet{Staff2016b} argued that the gravitational drag as parameterised analytically by \citet{Iben1993} and \citet{Ostriker1999} is of the same order of magnitude as the drag force experienced in their simulations.

\citet{Staff2016b} showed the challenge to evaluate the gravitational drag during the rapid in-spiral in our CE simulations because the procedure to measure the force experienced by the companion is very noisy. Here we chose to trust that the gravitational drag analytical representation of \citet{Iben1993} is reasonably accurate and use it to appraise some aspects of our simulations. To do so we need to estimate quantities, such as density and relative velocity, within a Bondi radius of the companion. These are between 3 and 10~\rs\ for SIM5-10 and between 2 and 6~\rs for SIM6-10. We have used spheres centred around the companion with a radius of 2.5~\rs, but we have checked that the results do not change with larger spheres, up to 20~\rs (see also Iaconi et al. 2017). The Bondi radii are well resolved in our simulations where the innermost cell near the companion is 0.84~\rs. This is important considering that Staff et al. (2016b) showed that under-resolving the Bondi radius increases the strength of the drag force in their simulations.

According to the expression by \citet{Iben1993}, the drag goes to zero when the companion velocity is the same as that of the surrounding gas. In SIM4 and SIM5 the companion and gas velocities become approximately the same after the stabilisation of the orbit. However, this is not so in other simulations, showing that even after stabilisation of the orbit there is a considerable velocity contrast that should contribute to a continued drag. Bringing the gas into co-rotation may be something that is only achieved by more massive companions with lower mass envelopes. No co-rotation is observed in SIM6-10 by the time the in-spiral is halted by reaching the smoothing length, although SIM10, with the heaviest companion is close.

The density does decrease in the vicinity of the companion at approximately the same time as the slow down of the in-spiral in simulations SIM2-5 and SIM9-10, something that would contribute to a stabilisation of the orbit, but this does not happen in the other simulations. In particular the simulations with light companions show almost no change in the density around the companion because there is very little outflow of the envelope, indicating that the in-spiral of these objects is not halted by an evacuation of the orbit. 

Finally, the Mach number is effectively lower than unity in most simulations with only SIM6 and SIM7 showing variable Mach values peaking above unity in the very early part of the in-spiral. {There is no clear relation between the companion's velocity becoming subsonic and the slowdown of the orbit in any of our simulations, as was instead the case in the simulations of Staff et al. (2016b).}

\begin{figure*}
\centering     
\includegraphics[scale=0.23, trim=0.55cm 0.cm 1.0cm 0.cm, clip]{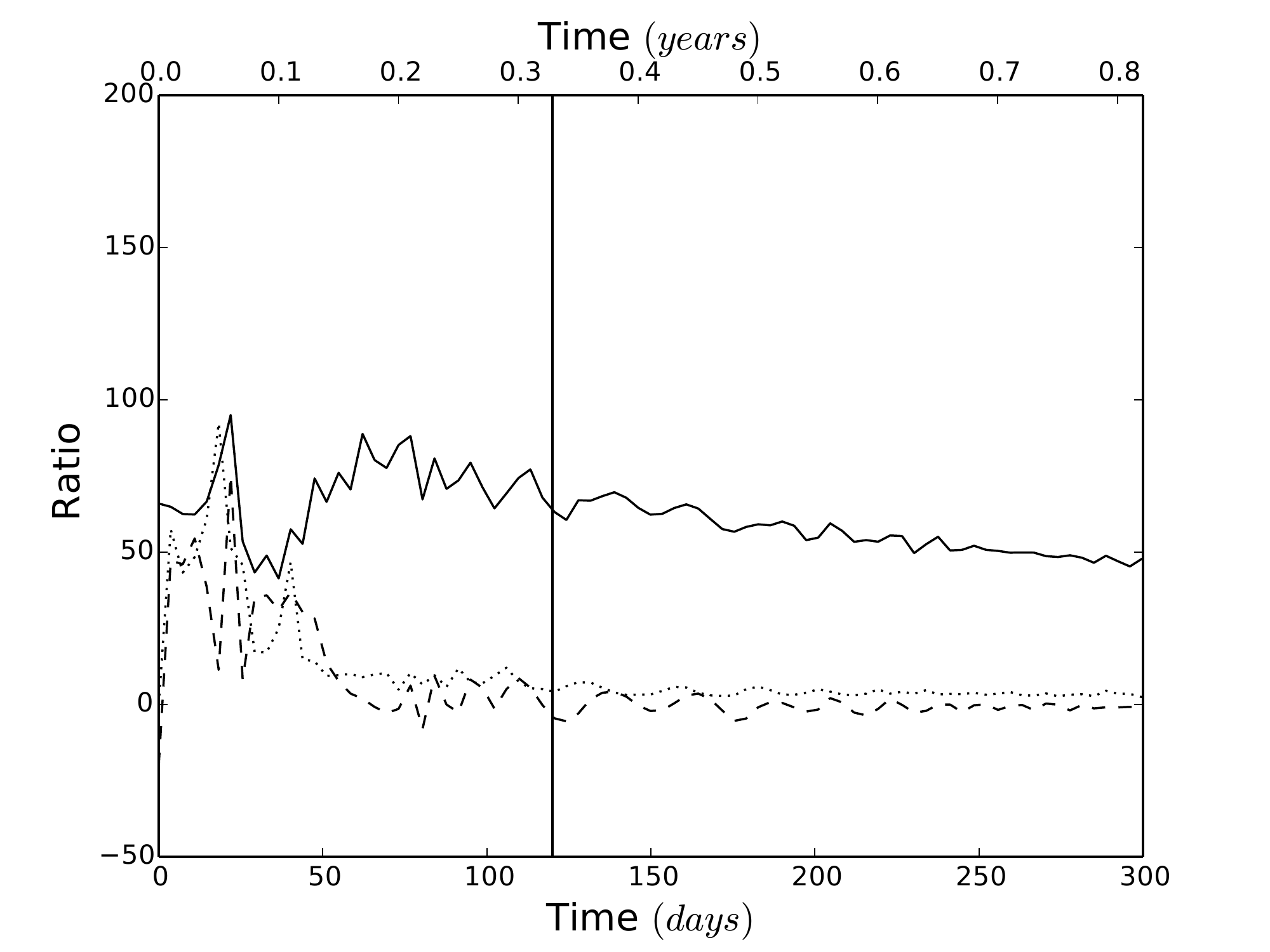}
\includegraphics[scale=0.23, trim=0.9cm   0.cm 1.0cm 0.cm, clip]{gravodrag_efficiency_fractional_quantities_3sphere_2MsunPrimary_128_2lev_2AUbox_M2_0p6.pdf}
\includegraphics[scale=0.23, trim=0.55cm 0.cm 1.0cm 0.cm, clip]{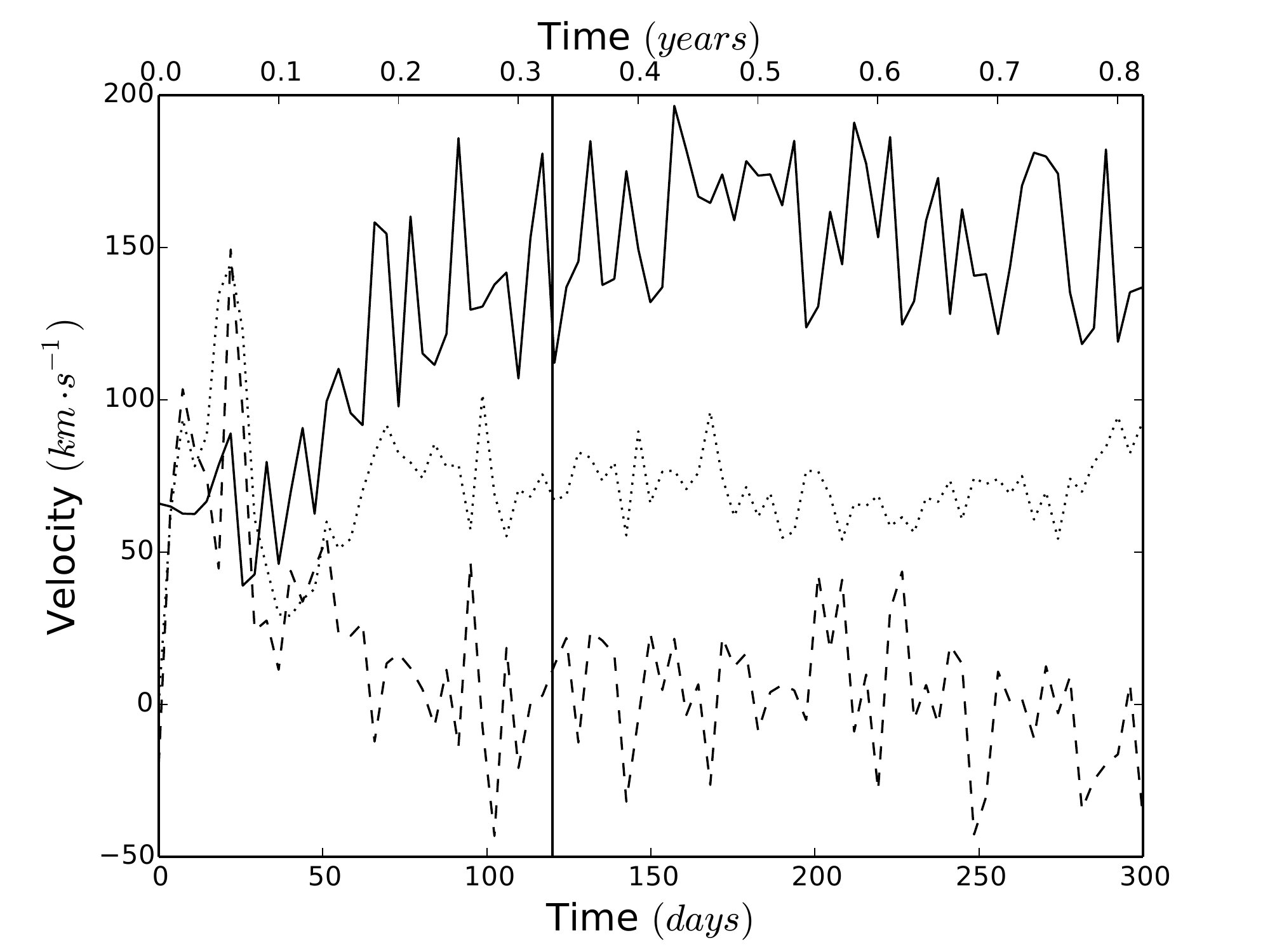}
\includegraphics[scale=0.23, trim=0.9cm   0.cm 1.0cm 0.cm, clip]{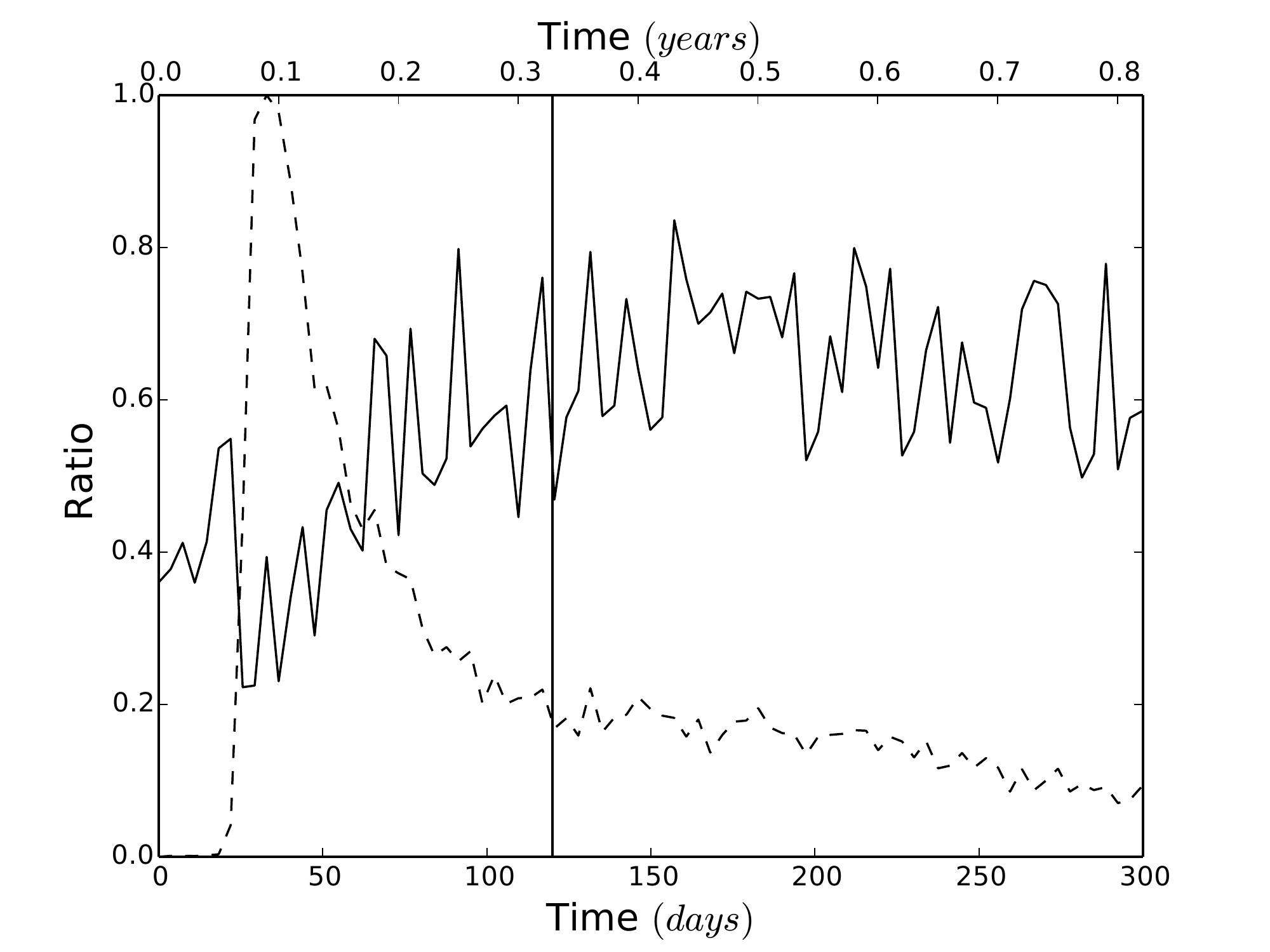}
\caption{\protect\footnotesize{{First column:} companion velocity (solid line), local average gas velocity projected on the direction of the companion velocity ($\left< v_{\mathrm{gas,\|}} \right>$, dashed line) and local average gas velocity perpendicular to the direction of the companion velocity ($\left< v_{\mathrm{gas,\bot}} \right>$, dotted line) for SIM9 (top: heavy primary with 0.6~\ms\ companion) and SIM11 (bottom: resolution twice as high).
{Second column:} companion Mach number (solid line) and normalised average gas density in the companion's proximity (dashed line) for SIM9 and SIM11.
The vertical lines approximately represent the points where the rapid in-spiral terminates.}}
\label{fig:gravodrag-resolution}
\end{figure*}

In Figure~\ref{fig:gravodrag-resolution} we display the same quantities shown in Figure~\ref{fig:gravodrag} but for SIM9 and SIM11 ($M_1 = 2.0$~\ms, $M_2=0.6$~\ms), where the only difference is a higher resolution in SIM11. The behaviours are similar for the first 40 days (as also expected by inspecting the in-spiral comparison in Figure~\ref{fig:resolution_study}). Then the companion velocity increases much more in SIM11 because the companion in-spirals deeper. The gas is not brought into co-rotation at either resolution, but at higher resolution the perpendicular gas velocity (outflow) is far more substantial, justifying the larger unbound mass fraction (Table~1). The density contrast at the time of orbital stabilisation is much lower for the more resolved simulation, possibly explaining the stabilisation itself, but also once again bearing witness to the much more efficient ejection of material for higher resolution simulations.

In conclusion while for SIM1-2 and SIM6-10 the final separation is of the order of the smoothing length, which shows the orbital stabilisation has a numerical cause, for SIM3-5 the primary cause of orbital slow down seems to be that the gas is close to co-rotation with the orbit.

\section{Discussion}
\label{sec:discussion}

\subsection{When should stars merge inside a common envelope?} 
\label{ssec:roche_lobe_radii_and_roche_radii}

SIM6-SIM10 end up in a very compact configuration and at higher resolution the final separation is even smaller (SIM11, SIM12). A post-CE binary survives as such if the final orbital separation does not allow any mass transfer, i.e., the two stars are smaller than their respective Roche lobe radii, and if the two stars are not tidally disrupted by each other's gravity. It is therefore important to determine whether the simulations point to the fact that these systems should merge.

The values of the final separation obtained may be considered upper limits on the grounds that they are all comparable to the size of the smoothed potential. If so, then we would tend to conclude that these system may merge. On the other hand, the uncertainties over angular momentum conservation may introduce a source of error, whereby we may suspect that the final separations are {\it lower} limits, invalidating the upper limit argument just made. However, if the entire non-conservation budget were to be attributed to the companion having in-spiralled too much in the ``lossy" simulation, then we would conclude that a 20\% loss of total angular momentum mean that the orbit should be 35\% larger than the simulations tell us. Increasing all the final separation values ($a_f$) for SIM6-10 by 35\%, does not increase the values significantly, and leaves them all still close to the respective smoothing lengths (in particular SIM7, SIM8 and SIM9 all have final separations that are smaller than the smoothing length even if augmented by 35\%). We leave a final conclusion to additional supporting simulations. However, the smaller final configurations typical of simulations with primaries with more bound envelopes still point to the possibility of a merger.

In partial support of the latter consideration, we notice that the final separations in similar simulations carried out by \citet{Nandez2016} point to similar, or even smaller final separations and we trust that their SPH code conserves angular momentum well. In further support of our conclusion that the final separations are indeed very small, we point to the fact that our side-by side grid vs. SPH comparisons have never given us reason to believe that the 10-15\%  lack of angular momentum conservation in grid simulations is a reason to discredit the final separations. This said, one must remain alert to the possible problems caused by these  issues with these simulations. 

Next we determine whether the companion would overflow its Roche lobe if it is at a distance $a_f$ from the core of the primary. For this purpouse, we assume that the companion is a low mass main sequence star. This is the simplest configuration: an AGB giant has suffered one common envelope phase with a less massive and less evolved companion. In this case the companion will be larger than the primary after the common envelope and it would be the first to overflow its Roche lobe radius. For this purpose we computed the Roche lobe radii of the companions ($R_{\mathrm{RL,2}}$) at the end of simulations SIM6-SIM12 using the approximation of \citet[][Table~\ref{tab:roche_lobe_radii_and_roche_radii}]{Eggleton1983}:

\begin{equation}
 \frac{R_{\mathrm{RL,2}}}{a_{\mathrm{f}}} = \frac{0.49 (M_2 / M_1)^{2/3}}{0.6 (M_2 / M_1)^{2/3} + \ln[1 + (M_2 / M_1)^{1/3}]} \ .
 \label{eq:roche_lobe_radius}
\end{equation} 

\noindent Additionally, we computed the Roche limit around the primary for companion disruption ($R_\mathrm{lim,1}$) using the analytic formula from \citet{Carroll2006}:

\begin{equation}
 R_\mathrm{lim,1} = 2.456 \big( \frac{M_1}{M_2} \big)^{1/3} R_2 \ ,
 \label{eq:roche_limit}
\end{equation}

\noindent where we have assumed the companions to be main sequence stars and estimated their radii by using the mass-radius relation for low mass main sequence stars ($R_2 = M_2^{0.8}$ all in solar units; \citealt{Torres2010}). Results are shown in Table~\ref{tab:roche_lobe_radii_and_roche_radii}.

The companions' radii, $R_2$, are smaller than their Roche lobe radii, $R_{\mathrm{RL,2}}$, for SIM6-10. However, for the more resolved SIM11 and SIM12, $R_2$ values are larger than the radius of the corresponding Roche lobe, even if we increased the final separation by 35\% to account for a 20\% loss of total angular momentum. Moreover, at such low final orbital separations, the companion is also well within the Roche limit of the primary's core, which would result in its disruption. According to these simulations, this companion should merge with the giant core inside the CE. Energetic considerations would suggest that a 0.6~\ms\ companion that in-spirals into the envelope of a 2~\ms, 66~\rs\ giant to a final orbital separation such that the companion does not quite fill its Roche lobe (companion fills its Roche lobe at a separation of 1.6~\rs, while the final separations derived for an efficiency parameter in the range $\alpha_{CE}= 0.5 -- 1.0$ would be 1.7 - 3.4~\rs) would deliver enough orbital energy to unbind the envelope. The simulations, however, tell us that the orbital in-spiral would go deeper and bring the companion to merging with the core. 

The simulations of \citet{Nandez2016} result in smaller orbital separations overall compared to ours. Although the secondary would overflow its Roche lobe if it were a main sequence star, their work is aimed at understanding compact double white dwarfs. If the secondary is a white dwarf in their simulation (or in ours), then it would fit comfortably within its Roche lobe and a merger would be avoided. 

Since companions in CE simulations are always simulated as point masses with no size nor structure, we can use the simulations of  \citet{Nandez2016} to reinforce our conclusions. Their primaries are in the range 1.2-1.8~\ms\ with envelope radii smaller than those used here (in the range 20-60~\rs), while their companion masses are in the range 0.32-0.40~\ms. Among these simulations the 1.8~\ms, 40~\rs\ simulations are the most similar to our 2~\ms\ ones. Their final separation for that simulation is 1.182~\rs, similar to the separations reached by our SIM6-10. If their companion were considered to be a main sequence star, it would likely be in Roche lobe contact ($R_{\rm 2,RL}$ = 0.43~\rs\ at that separation and its radius would be $\sim$0.4~\rs), suggesting a merger. This is even more so for their simulations with a more compact, 16~\rs\ giant. If these results are correct then their simulations, similarly to ours, suggest that ~2~\ms\ giants may readily merge with a main sequence companion as massive as 0.3 or even 0.6~\ms.  

\begin{table}
\begin{center}
\begin{adjustbox}{max width=\textwidth}
\begin{tabular}{cccccc}
\hline
ID & $a_{\mathrm{f}}$  &  $M_2$ & $R_2$ & $R_\mathrm{lim,1}$ & $R_{\mathrm{RL,2}}$ \\
 & (\rs) & (\ms)   & (\rs) & (\rs) & (\rs) \\
\hline
SIM6  & 2.2 &  0.1  & 0.16 & 0.62 & 0.59 \\
SIM7  & 1.8 &   0.15 & 0.21 & 0.71 & 0.54 \\
SIM8  & 1.6 &  0.3  & 0.38 & 1.02 & 0.57 \\
SIM9  & 1.8 &   0.6  & 0.66 & 1.40 & 0.75 \\
SIM10 & 2.8 &   0.9  & 0.92 & 1.71 & 1.27 \\
\hline
SIM11 & 0.67&  0.6  & 0.66 & 1.40 & 0.28 \\
SIM12 & 0.87&  0.6  & 0.66 & 1.40 & 0.36 \\
\hline
\end{tabular}
\end{adjustbox}
\end{center}
 \begin{quote}
  \caption{\protect\footnotesize{Roche limit of the primary's core ($R_\mathrm{lim,1}$) and Roche lobe radius of the companion ($R_{\mathrm{RL,2}}$) at the end of SIM6-SIM12 ($M_1=2.0$~\ms) compared to the final orbital separation ($a_{\mathrm{f}}$) and the mass and radius of the companion ($M_2$ and $R_2$). The higher resolution SIM11 and SIM12 could result in a merger.}} \label{tab:roche_lobe_radii_and_roche_radii}
 \end{quote}
\end{table}

\subsection{How bound is the bound gas?}
\label{ssec:bound_mass}
In this work we have increased the binding energy of the envelope with the aim of increasing the strength of the gravitational interaction. However, there is clearly a feedback mechanism: the stronger the interaction, the more the envelope is lifted out of the region of interaction (though not necessarily ejected), thereby weakening the gravitational interaction and reducing the ability to mine additional orbital energy. Although the orbit may have delivered enough energy to match or surpass the total initial envelope binding energy, inefficiencies clearly leave much mass still bound. Inefficiencies in our code, which is adiabatic and does not radiate, equate to having ejected some mass with  velocities in excess of the escape velocity and having increased the total energy of bound gas by heating and increasing its kinetic energy. {\it In other words the average energy of the envelope can be zero after the in-spiral,  but not all gas parcels have zero total energy and are therefore unbound. }

The percentage value of unbound mass listed in Table~\ref{tab:simulation_parameters} does not fully describe how a given simulation may have come close to unbinding the envelope: ninety per cent of the mass may still be bound, but only just. Here we investigate how close to being unbound, the {\it bound part} of the envelope is when the rapid in-spiral is completed, by comparing the binding energy of the {\it bound} envelope at the beginning and end of the simulations (Table~\ref{tab:envelope_binding_status}). 

Computing the total energy of the gas at the beginning of the simulations is trivial, since it is all contained inside the computational domain\footnote{We do not include the effect of the companion in calculating the binding energy of the envelope at time zero, column 2, Table~\ref{tab:envelope_binding_status}, but we do check the effect that the companion would have, column~3, Table~\ref{tab:envelope_binding_status}.}. On the other hand, calculating any quantities that involve mass that has left the grid is approximate. Following the method devised by Iaconi et al. (2017), we determine the total energy of the {\it bound} gas as it crosses the domain boundary. In this way we calculate the total energy of the bound gas that left the computational domain by the end of the simulation. The binding energy of the bound gas at the end of the simulation calculated in this way is listed in Table~\ref{tab:envelope_binding_status}, Column~4, while listed in Columns 5 and 6 are the total energies inside and outside the computational domain, respectively, at 300 days (here we see that even a miscalculation of the energy of the bound gas that leaves the domain would not affect our conclusion). In Column~7 we can see that by the end of the simulation the binding energy of the bound gas is still a substantial fraction of the initial binding energy of the entire envelope. Hence we conclude that at the end of the simulation the bound gas is {\it tightly} bound rather than loosely bound. 

\begin{table*}
\begin{center}
\begin{adjustbox}{max width=\textwidth}
\begin{tabular}{ccccccc}
\hline
ID & $E_{\mathrm{i}}$ &  $E_{\mathrm{i,comp}}$ & $E_{\mathrm{f,tot}}$ & $E_{\mathrm{f,in}}$ & $E_{\mathrm{f,out}}$  & $\frac{|E_{\mathrm{f,tot}}|}{|E_{\mathrm{i}}|}$ \\
& no comp. &  w/ comp. &w/ comp.& w/ comp.& w/ comp.& \\
& (erg) &  (erg) & (erg) & (erg) & (erg) & (\%) \\
\hline
SIM1 & $-2.31 \times 10^{46}$ &  $-2.46 \times 10^{46}$ & $-1.35 \times 10^{46}$ & $-1.32 \times 10^{46}$ & $-2.81 \times 10^{44}$ & 58 \\
SIM2 & $-2.31 \times 10^{46}$ &  $-2.56 \times 10^{46}$ & $-9.74 \times 10^{45}$ & $-9.02 \times 10^{45}$ & $-7.18 \times 10^{44}$ & 42 \\
SIM3 & $-2.31 \times 10^{46}$ &  $-2.83 \times 10^{46}$ & $-7.96 \times 10^{45}$ & $-6.90 \times 10^{45}$ & $-1.06 \times 10^{45}$ & 34 \\
SIM4 & $-2.31 \times 10^{46}$ &  $-3.31 \times 10^{46}$ & $-8.84 \times 10^{45}$ & $-6.97 \times 10^{45}$ & $-1.87 \times 10^{45}$ & 38 \\
SIM5 & $-2.31 \times 10^{46}$ &  $-3.74 \times 10^{46}$ & $-1.20 \times 10^{46}$ & $-8.94 \times 10^{45}$ & $-3.05 \times 10^{45}$ & 52 \\
\hline
SIM6 & $-1.32 \times 10^{47}$ &  $-1.41 \times 10^{47}$ & $-1.30 \times 10^{47}$ & $-1.30 \times 10^{47}$ & $-2.57 \times 10^{44}$ & 98 \\
SIM7 & $-1.32 \times 10^{47}$ &  $-1.45 \times 10^{47}$ & $-1.28 \times 10^{47}$ & $-1.27 \times 10^{47}$ & $-5.03 \times 10^{44}$ & 97 \\
SIM8 & $-1.32 \times 10^{47}$ &  $-1.58 \times 10^{47}$ & $-1.21 \times 10^{47}$ & $-1.20 \times 10^{47}$ & $-1.16 \times 10^{45}$ & 92 \\
SIM9 & $-1.32 \times 10^{47}$ &  $-1.80 \times 10^{47}$ & $-1.08 \times 10^{47}$ & $-1.04 \times 10^{47}$ & $-3.93 \times 10^{45}$ & 82 \\
SIM10 & $-1.32 \times 10^{47}$ &  $-2.01 \times 10^{47}$ & $-8.51 \times 10^{46}$ & $-7.88 \times 10^{46}$ & $-6.25 \times 10^{45}$ & 64 \\
\hline
\end{tabular}
\end{adjustbox}
\end{center}
 \begin{quote}
  \caption{\protect\footnotesize{Binding energy of the {\it bound} envelope at the beginning ($E_i$) and at 300~days from the beginning ($E_f$) of the simulations. We consider the contribution of the companion to the binding energy (columns 3, 4, 5, 6) and list the components of $E_f$ associated with bound gas inside and outside the computational domain (columns 5 and 6, respectively). Finally, we list the bound gas energy at the end of the simulations as a percentage of the initial value (column 7).}} \label{tab:envelope_binding_status}
 \end{quote}
\end{table*}

Any energy contribution that may unbind the bound gas should therefore be substantial. 
We also highlight that, although the percentage of unbound mass is similar between the simulations with a lighter or heavier envelope (Table~\ref{tab:simulation_parameters}), the lighter envelopes are closer to being unbound than the heavier ones. This is particularly evident for SIM6 and SIM7, where not only almost no mass is unbound, but also the amount of envelope removed from the initial volume of the giant is very low.

The idea of recombination energy produced in the optically thick zones of the expanding layers of the envelope as they cool down has shown promising results. \citet{Nandez2015} and \citet{Nandez2016} manage to unbind more than $99\%$ of the giant's envelope in  all their simulations. According to them, the budget of recombination energy available to be injected into the envelope ($E_{\mathrm{rec,ini}}$) ranged from $2.059 \times 10^{46}$~erg (for 1~\ms\ stars) to $4.676 \times 10^{46}$~erg (for 2~\ms\ stars) and becomes available after the in-spiral. We can use their formula to gauge whether recombination energy would unbind the bound gas in our simulations (section~3 of \citet{Nandez2016}):

\begin{equation}
E_{\mathrm{rec,ini}} = \eta (M_1 - M_{\mathrm{c}}) \ ,
\label{eq:initial_recombination_energy}
\end{equation}
with the same value of $\eta = 1.6 \times 10^{13}$~erg~g$^{-1}$, hence assuming that our stars have a chemical composition similar to those of \citet{Nandez2016}, we obtain $E_{\mathrm{rec,ini}}$  equal to $1.55 \times 10^{46}$~erg for the primary we use in SIM1-SIM5 and $5.03 \times 10^{46}$~erg for the one of SIM6-SIM10. These energies are large  enough to unbind the remaining portions of the envelope in SIM1-SIM5, but insufficient in the case of SIM6-SIM10.
We additionally highlight that for increasing companion mass we obtain values of $E_{\mathrm{bin,f}}$ increasingly closer to $E_{\mathrm{rec,ini}}$. This is not unexpected, since a more massive companion transfers more orbital energy and angular momentum to the envelope gas.

\subsection{Unbinding efficiency}
\label{ssec:unbinding_inefficiency}
Population synthesis codes use an energy balance method, mediated by an efficiency parameter, $\alpha_{\rm CE}$, to determine the final separation after the in-spiral (for a comprehensive list of population codes using common envelope formalisms see \citealt{DeMarco2017}). Simulations could in principle be used to determine $\alpha_{\rm CE}$, because the initial and final parameters are known. However, it is not meaningful to calculate $\alpha_{\rm CE}$ if the simulations do not achieve complete envelope unbinding. 

Here we use the energy formalism to show that even when enough orbital energy is delivered this does not necessarily mean that the envelope is unbound. We show here that the part of the envelope that is unbound, leaves the system with velocity well in excess of the escape value and the part that is still bound, has nonetheless received energy and is, but the end of the simulation less bound than it was at the beginning. In the case of our simulations, and of all simulations that do not make use of recombination energy, the lack of total unbinding is due to this form of inefficiency. 

Here, we take SIM10 as an example. It is clear that the delivered orbital energy: 

\begin{equation}
\Delta E_{\mathrm{orb}} = \frac{G M_{\mathrm{c}} M_2}{2a_{\mathrm{f}}} - \frac{G M_1 M_2}{2a_{\mathrm{i}}} = 1.89 \times 10^{47}~erg\ ,
\label{eq:orbital_energy_released}
\end{equation}

\noindent is sufficient to overcome the envelope binding energy of $-1.32 \times 10^{47}$~erg, calculated as the sum of potential, thermal and kinetic energies. So why is the envelope not unbound? The reason is that the orbital energy is expended to eject a fraction of the envelope at faster than the escape speed and to increase the energy of the remaining fraction, but not enough to unbind it. The delivered orbital energy for SIM10 must be equal to the change in envelope energy between the beginning and end of the simulation. To carry out the comparison we split the envelope gas into the bound and unbound parts:

\begin{equation}
\Delta E_{\rm orb} = (E_{\rm f, bound} + E_{\rm f, unbound}) - E_{i} + E_{\rm non-cons}
\end{equation}

\noindent or, in words, the energy delivered to the envelope by the in-spiralling masses ($\Delta E_{\rm orb} = \simeq 1.89 \times 10^{47}$~erg; data provided in Table~\ref{tab:simulation_parameters}) must be the same as the final total energy of the bound gas ($E_{\rm f, bound}$=$-8.51 \times 10^{46}$~erg; Table~4) {\it plus} the final energy of the unbound gas ($E_{\rm f, unbound}$), minus the initial energy of the entire envelope gas ($E_i = -1.32 \times 10^{47}$~erg; Table~4), minus whatever energy non conservation has taken place ($E_{\rm non-cons}$ = $1.98 \times 10^{46}$~erg; using the the upper limit percentage of energy non conservation from SIM14, 15 per cent, multiplied by the initial envelope binding energy of SIM10). The final energy of the bound part of the envelope is easy to measure, because most of the bound mass is still inside the simulation box. What is impossible to calculate is the final energy of the unbound material because this material is almost entirely outside the box. 
Using the equation above, we know the value of the total energy of the unbound material must be $1.22 \times 10^{47}$~erg.

If we approximate the final total energy of the unbound material by its kinetic energy, we can calculate the mass-averaged velocity of the unbound material (unbound mass is 0.16~\ms; Table~1)\footnote{The potential energy of the unbound material at the end of the simulation should be very close to zero, as the material should find itself at quite a large distance from the centre of the system. If the thermal component of the total energy is significant our calculation would result in an upper limit to the determined ejection velocity.}. This value is $\sim$277~km~s$^{-1}$, which is approximately the velocity of the bulk of the unbound gas as it leaves the computational domain. In conclusion, the orbital energy is delivered inefficiently: it unbinds material, by accelerating it to velocities well above the escape velocity and it also increases the energy of the remaining gas but fails to unbind it.
\section{Summary and conclusions}
\label{sec:conclusions}

In this paper we have continued the systematic investigation  (P12, \citealt{Staff2016a}, \citealt{Staff2016b}, 
and \citealt{Iaconi2017}) of the reliability of numerical simulations of the CE interaction by focussing on the effect of the envelope binding energy, and on resolution. We have performed two sets of simulations with the grid-code {\sc Enzo} using AMR and a new gravity solver. The first set is similar to the simulations carried by P12 with {\sc enzo} in unigrid mode, while the second set models a more massive and compact primary, but with the same core mass as used by P12 and the same companions. 

The new AMR simulations (SIM1-5) mimic those carried out by P12, but result in smaller final orbital separations, as well as in a stronger interaction causing a steeper fast in-spiral (Section~\ref{ssec:final_separation}). We believe that these differences are due to the higher resolution achieved near the core and near the companion by the AMR technique. 
In particular, at higher resolution the giant's stellar envelope is more bound because the density profiles are better reproduced near the core where the density is high. The trend whereby lighter companions in-spiral deeper but are less effective in unbinding the envelope is observed here as was in P12.

The simulations with a more massive envelope (SIM6-10) result in systematically smaller final separations. However, all final separations, irrespective of the companion mass, are similar to the value of the smoothing length, which is a numerical artefact, indicating that the final separations could be smaller. Indeed, by increasing the resolution for the simulations with a 0.6~\ms\ companion leads to a smaller final separation also of the order of the smoothing length (Section~\ref{ssec:final_separation}). Even increasing the values of the final separations to account by the 20\%-level angular momentum loss, leaves us with small final separations that are close to the value of the smoothing length.  We conclude that there is a chance that common envelope interactions between a 2-\ms\ RGB star and a companion as heavy as 0.6~\ms, may still result in a merger. 

None of our simulations unbinds the envelope, as we have now come to expect of simulations that do not include recombination energy. We have estimated that even the addition of recombination energy will not lead to unbinding the envelope in the ``heavy star" simulations (SIM6-10), though it might in the case of the ``light star" simulations (SIM1-5). The timing and location of the injection of recombination energy as well as how much of it can be used to do work (as opposed to being radiated away) will change the picture of how the envelope becomes unbound. However, work carried out thus far \citep{Nandez2016,Ivanova2016} shows that the morphology of the fast in-spiral is not dramatically altered by the ejection of the envelope at the hand of recombination energy. The ejection happens primarily after the in-spiral has slowed down.

In a follow up paper (Iaconi \& De~Marco, in preparation) we will carry out a more in depth analysis of the simulations and of the post-CE observations available in order to determine what constraints, if any, can current observations provide for the simulations.
\section*{Acknowledgments}
\label{sec:acknowledgments}
RI is grateful for financial support provided by the International Macquarie University Research Excellence Scholarship. RI would also like to thank Francesco D'Eugenio for the useful discussion on angular momentum conservation and Birendra Pandey for his essential suggestions. OD gratefully acknowledges support from the Australian Research Council Future Fellowship grant FT120100452. JES acknowledges support from Australian Research Council Discovery Project DP12013337 and NASA grant NNX15AP95A.This research was undertaken, in part, on the National Computational Infrastructure facility in Canberra, Australia, which is supported by the Australian Commonwealth Government and on the swinSTAR supercomputer at Swinburne University of Technology. Computations described in this work were performed using the ENZO code (http://enzo-project.org), which is the product of a collaborative effort of scientists at many universities and national laboratories. Finally, we thank two anonymous referees for thorough reports. All simulation outputs are available upon request by e-mailing orsola.demarco@mq.edu.au.
\bibliographystyle{aa}
\bibliography{bibliography}{}
\appendix
\section{Numerical considerations}
\label{sec:numerical_caveats}
In this Appendix we give further thought to two aspects of our simulations. The first concerns the use of a high pressure, low-density, hot {\it vacuum} surrounding the star at time zero. The second concerns a distinction that should be made when discussing resolution testing.

\subsection{The artificial, hot {\it vacuum}}
\label{ssec:vacuum}
Grid simulations (e.g., Sandquist et al. 1998) need to use a very low density, low mass medium around the star so as to maintain the outer layers stable. To do so, the medium has to match the stellar pressure at the surface, and because of the low density, the temperature of the medium must be large.  The question is whether this hot {\it vacuum} affects the dynamics of the simulation (in-spiral pattern, timescale and unbinding of the envelope gas). As we argue below the answer to this question is that the impact of the {\it vacuum} is negligible. 

The first argument is that every time a comparison has been carried out between grid (with {\it vacuum}) and SPH (without) simulations the results have been very comparable (see Passy et al. 2012 and Iaconi et al. 2017). The second argument is that the shapes of the in-spiral, final separations and the unbound masses are very comparable between SIM4 and the most equivalent, large-domain simulations (SIM14 and SIM15), despite a far larger amount of hot vacuum in the latter two. Below we explain why this similarity comes about.

The last column in Table~\ref{tab:energy_initial_parameters} lists the mass of {\it vacuum} that fills the computational domain outside the primary, its mass relative to the primary's mass and the maximum work required to push the {\it vacuum} out of the computational domain. For similar giants at different resolutions its mass only depends on the size of the computational domain (see, e.g., SIM6-SIM10, SIM11, SIM12) and it is negligible for all simulations except the ones carried out with a large box (SIM13, SIM14 and SIM15), where the mass of the {\it vacuum} accounts for $\simeq 10\%$ of the primary's mass. 

To estimate how much the {\it vacuum} can dynamically affect the expansion of the envelope during the rapid in-spiral phase, we evaluate the maximum amount of work necessary to push the {\it vacuum} out of the computational domain. From Table~\ref{tab:energy_initial_parameters} we see that the potential energy of the {\it vacuum} is negligible compared to its thermal energy, so we have ignored the gravity pull on the {\it vacuum} in our estimate. The {\it vacuum} has constant density and thermal energy, in addition it resides entirely on the coarser AMR level (i.e., cell size is constant). Hence using the ideal gas equation of state, an approximate expression for an upper limit to the total work necessary to remove this gas ($W_{\mathrm{vac}}$) is:

\begin{equation}
 W_{\mathrm{vac}} = \Big[ \rho_{\mathrm{vac}} (\gamma -1) \frac{E_{\mathrm{therm,vac}}}{M_{\mathrm{vac}}} \Big] (n_{\mathrm{cell,vac}}6l^2) (R_{\mathrm{box}} - R_1) \ ,
 \label{eq:vacuum_total_force}
\end{equation} 

\noindent where $\rho_{\mathrm{vac}}$ is the {\it vacuum} density, $\gamma$ is the adiabatic index of $5/3$, $E_{\mathrm{therm,vac}}$ is the total thermal energy of the {\it vacuum}, $M_{\mathrm{vac}}$ is the total mass of the {\it vacuum}, $n_{\mathrm{cell,vac}}$ is the number of cells occupied by the {\it vacuum}, $l$ is the cell linear length ($6l^2$ is therefore the surface of a single cell) and $R_{\mathrm{box}} - R_1$ is the difference between the diagonal of the box and the radius of the primary. 
In other words, to obtain the work done, we multiply the pressure exerted by the {\it vacuum}, calculated via the equation of state, by the total surface area of all cells occupied by the {\it vacuum} and then we multiply this force by the maximum displacement over which the vacuum has to be move in order to leave the box.
If we apply this formula to our set of simulations using $\rho_{\mathrm{vac}} \simeq 10^{-11}$~g~cm$^{-3}$ for our 1~\ms \ giant (SIM1-SIM5, SIM13-SIM15) and $\rho_{\mathrm{vac}} \simeq 10^{-12}$~g~cm$^{-3}$ for the 2~\ms \ ones (SIM6-SIM10, SIM11, SIM12) we obtain the values listed in Table~\ref{tab:energy_initial_parameters}. The work necessary to push the {\it vacuum} out of the computational domain is therefore several orders of magnitudes smaller than the typical energy budgets of the primaries used for all the simulations, albeit higher for SIM13-SIM15. 

In all simulations, but more so in SIM13-SIM15 we additionally see over time that the lowest density layers ejected during the rapid in-spiral are warmed up by the {\it vacuum}. However, the mass of these layers is very low. This effect is fully analysed for a simulation identical to SIM4 by \citet{Galaviz2017}, who study the optical properties of the CE.

An analysis of the energetics of the envelope and of the {\it vacuum} at the end of the rapid in-spiral is impossible because of the mixing between envelope and {\it vacuum} gas over time that prevents us from separating the envelope gas from the {\it vacuum}  after the initial configuration. Additionally, the gas leaving the computational domain is lost and only approximate quantities can be reconstructed, thus preventing a full quantification of the {\it vacuum} at later times. 

We conclude that the {\it vacuum} does not alter the results beyond the precision required for this analysis. This said, we note that we do not use the simulations carried out with a large domain, where the {\it vacuum} accounts for 10\% of the total mass in the simulation, except to consider energy conservation aspects of the calculation.

\begin{table*}
\begin{center}
\begin{adjustbox}{max width=\textwidth}
\begin{tabular}{ccccccccc}
\hline
ID                    & $E_{\mathrm{pot,1}}$   & $E_{\mathrm{therm,1}}$  & $E_{\mathrm{pot,vac}}$ & $E_{\mathrm{therm,vac}}$ & $M_{\mathrm{vac}}$ & $M_{\mathrm{vac}}/M_1$ & $W_{\mathrm{vac}}$\\
                      & (erg)                  & (erg)                   & (erg)                     & (erg)                       & 
(\ms)                 &                       & (erg)\\
\hline                
1~\ms (1D {\sc evol}) & $-2.53 \times 10^{49}$ & $1.30  \times 10^{49}$  & -- & --  &-- & -- & --\\
2~\ms (1D {\sc mesa}) & $-2.37 \times 10^{49}$ & $1.22  \times 10^{49}$  & -- & -- &-- & --& --\\                                                                           
\hline                
SIM1-5             & $-4.45 \times 10^{46}$ & $2.14  \times 10^{46}$  & $-1.45 \times 10^{42}$    & $4.31 \times 10^{45}$       & 
$1.02 \times 10^{-4}$ & $10^{-4}$                & $7.20  \times 10^{41}$\\
\hline                
SIM6-10            & $-2.63 \times 10^{47}$ & $1.31 \times 10^{47}$   & $-4.72 \times 10^{41}$    & $7.77 \times 10^{44}$       & $2.92 \times 10^{-5}$ & $10^{-5}$                     & $4.89  \times 10^{40}$\\
\hline                
SIM11           & $-2.73 \times 10^{47}$ & $1.35  \times 10^{47}$  & $-6.99 \times 10^{41}$    & $7.77 \times 10^{44}$       & 
$2.92 \times 10^{-5}$ & $10^{-5}$                     & $4.89  \times 10^{40}$\\
SIM12           & $-3.02 \times 10^{47}$ & $1.40  \times 10^{47}$  & $-6.99 \times 10^{41}$    & $7.77 \times 10^{44}$       &
$2.92 \times 10^{-5}$ & $10^{-5}$                     & $4.89  \times 10^{40}$\\
\hline                
SIM13           & $-3.14 \times 10^{46}$ & $1.55  \times 10^{46}$  & $-1.35 \times 10^{44}$    & $4.44 \times 10^{48}$       & 
$9.40 \times 10^{-2}$ & 0.1                        & $1.08  \times 10^{45}$\\
SIM14           & $-4.17 \times 10^{46}$ & $1.93  \times 10^{46}$  & $-1.33 \times 10^{44}$    & $4.44 \times 10^{48}$       & 
$9.40 \times 10^{-2}$ & 0.1                        & $1.08  \times 10^{45}$\\
SIM15           & $-4.72 \times 10^{46}$ & $2.23  \times 10^{46}$  & $-1.33 \times 10^{44}$    & $4.44 \times 10^{48}$       & 
$9.40 \times 10^{-2}$ & 0.1                        & $1.08  \times 10^{45}$\\
\hline
\end{tabular}
\end{adjustbox}
\end{center}
 \begin{quote}
  \caption{\protect\footnotesize{Energies of the primary star and of the {\it vacuum} at the beginning of the simulation (the kinetic energies are at this time all zero). The 1D stellar evolution models are followed by the same models mapped into the {\sc Enzo} computational domain at different resolutions. 
  $E_{\mathrm{pot,1}}$ is the primary's potential energy, $E_{\mathrm{therm,1}}$ is the primary's internal energy, $E_{\mathrm{pot,vac}}$ is the potential energy of the {\it vacuum} and $E_{\mathrm{therm,vac}}$ is the thermal energy of the {\it vacuum}. We additionally report the mass of the {\it vacuum} ($M_{\mathrm{vac}}$), the mass ratio vacuum/primary ($M_{\mathrm{vac}}/M_1$) and the maximum work needed to push the {\it vacuum} out of the computational domain ( $W_{\mathrm{vac}}$).}} \label{tab:energy_initial_parameters}
 \end{quote}
\end{table*}

\subsection{Resolution testing}
\label{ssec:resolution_testing}
When carrying out resolution studies we check that different resolutions do not affect the simulations' outcomes and the results we draw from them. The reasons why resolution affects the simulations are many and include the fact that the initial setup is different at different resolutions. For example, the 1D stellar models, calculated with large resolution in 1D are poorly reproduced by 3D mapping, worse so at lower resolutions. 

In  Table~\ref{tab:energy_initial_parameters} we list the energetics of the 1D {\sc evol} and {\sc mesa} models (note that we did not report the values of kinetic energy because they are negligible) and those of the stars as mapped in the {\sc Enzo} computational domain at various resolutions. As is also obvious from in Figure~\ref{fig:star_profiles}, where we plotted density as a function of radius for all the initial stellar models and their 1D counterparts, resolution affects particularly the inner part of the star, where most of the mass resides. Even the highest 3D resolution is not able reproduce the core of the star (Figure~\ref{fig:star_profiles}), which we replace with a point-mass interacting only gravitationally with the rest of the gas. For the three resolutions of SIM6-10, SIM11 and SIM12, the stellar mass, once mapped into the computational domain is larger for the highest resolution. As a result the core particle has a lower mass (so as to conserve total mass). However this difference is not very substantial, with core masses for the three resolutions at 0.3887, 0.3872 and 0.3852~\ms, respectively. 

The gas near the stellar core particle holds a substantial part of the thermal and potential energies and even ensuring the same total mass, the difference in the gas distribution near the core for different resolutions, induces a difference in initial conditions. This leads to differences listed in Table~\ref{tab:energy_initial_parameters} which are 14 and 7 per cent, for potential and thermal energies, respectively, between lowest and highest resolution.

As a result of this effect, the binding energy of the primary star increases with resolution, making, in principle, more resolved primaries  harder to unbind. However, this does not affect our simulations, because all the unbinding takes place at the beginning of the rapid in-spiral and is in the outer layers of the envelope (\citet{Iaconi2017}, Section~\ref{ssec:unbound_mass}). Therefore, this gas is minimally affected by the increase in binding energy and will require a similar amount of work to be unbound. In other words, ejecting the same amount of mass in two simulations with the same star at different resolutions takes similar amounts of energy.

Indeed from Table~\ref{tab:simulation_parameters} we see that for SIM9, SIM11 and SIM12 there is only a small increase of unbound mass with increasing resolution. The small increase is caused by higher resolution around the companion that results in a more realistic, stronger gravitational drag, but has nothing to do with the larger binding energy of the giant. If recombination energy were included in these simulations and more unbinding were to take place as a result, it is possible that the larger binding energy of more resolved stars would impact the unbound mass.

This said, it is fair to recognise that testing resolution in the way we have done is not the same as a convergence test. Rather increasing the resolution is a way to trigger a series of inter-related effects, which can then be studied in order to understand the reliability of the simulations.

\bsp
\label{lastpage}
\end{document}